\documentclass{aa}  

\usepackage{graphicx}
\usepackage{txfonts}
\usepackage{dsfont}
\usepackage{xcolor}
\usepackage{subcaption}

\usepackage[colorlinks = true,citecolor = blue, linkcolor = blue, urlcolor = blue]{hyperref}

\begin{document}

   \title{YOLO-CIANNA: Galaxy detection with deep learning in radio data}

   \subtitle{I. A new YOLO-inspired source detection method applied to the SKAO SDC1}
   
   \titlerunning{YOLO-CIANNA: Galaxy detection with deep learning in radio data}
   
   \author{D. Cornu
          \inst{1}, P. Salomé \inst{1}, B. Semelin \inst{1}, A. Marchal \inst{2,3}, J. Freundlich \inst{4},  
          S. Aicardi \inst{5}, \break X. Lu \inst{6}, G. Sainton \inst{1}, F. Mertens \inst{1}, F. Combes \inst{1,7}, C. Tasse \inst{8,9}
          }
    \authorrunning{Cornu et al.}

   \institute{$^{1}$LERMA, Observatoire de Paris, Université PSL, Sorbonne Université, CNRS, 75014, Paris, France\\
$^{2}$Canadian Institute for Theoretical Astrophysics, University of Toronto, 60 St. George Street, Toronto, ON M5S 3H8 \\
$^{3}$Research School of Astronomy \& Astrophysics, Australian National 
University, Canberra ACT 2610 Australia \\
$^{4}$Université de Strasbourg, CNRS UMR 7550, Observatoire astronomique de Strasbourg, 67000 Strasbourg, France\\
$^{5}$DIO, Observatoire de Paris, CNRS, PSL, 75014, Paris, France\\
$^{6}$IDRIS, CNRS, F-91403 Orsay, France\\
$^{7}$Collège de France, 11 Place Marcelin Berthelot, 75005, Paris, France\\
$^{8}$GEPI, Observatoire de Paris, CNRS, Université Paris Diderot, 5 Place Jules Janssen, 92190, Meudon, France\\
$^{9}$Department of Physics \& Electronics, Rhodes University, PO Box 94, Grahamstown, 6140, South Africa\\
}

   \date{Received February 8, 2024; accepted August 19, 2024}

 
  \abstract
   {The upcoming Square Kilometer Array (SKA) will set a new standard regarding data volume generated by an astronomical instrument, which is likely to challenge widely adopted data-analysis tools that scale inadequately with the data size.}
   {\textcolor{black}{The aim of this study is to develop a new source detection and characterization method for massive radio astronomical datasets based on modern deep-learning object detection techniques. For this, we seek to identify the specific strengths and weaknesses of this type of approach when applied to astronomical data.}}
   {\textcolor{black}{We introduce YOLO-CIANNA, a highly customized deep-learning object detector designed specifically for astronomical datasets. In this paper, we present the method and describe all the elements introduced to address the specific challenges of radio astronomical images.} We then demonstrate the capabilities of this method by applying it to simulated 2D continuum images from the SKA observatory Science Data Challenge 1 (SDC1) dataset.}
    {\textcolor{black}{Using the SDC1 metric, we improve the challenge-winning score by +139\% and the score of the only other post-challenge participation by +61\%. Our catalog has a detection purity of 94\% while detecting 40 to 60 \% more sources than previous top-score results, and exhibits strong characterization accuracy. The trained model can also be forced to reach 99\% purity in post-process and still detect 10 to 30\% more sources than the other top-score methods. It is also computationally efficient, with a peak prediction speed of 500 images of $512{\times}512$ pixels per second on a single GPU.}}
   {YOLO-CIANNA achieves state-of-the-art detection and characterization results on the simulated SDC1 dataset and is expected to transfer well to observational data from SKA precursors.}


   \maketitle
%
\section{Introduction}
\label{sec:introduction}

Modern astronomical instruments generate ever-increasing data volumes in response to the need for better resolution, sensitivity, and wider wavelength intervals. Astronomical datasets are often highly dimensional and require precise encoding of the measurements due to a high dynamic range. Also, it is often necessary to preserve the raw data so they can be re-analyzed with new versions of the processing pipelines. Radio-astronomy is strongly affected by this explosion of data volumes, especially regarding giant radio interferometers. The upcoming Square Kilometer Array \citep[SKA, ][]{paper:ska_ref} is expected to have an unprecedented real-time data-production rate, and will provide 700 PB of archived data per year \citep{paper:scaife_ska}. This instrument is foreseen to have the necessary sensitivity to set constraints on the cosmic dawn and the epoch of reionization and to trace the evolution of astronomical objects over cosmological times. Faced with data of such volume and complexity, some classical analysis methods and tools employed in radio astronomy will struggle to keep up due to their limited scalability.

In this context, the SKA Observatory (SKAO) began organizing recurrent Science Data Challenges (SDCs) in order to gather astronomers from the international community around simulated datasets that resemble future SKA data products. The objective is to evaluate the suitability of existing analysis methods and encourage the development of new ones. It is also an opportunity for astronomers to become familiar with the nature of such datasets and to gain experience in their exploration.

\begin{figure}
    \centering
    \includegraphics[width=1.0\hsize]{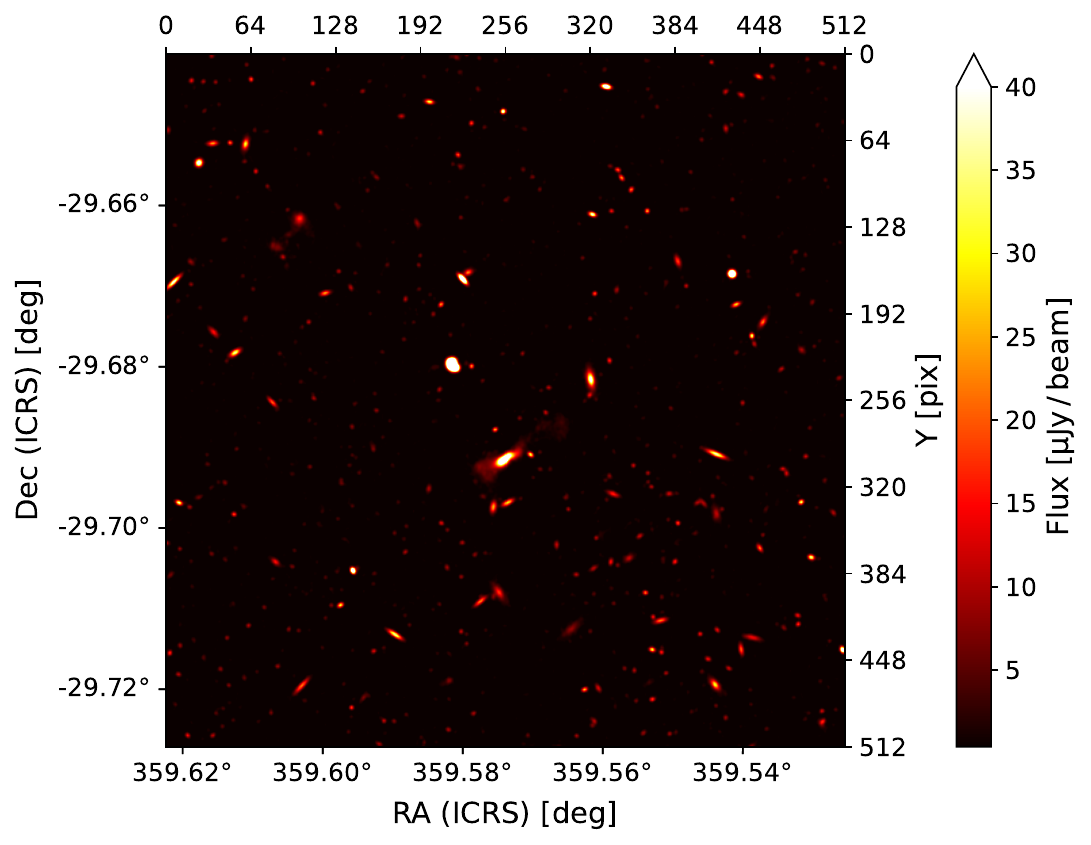}
    \caption{Cutout of 512 square pixels in the SDC1 560 MHz 1000h simulated field. Minimum and maximum cutting values are those used for our object detector, and the displayed intensity is the raw image flux.}
    \label{fig:example_field}
\end{figure}

The first edition, SDC1 \citep{paper:sdc1}, focused on a source detection and characterization task in simulated continuum radio images at different frequencies and integration times. Figure~\ref{fig:example_field} shows a cutout from one of the SDC1 images, illustrating the source density and high dynamic range. Source-finding is a common task in astronomy and is often the first analysis carried out on a newly acquired image product; it is already performed by a variety of classical methods, \textcolor{black}{such as Source-Extractor \citep{paper:sextractor}, SFIND \citep{paper:sfind}, CUTEX \citep{paper:cutex}, BLOBCAT \citep{paper:blobcat}, DUCHAMP \citep{paper:duchamp}, SELAVY \citep{paper:selavy},  AEGEAN \citep{paper:aegean}, PyBDSF \citep{soft:pybdsf}, PROFOUND \citep{paper:profound}, PySE \citep{paper:pyse}, CAESAR \citep{paper:caesar}, and CERES \citep{paper:ceres}.} The obtained source catalogs can then be augmented with characterization information and used as primary data for subsequent studies. This task is especially affected by increases in volume and dimensionality, making it a good probe of the upcoming data-handling challenges.

The past decade has seen a rapid increase in the use of machine learning (ML) methods in all fields, including astronomy and astrophysics \citep{paper:astro_review_deepl}. One of the advantages of ML methods is their superior scaling with data size and dimensionality. There are a considerable variety of ML approaches, but we focus here on methods based on deep artificial neural networks \citep{paper:deep_learning}. These approaches have been extensively used for computer vision tasks, including object detection in everyday-life images \citep{paper:imagenet, paper:pascal_voc, paper:coco_2014}. While detection models have been used in other domains for several years, they are not yet widely adopted in the astronomical community. 

Deep-learning object-detection methods are usually separated into three families \citep{paper:object_detection_review}: segmentation models, region-based detectors, and regression-based detectors. The main advantage of the segmentation models is their ability to identify all the pixels that belong to a given class or even individual objects. They can also be used as a convenient structure for denoising tasks. Their main drawback is their symmetric structure (encoder and decoder) and the high level of expressivity required at near-input resolution, making them computationally intensive for high-resolution images. This family is mainly represented by the U-Net \citep{paper:u-net} method. Given their proximity with classical source detection approaches, they have been employed for a variety of astronomical applications \citep[e.g.,][]{paper:akeret_2017, paper:deep_source, paper:convo_source, paper:maximask, paper:bianco, paper:makinen, paper:deepl_source_finding, paper:forska_sdc2}. 

The second family, the region-based detectors, mostly comprise multi-stage neural networks that split the detection task into a region-proposal step and a detection-refinement step. They are the most popular method for mission-critical tasks due to their accuracy. While faster than segmentation methods, these high-detection-accuracy models are computationally intensive due to the multi-stage process. This family is mainly represented by the R-CNN method \citep{paper:rcnn} and all its derivatives (e.g., Fast R-CNN, Faster R-CNN). Examples of astronomical applications with these methods are more limited \citep[e.g.,][]{paper:claran, paper:jia_rcnn, paper:hetu, paper:jlrat2, paper:deepl_source_finding}. There is also a subfamily that combines the region-based detection formalism with a mask prediction used to perform instance segmentation. This subfamily is mainly represented by the Mask R-CNN method \citep{paper:mask_rcnn}, which is increasingly used in astronomy \citep[e.g.,][]{paper:astro_rcnn, paper:mask_galaxy, paper:riggi_mask_rcnn,paper:deepl_source_finding}. We note that region-based methods are commonly combined with some flavor of pyramidal feature hierarchy construction \citep{paper:fpn_det}, which helps represent multiple scales in the detection task. 

The last family, the regression-based detectors, are often based on single-stage neural networks, making them computationally efficient. Therefore, they are frequently used for real-time object detection. The most popular regression-based detector method is the You Only Look Once (YOLO) method and its descendants \citep{paper:yolo_v1, paper:yolo_v2, paper:yolo_v3}, but we can also cite the Single Shot Detector (SSD) method \citep{paper:ssd}. There have only been a few astronomical applications of regression-based detectors, mainly in the visible domain \citep{paper:yolo_sdss, paper:yolo_sdss2, paper:yolo_fr1_fr2, paper:yolo_cl, paper:yolo_lsb}. 

We highlight that methods based on transformers \citep{paper:transformer} are now common in computer vision \citep{paper:detr}, and astronomical applications are just starting to be published \citep{paper:gal_detr, paper:astro_yolo_detr}. We also note that some methods include deep learning parts in more classical source detection tools, which can improve the detection purity and the source characterization \citep[e.g.,][]{paper:lisa}. Further references regarding deep learning methods for source detection can be found in \citet{paper:deepl_source_finding} and \citet{paper:deepl_det_astro_review_2}.

\textcolor{black}{The present paper is the first of a series, the aim of which is to present a new source detection and characterization method called YOLO-CIANNA. It was developed and used in the context of the MINERVA (MachINe lEarning for Radioastronomy at Observatoire de Paris) team participation in the SDC2 \citep{paper:sdc2}, where we obtained first place. The primary objective of this first paper is to provide a complete description of the method and to justify several design choices regarding the specific properties of astronomical images. We then illustrate its capability by presenting the results of its application to simulated 2D continuum images from the SDC1 dataset. A second paper will present an application to simulated 3D cubes of HI emission using the SDC2 dataset. The series will then continue by applying the method to observational data from surveys of several SKA precursors.}

\textcolor{black}{The present paper is organized as follows: 
\begin{itemize}
\item In Sect.~\ref{sec:method}, we present the YOLO-CIANNA method in a complete and comprehensive manner, targeting readers unfamiliar with deep-learning object detectors. Details regarding the most complex aspects of the method and the in-depth differences from a classical YOLO implementation are given in Appendix~\ref{sec:appendix:in_depth_method} along with details adapted to readers familiar with these families of methods.
\item In Sect.~\ref{sec:dataset_and_training}, we present the SDC1 dataset, which is composed of comprehensive 2D images, and describe how we used it as a benchmark to evaluate the detection and characterization capabilities of  YOLO-CIANNA. 
\item In Sect.~\ref{sec:results}, we present the detection results of our method, as well as a detailed analysis of the detection catalog we obtained from the SDC1.
\item In Sect.~\ref{sec:discussion}, we use these results to highlight the strengths and weaknesses of our source-detector and also discuss the impact of some design choices of the SDC1. We then elaborate on how this new approach could be applied to real observational data from SKA precursor instruments.
\item There are four Appendices. In Appendix~\ref{sec:appendix:in_depth_method}, we detail the differences between our method and the classical YOLO implementation, as well as benchmarks over classical computer vision datasets. In Appendix~\ref{sec:appendix:yolo_v2_arch}, we evaluate the performance of a classical YOLO backbone architecture on the SDC1 and compare it to our custom backbone. In Appendix~\ref{sec:appendix:training_area}, we present an alternative training area definition for the SDC1. In Appendix~\ref{sec:appendix:characterization_impact}, we evaluate whether performing the detection and characterization using a single unified network is beneficial or detrimental to detection-only performance. 
\end{itemize}
}

\section{Method}
\label{sec:method}

\textcolor{black}{Our method finds its inspiration in the classical YOLO \citep{paper:yolo_v1,paper:yolo_v2,paper:yolo_v3} object detector, but it is also very similar to the SSD method \citep{paper:ssd}, both regression-based deep learning object detector. While region-based approaches like R-CNN \citep{paper:rcnn} are often considered the most accurate object detectors, regression-based methods present a straightforward single network architecture, making them more computationally efficient at a given detection accuracy. Both families can reach state-of-the-art accuracy depending on implementation details and architecture design, but regression methods are usually preferred for real-time detection applications. In this context, our choice to design a regression-based approach was driven by i) fewer implementation constraints, ii) a strong emphasis on computational performance considering the data volume of current and future radio-astronomical surveys, and iii) the single-stage regression-based network structure on top of which it is easier to add more predictive capabilities.}

\textcolor{black}{
In this section, we present the main design of our method. Despite being depicted for an astronomical application, our method remains suitable for general-purpose object detection (Appendix~\ref{sec:appendix:yolo_cianna_benchmarks}). Instead of listing all the subtle differences with various descendant versions of the classical YOLO approach or other detectors, we describe the main components of our method from scratch in a comprehensive manner. The description of the most technical parts of the method is done in the dedicated Appendix~\ref{sec:appendix:in_depth_method}, which also discusses the differences with the classical YOLO approach. As a result, the main method description should remain accessible to readers unfamiliar with deep-learning object detectors. Even though our method differs significantly from the YOLO algorithm in some critical aspects, we chose to refer to our approach as YOLO-CIANNA for simplicity and as a legacy for its inspiration.}

\textcolor{black}{
The implementation was made inside the custom high-performance deep learning framework CIANNA \citep{soft:cianna}\footnote{CIANNA is open source and freely accessible through GitHub \href{https://github.com/Deyht/CIANNA}{https://github.com/Deyht/CIANNA}. The version used in this paper corresponds to the V-1.0 release \href{https://doi.org/10.5281/zenodo.12806325}{10.5281/zenodo.12806325}}. The implementation and usage details can be found on the CIANNA \href{https://github.com/Deyht/CIANNA/wiki}{wiki pages}. For reproducibility purposes, we provide example scripts for training and applying the method to the SDC1 dataset in the \href{https://github.com/Deyht/CIANNA}{CIANNA} git repository.}

To ease the understanding of the technical parts of the paper, we list a few ML-specific terms and the descriptions we have for them. The most common ML terms are not redefined as they can be found in any ML textbook or review \citep{paper:deep_learning}.
\begin{itemize}
\item {Bounding box:} in classical computer vision, the smallest rectangular box that includes all the visible pixels belonging to a specific object in a given image.
\item {Expressivity:} refers to the predictive strength of a network. The higher the expressivity, the more complex or diverse the predictions can be. The expressivity increases with the number of weights and layers in a network.
\item {Receptive field:} corresponds to all the input pixels that can contribute to the activation of a neuron at a specific point in the network. It represents the maximum size of the patterns that can be identified in the input space.
\item {Reduction factor:} the ratio between the input layer spatial dimension and the output layer spatial dimension.
\end{itemize}

\subsection{Bounding boxes for object detection}
\label{sec:bounding_boxes}

Our method uses a fully convolutional neural network (CNN) to construct a mapping from a 2D input image to a regular \textcolor{black}{output grid. Each output grid cell represents a fixed area of the input image with a size that depends on the reduction factor of the chosen CNN backbone. Each grid cell is tasked to detect all possible objects whose center is located inside the input region it represents.} To characterize an object, we rely on the bounding box formalism that encodes an object as a four-dimension vector composed of the box center and its size ($x,y,w,h$). The grid cells are tasked to predict these quantities. As a supervised learning approach, our method relies on a training set composed of images associated with a list of visible objects to be detected. Each object can be encoded as a target bounding box that the CNN is tasked to predict from the raw input image. A training phase is used to optimize the network parameters to minimize a loss function $\mathcal{L}$ that compares the target boxes with the predicted boxes from all grid cells at the current step. This loss should encompass all the object properties to be predicted. To ease the method description, we first write an abstract loss as
\begin{align}
    \mathcal{L} = \mathcal{L}_{\textrm{pos}} + \mathcal{L}_{\textrm{size}} + \mathcal{L}_{\textrm{prob}} + \mathcal{L}_{\textrm{obj}} +  \mathcal{L}_{\textrm{class}} + \mathcal{L}_{\textrm{param}}.
\label{eq:simplified_loss}
\end{align}
The aim of the Sects.~\ref{sec:bounding_boxes} to~\ref{sec:add_pred} is to describe all of the loss subparts. Our complete detailed loss function is presented in Sect.~\ref{sec:method_loss} with Eq.~\ref{eq:method_loss}.

For now, we only describe the case of a single box prediction per grid cell. The more realistic case of multiple objects per grid cell is presented in Sect.~\ref{sec:multiple_boxes}. \textcolor{black}{We consider that the grid is composed of $g_w$ columns and $g_h$ lines and that each grid cell is represented by its coordinate in the grid ($g_x$,$g_y$).} To represent a bounding box, a grid cell predicts a 4-element vector ($o^x, o^y, o^w, o^h$) that maps to the geometric properties of the box following
\begin{align}
        x & = o^x + g_x, \label{eq:box_coord:1}\\
        y & = o^y + g_y, \label{eq:box_coord:2}\\
        w & = p_w \mathrm{e}^{(o^w)}, \label{eq:box_coord:3}\\
        h & = p_h \mathrm{e}^{(o^h)}. \label{eq:box_coord:4}
\end{align}

\begin{figure}
    \centering
    \includegraphics[width=1.0\hsize]{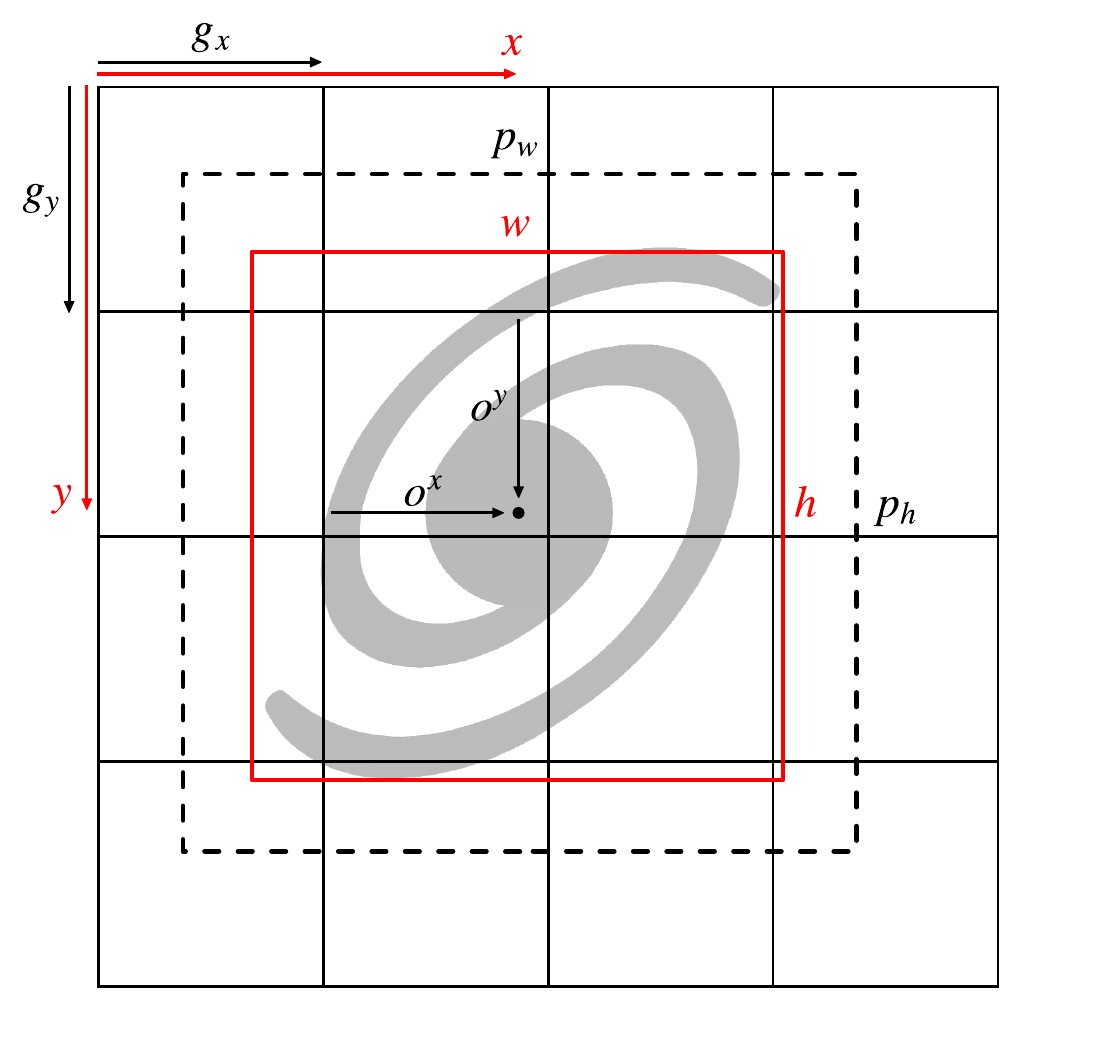}
    \caption{Illustration of the YOLO bounding box \textcolor{black}{representation following Eqs.~\ref{eq:box_coord:1} to~\ref{eq:box_coord:4}}. The dashed black box represents the theoretical prior ($o_w=o_h=0$), while the red box is the scaled predicted size.}
    \label{fig:yolo_grid_illustration}
\end{figure}

Each grid cell is only tasked to position the object center inside its dedicated area using two sigmoid-activated values ($o^x, o^y$). \textcolor{black}{The position of the grid cell in the image ($g_x, g_y$) is added to obtain the relative global position of the object. These coordinates must then be multiplied by the mapped width and height of a grid cell, corresponding to the reduction factor of the backbone network, to obtain pixel coordinates in the input image.} Object size is obtained by an exponential transform of the predicted values ($o^w,o^h$) that acts as a scaling on a predefined size prior ($p_w,p_h$), which can be expressed in pixels directly. This is equivalent to an anchor-box formalism \citep{paper:faster_rcnn} as discussed in Sect.~\ref{sec:multiple_boxes}. The corresponding bounding box construction on the output grid is illustrated in Fig.~\ref{fig:yolo_grid_illustration}.

\textcolor{black}{With this formalism, it is possible to construct a detector with an output of size $\left<g_w, g_h, 4\right>$ that can position and scale one bounding box per output grid cell.} For each prediction-target pair, we use a sum-of-square error to compute the loss function for center coordinates and sizes (Sect.~\ref{sec:method_loss}). The error is not computed on the sigmoid-activated positions but on the raw output for the sizes after target conversion using $\hat{o}^w = \log(w/p_w)$ and $\hat{o}^h = \log(h/p_h)$. This results in the following loss terms
\begin{align}
 \mathcal{L}_{\textrm{pos}}  & =  \sum_{i=0}^{N_g} \mathds{1}_{i}^{\textrm{match}} \left( (o^x_{i} - \hat{o}^x_{i})^2 + (o^y_{i} - \hat{o}^y_{i})^2\right), \label{eq:loss_pos}\\
 \mathcal{L}_{\textrm{size}} & = \sum_{i=0}^{N_g} \mathds{1}_{i}^{\textrm{match}} \left((o^w_{i} - \hat{o}^w_{i})^2 + (o^h_{i} - \hat{o}^h_{i})^2\right),
 \label{eq:loss_size}
\end{align}
where the hat values represent the target for the corresponding predicted value, the sum over $i$ represents all the grid cells with $N_g = g_w {\times} g_h$, and $\mathds{1}_{i}^{\textrm{match}}$ is a mask to identify the predicted boxes that have an associated target box (Sect.~\ref{sec:association_function}). The grid cells that do not contain any object have no contribution to these loss terms. \textcolor{black}{We discuss the possible limitations of using bounding boxes to describe astronomical objects in Sect.~\ref{sec:methode_improvement_disc:method_lim}.}

\textcolor{black}{A common misconception about grided detection is that the predicted size can only be as large as a grid element, which is wrong. The predicted size can be as large as necessary up to the size of the full image. Each grid cell receives information from a large area corresponding to the backbone network receptive field. The receptive fields of nearby grid cells usually overlap, but a target box center can only lie in one grid cell, hence the attribution of the detection to a single grid cell. Due to the fully convolutional structure required for our method, each grid cell performs a localized prediction using identical weights but using a different subpart of the image as input. More details} about the effect of the fully convolutional architecture and the output grid encoding are provided in Appendix~\ref{sec:appendix:conv_net_yolo} and~\ref{sec:appendix:grid_res}.

\subsection{Detection probability and objectness score}
\label{sec:probability_objectness}

\textcolor{black}{Only predicting bounding boxes is insufficient to obtain an object detector. We also need to evaluate the chances that they contain an object}. For this, we add a self-assessed detection probability prediction $P$ to each grid cell, which is constrained during training. This term uses a sigmoid activation and adds a sum-of-square error contribution to the loss. \textcolor{black}{Due to our grid structure, the detector always outputs box properties for every grid cell. In a context where only a few target boxes are present in the image, most grid cells only map irrelevant background regions.} During training, we identify the predicted boxes that best represent each target box and \textcolor{black}{attribute them a target probability of $\hat{P}^{\, \textrm{match}} = 1$. For all the remaining empty predicted boxes, we attribute them a target probability of $\hat{P}^{\, \textrm{void}} = 0$. To compensate for the likely imbalance between the number of matching and empty predictions, we must define a $\lambda_{\textrm{void}}$ factor to apply to the loss term corresponding to the empty case.} \textcolor{black}{The resulting loss term can be written as 
\begin{align}
\begin{split}
\mathcal{L}_{\textrm{prob}} & = \sum_{i=0}^{N_g} \mathds{1}_{i}^{\textrm{match}} \left(P_i - \hat{P}_{i}^{\, \textrm{match}}\right)^2 +  \mathds{1}_{i}^{\textrm{void}} \lambda_{\textrm{void}} \left(P_i - \hat{P}_{i}^{\, \textrm{void}}\right)^2 \\
& = \sum_{i=0}^{N_g} \mathds{1}_{i}^{\textrm{match}} \left(P_i - 1 \right)^2 +  \mathds{1}_{i}^{\textrm{void}} \lambda_{\textrm{void}} \left(P_i - 0 \right)^2,
\label{eq:loss_prob}
\end{split}
\end{align}
where the sum over $i$ represents all the grid cells,} $\mathds{1}_{i}^{\textrm{match}}$ is a mask to identify the predicted boxes that match a target box, and $\mathds{1}_{i}^{\textrm{void}}$ a mask to identify the empty predicted boxes. Through the stochasticity of the training process, the network will learn to predict a continuous probability score that reflects the detection confidence. At prediction time, it is used to identify the grid cells that should contain an object.

\textcolor{black}{One limit of this probability definition is that it contains no information about the quality of the predicted box. For this, we must define a metric that measures the proximity and resemblance between two bounding boxes. The classical object detection metric for this is the intersection over union \citep[IoU,][]{paper:pascal_voc, paper:coco_2014}.} It is defined as the surface area of the intersection between two boxes, A and B, divided by the surface area of their union, which is expressed as
\begin{equation}
{\rm IoU} = \frac{A \cap B}{A \cup B}.
\label{eq:iou}
\end{equation}
This quantity takes values between 0 and 1 depending on the amount of overlap. \textcolor{black}{This classical IoU is the most commonly used in computer vision, but it presents some weaknesses for astronomical applications. We present a few alternative matching metrics better suited to our application case in Appendix~\ref{sec:appendix:association_function:match_metric}. Since several hyperparameters of our method depend on this choice of metric, we will use a generic ${\rm fIoU}$ term} that can be replaced by the selected matching metric in all the following equations. The default choice for our detector is the distance-IoU \citep[DIoU,][]{paper:diou}, as it includes information about the distance between the center of the two boxes to compare. \textcolor{black}{For the unicity of the equations, the selected match metric function is always considered linearly rescaled in the 0 to 1 range.}

From this, we add a self-assessed score called objectness, $O,$  to each predicted box, which is also constrained during training. The objectness is defined as the combination of an object presence probability $P$ and the ${\rm fIoU}$ between the predicted box and the target box, expressed as
\begin{equation}
O = P\times {\rm fIoU}.
\label{eq:objectness}
\end{equation}
This term also uses a sigmoid activation and adds a sum-of-square error contribution to the loss. During training, the objectness is constrained like the probability by considering that $\hat{P}^{\, \textrm{match}} = 1$ for prediction-target matches, while $\hat{P}^{\, \textrm{void}} = 0$ for empty predictions. \textcolor{black}{Therefore, following Eq.~\ref{eq:objectness}, the target objectness for prediction-target matches is $\hat{O}^{\textrm{match}} = {\rm fIoU}$, using the ${\rm fIoU}$ between the target and predicted boxes. For predictions with no associated matches, the target objectness is $\hat{O}^{\textrm{void}} = 0$. The resulting loss term can be written as
\begin{align}
\begin{split}
\mathcal{L}_{\textrm{obj}} & = \sum_{i=0}^{N_g} \mathds{1}_{i}^{\textrm{match}} \left(O_i - \hat{O}_i^{\, \textrm{match}}\right)^2 + \mathds{1}_{i}^{\textrm{void}} \lambda_{\textrm{void}} \left(O_i - \hat{O}_i^{\, \textrm{void}}\right)^2\\
& = \sum_{i=0}^{N_g} \mathds{1}_{i}^{\textrm{match}} \left(O_i - {\rm fIoU}\right)^2 + \mathds{1}_{i}^{\textrm{void}} \lambda_{\textrm{void}} \left(O_i - 0\right)^2,
\label{eq:loss_obj}
\end{split}
\end{align}
using the same notations as Eq.~\ref{eq:loss_prob}.} We stress that ${\rm fIoU}$ is used as a scalar in this equation. Therefore, the derivative of the corresponding matching function is not computed for gradient propagation, so $\mathcal{L}_{\textrm{obj}}$ does not contribute to updating the position and the size. \textcolor{black}{After training, we obtain a continuous objectness prediction representing a global detection score that accounts for the predicted geometrical box quality. Probability and objectness can be used independently or in association to construct advanced prediction filtering conditions (Sect.~\ref{sec:filtering_and_nms}).}

With this formalism, we formulate only two states for a predicted box, either a match or empty. In practice, multiple predicted boxes can try to represent the same target simultaneously. This is common if the target box center is positioned at the edge of a grid cell or if the boxes are large. This will be even more common with multiple detections per grid cell (Sect.~\ref{sec:multiple_boxes}). In such a case, only the best-predicted box will be considered a match. The remaining plausible detections are called good-but-not-best (GBNB) predictions. The previous formalism would result in a loss that lowers the objectness of these GBNB predictions, actively forcing relevant features to fade. To prevent this, we define a representation quality threshold ${\rm fIoU} \geq L^{{\rm fIoU}}_{\textrm{GBNB}}$ above which the corresponding boxes are excluded from both $\mathds{1}_{i}^{\textrm{match}}$ and $\mathds{1}_{i}^{\textrm{void}}$ masks. In summary, there are three types of contribution to the loss: i) the best detection for each target updates its box position and size while increasing its probability and objectness, ii) the background boxes lower their probability and objectness, and iii) the GBNB boxes are ignored.

\subsection{Classification}
\label{sec:classification}

The detected box can be enriched with a classification capability. This can be done by adding $N_c$ components, corresponding to all the possible classes, to the output vector of the detected boxes (Fig.~\ref{fig:yolo_out_vect}). The activation of these components can either be i) a sigmoid for all classes using a sum-of-square error, which allows multi-labeling, or ii) a soft-max activation, which corresponds to exponentiating all the outputs and normalizing them so their sum is equal to 1, with a cross-entropy error. These two options are available in our method. In both cases, only the best detection for each target box updates its classes by comparing the target class vector with the predicted one. There is no contribution to the class loss from either GBNB or background predictions. The resulting loss term for a soft-max activation with a cross-entropy error can be written as
\begin{equation}
\mathcal{L}_{\textrm{class}} = \sum_{i=0}^{N_g} \mathds{1}_{i}^{\textrm{match}} \sum_{k}^{N_C} \left( -\hat{C}_{i}^k\log(C_{i}^k) \right),
\label{eq:loss_class}
\end{equation}
where the sum over $k$ represents all the classes for a given predicted box, and $C_{i}^k)$ is the corresponding class output for the $k$-th class of the predicted box $i$.

\subsection{Additional parameters prediction}
\label{sec:add_pred}

 For astrophysical applications, we often need to predict the characteristics of the sources, such as the flux or some geometric properties not described by a bounding box formalism. For this, we propose to add $N_p$ components to the output vector of the detected boxes, corresponding to all the additional parameters to predict. The activation of these components is linear with a sum-of-square error contribution to the loss. The respective contribution of these parameters to the loss can be scaled with a set of $\gamma^p$ factors. The resulting loss term can be written as
\begin{equation}
\mathcal{L}_{\textrm{param}} = \sum_{i=0}^{N_g} \mathds{1}_{i}^{\textrm{match}} \sum_{k=0}^{N_p} \gamma^k \left( p^k_{i} - \hat{p}^k_{i} \right)^2,
\label{eq:loss_param}
\end{equation}
where the sum over $k$ represents all the independent parameters for a given predicted box, and $p_{i}^k$ is the corresponding parameter output for the $k$-th parameter of the predicted box $i$. This is a strong added value of our YOLO-CIANNA method compared to other approaches, allowing it to predict an arbitrary number of additional properties per detection for any application while preserving the one-stage formalism specific to regression-based object detectors.

\subsection{Multiple boxes per grid-cell and detection unit definition}
\label{sec:multiple_boxes}

\begin{figure*}
    \centering
    \includegraphics[width=0.95\hsize]{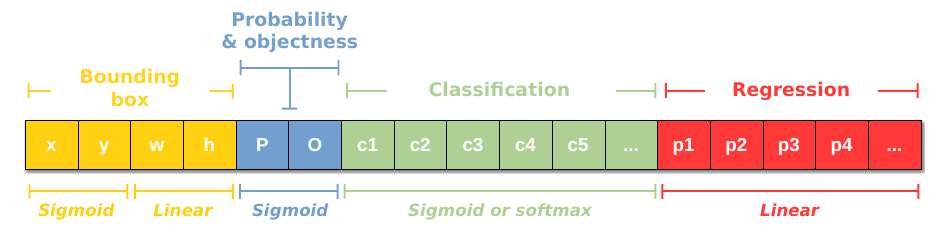}
    \caption{Illustration of the output vector of a single detection unit. The elements are colored according to the corresponding loss subpart. For multiple detection units per grid cell, this vector structure is repeated on the same axis (Sect.~\ref{sec:multiple_boxes}).}
    \label{fig:yolo_out_vect}
\end{figure*}

With the present definition, the detector output would have a shape of $\left< g_w, g_h, (6+N_c+N_p)\right>$, where $g_w$ and $g_h$ are the grid dimensions, the six static parameters are the box coordinates, probability, and objectness ($x,y,w,h,P,O$), $N_c$ is the number of classes, and $N_p$ is the number of additional parameters (Fig.~\ref{fig:yolo_out_vect}). \textcolor{black}{While the geometric and detection score outputs are always predicted, both $N_c$ and $N_p$ are problem-dependent and user-defined.}

\textcolor{black}{In theory, the network reduction factor could be adjusted to the average object density of the application to have no more than one target object in each grid cell. However, for many use cases, it would result in an output grid resolution close to the input resolution, which is very computationally intensive (Appendix~\ref{sec:appendix:grid_res}). To overcome this, we can expand the output vector at each grid cell to contain multiple boxes by stacking their independent vector as a longer 1D vector. The new output shape is then $\left< g_w, g_h, N_b{\times} (6+N_c+N_p)\right>$, with $N_b$ the number of independent boxes predicted at each grid cell, which is a hyperparameter of the detector. For each possible box in a grid cell, we define an individual size-prior ($p_w, p_h$), which impacts the size scaling in Eqs.~\ref{eq:box_coord:3} and~\ref{eq:box_coord:4}. This definition helps to distribute objects over the available boxes on a given grid cell based on box sizes and shapes. The prior list is the same for all grid cells as they only represent different positions at which the same detector is applied through the fully convolutional architecture (Appendix~\ref{sec:appendix:conv_net_yolo}). The size priors are user-defined hyperparameters but are often automatically defined through clustering of the size distribution of targets from the training sample.} In the latter, we refer to the independent predictive elements for a single grid cell as detection units. For example, a detector set capable of predicting up to six independent boxes per grid cell comprises six detection units. Some detection units can have the same size prior, but they still represent independent predictions.

Because scales or shapes are unlikely to be evenly represented in the training sample, the detection units should adapt their positive-to-negative detection ratio to rebalance their probability and objectness distribution (Sect.~\ref{sec:probability_objectness}). For this, a $\lambda_{\textrm{void}}^b$ factor is defined for each detection unit. \textcolor{black}{These factors must be adjusted so the objectness and probability responses are similar enough for all detection units to be comparable during prediction filtering (Sect.~\ref{sec:filtering_and_nms}).}

\subsection{Target-prediction association function}
\label{sec:association_function}

With multiple box predictions per grid cell, deciding which detection unit should be associated with each target box is critical. \textcolor{black}{With our YOLO-CIANNA method, we introduce a prediction-aware association process. Our approach differs significantly from the one used in the classical YOLO formalism, which only uses size priors to define theoretical best matches. Our method is expected to be more robust for images with a high density of small objects, which is typical for astronomical data products. An in-depth discussion about the motivation behind our implementation and a comparison with the classical YOLO association process are both presented in Appendix~\ref{sec:appendix:association_function}.}

The main objective of our association process is to find the best target-prediction pairs regarding a specific ${\rm fIoU}$ matching metric. We start by setting $\mathds{1}^{\textrm{void}} = 1$ for all detection units. Then, we identify all predicted boxes that are a good enough representation of at least one target regarding our ${\rm fIoU} \geq L^{{\rm fIoU}}_{\textrm{GBNB}}$ threshold. This comparison is made for all detection units and all targets regardless of their center position. All objects that respect this criterion are removed from $\mathds{1}^{\textrm{void}}$. The rest of the association algorithm aims to find the best match for each target box through an iterative process. First, matching scores for all possible target-prediction pairs in a grid cell are stored in a scoring matrix. Then, the best current score in the matrix is used to define a new target-prediction pair that is added to the $\mathds{1}^{\textrm{match}}$ mask and removed from $\mathds{1}^{\textrm{void}}$ if it was not already the case. The full row and column corresponding to the target and detection unit of the best match are masked in the score matrix. This search process is repeated until the score matrix is empty or fully masked. A full example of the association a in a grid cell with the evolution of the score matrix is presented in Fig.~\ref{fig:association_ordering}, associated with Appendix~\ref{sec:appendix:association_ordering}. \textcolor{black}{Interestingly, we observed that our approach presents many similarities with the Kuhn–Munkres algorithm \citep{paper:kuhn, paper:munkres} that tackles the problem of optimal score-based association, which was not anticipated.} 

\begin{figure*}
    \centering
    \includegraphics[width=0.98\hsize]{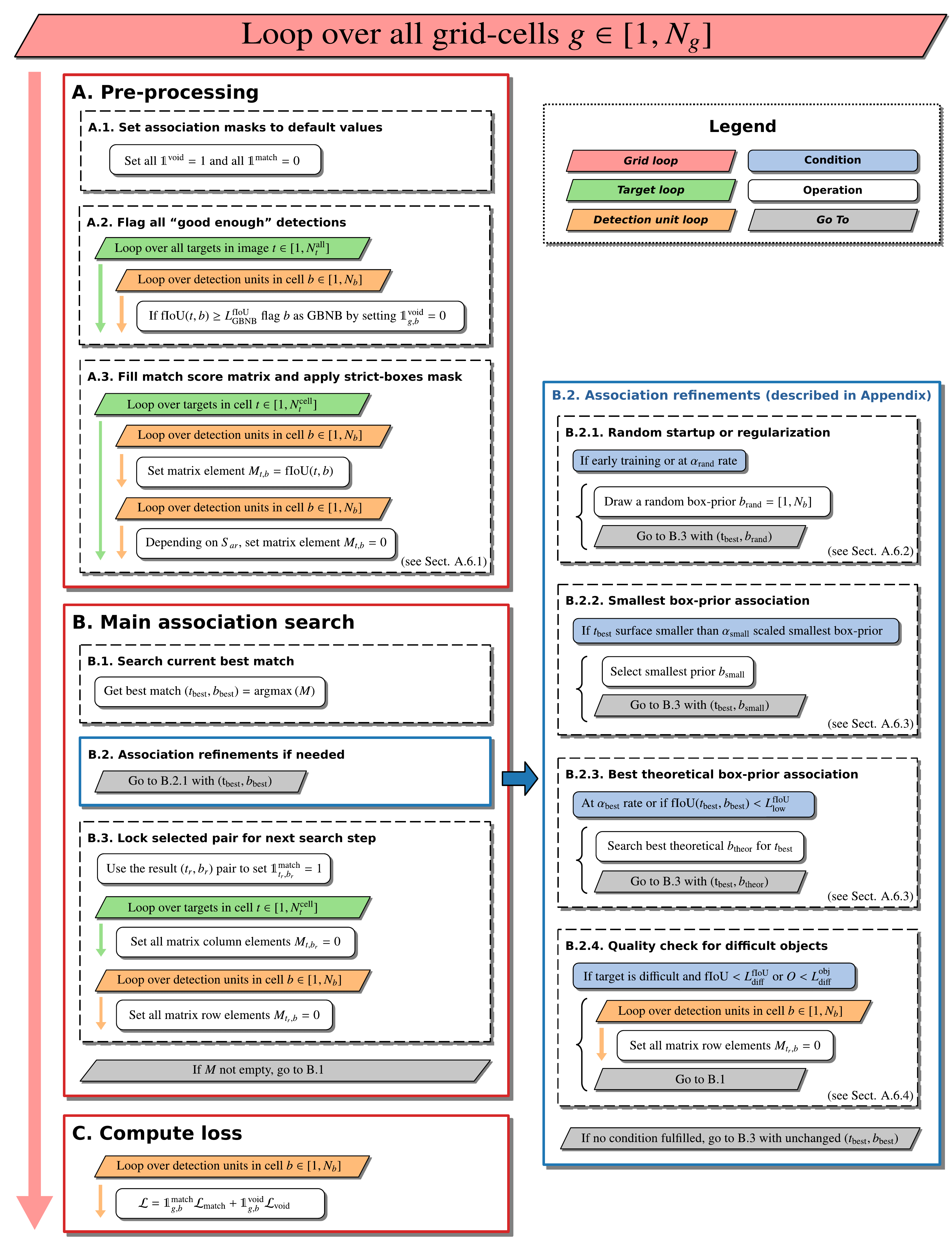}
    \caption{Summary of the target-prediction association algorithm of YOLO-CIANNA. All steps are performed independently by \textcolor{black}{each grid cell. All the elements are executed in order from top to bottom. The A, B, and C blocks represent the general case for the association. All the refinement steps of the association process} corresponding to the B.2 block are described in Sect.~\ref{sec:appendix:association_function:association_tricks}.}
    \label{fig:association_algorithm}
\end{figure*}

As presented in Sect.~\ref{sec:probability_objectness}, the best match associations contribute to all subparts of the loss, while the detection units that remained in $\mathds{1}^{\textrm{void}}$ contribute only to the probability and objectness following Eq.~\ref{eq:loss_obj}. The remaining GBNB detection units not being part of either $\mathds{1}^{\textrm{match}}$ or $\mathds{1}^{\textrm{void}}$ do not contribute to the loss. To account for edge cases, size imbalance, or training instability, we added several specific refinements inside our new association process, which we detail in an appendix section dedicated to advanced subtleties of our method (Appendix~\ref{sec:appendix:association_function:association_tricks}). The global association function algorithm is presented in Fig.~\ref{fig:association_algorithm} in a way that separates the simplified association process and the advanced association with refinements.

\subsection{YOLO-CIANNA complete loss function}
\label{sec:method_loss}

\begin{figure*}
    \centering
    \includegraphics[width=0.95\hsize]{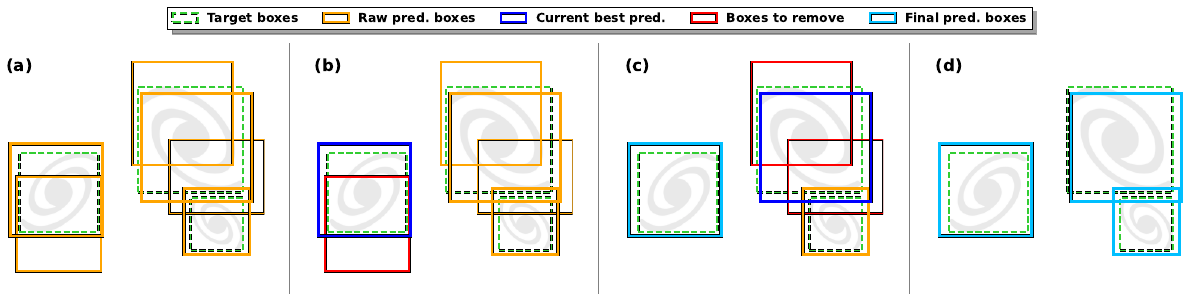}
    \caption{Illustration of the NMS process. Dashed boxes are the targets, and solid boxes are the predictions. The line width of each box is scaled according to its objectness score. The colors indicate the state of the box in the NMS process at different steps. Frame (a) shows the targets and the remaining detector predictions after objectness filtering. Frames (b) and (c) represent two successive steps of the NMS process with a different best current box. Frame (d) shows the selected boxes after the NMS.}
    \label{fig:nms_filtering}
\end{figure*}

Depending on the application, it might be necessary to balance the relative importance of the predicted quantities. {For example, when detecting small objects, the center coordinates become the main estimate of prediction quality, while a high precision of the predicted size becomes mostly irrelevant. Therefore, we use loss scaling factors for} the box position $\lambda_{\textrm{pos}}$, the box size $\lambda_{\textrm{size}}$, the probability $\lambda_{\textrm{prob}}$, the objectness $\lambda_{\textrm{obj}}$, the classes $\lambda_{\textrm{class}}$, and the extra parameters $\lambda_{\textrm{param}}$.

\textcolor{black}{While loss scaling balances the general importance of each predicted quantity, it does not allow the network to guide its expressivity regarding its current prediction quality dynamically. For example, adjusting the predicted class of a detection unit that does not yet properly position or detect the corresponding object is irrelevant and can result in reinforced wrong features or noisy training. Therefore, we chose to add prediction quality limits over some predicted properties directly into the loss function. While the position and the size are always updated for a match, we added quality conditions for the objectness $L^{{\rm fIoU}}_{O}$, the probability $L^{{\rm fIoU}}_{P}$, the classification $L^{{\rm fIoU}}_{C}$, and the extra-parameters $L^{{\rm fIoU}}_{p}$. Each loss subpart is set to zero if the current ${\rm fIoU}$ between the target and predicted boxes is below its specific quality threshold.}

This quality limit principle, combined with other association refinements, results in what we call a ``cascading loss'' that varies during training to guide the network expressivity toward the important aspects, not adjusting currently irrelevant properties. A complete description of this process and the effect it has on training performances and loss monitoring is given in Appendix~\ref{sec:appendix:cascading_loss}.

Combining all the previously introduced loss subparts, conditional masks, and scalings, we can define our complete YOLO-CIANNA loss function for one image as

\begin{align}
\begin{split}
    \mathcal{L} = & \sum_{i=0}^{N_g}\sum_{j=0}^{N_b} \mathds{1}_{ij}^{\textrm{match}}
    \begin{aligned}[t] \Biggl( & \lambda_{\textrm{pos}} && \left[ (o^x_{ij} - \hat{o}^x_{ij})^2 + (o^y_{ij} - \hat{o}^y_{ij})^2 \right] \\
    + & \lambda_{\textrm{size}}  &&\left[ (o^w_{ij} - \hat{o}^w_{ij})^2 + (o^h_{ij} - \hat{o}^h_{ij})^2 \right] \\
    + & \lambda_{\textrm{class}} &&\mathds{1}_{ij}^{C} \sum_{k}^{N_C} \left( -\hat{C}_{ij}^k\log(C_{ij}^k) \right) \\ 
    + & \lambda_{\textrm{param}} &&\mathds{1}_{ij}^{p} \sum_{k}^{N_p} \gamma^k \left( p^k_{ij} - \hat{p}^k_{ij} \right)^2 \\
    + &  \lambda_{\textrm{prob}} &&\mathds{1}_{ij}^{P} \left ( P_{ij} - 1 \right)^2 \\
    + & \lambda_{\textrm{obj}}   &&\mathds{1}_{ij}^{O} \left ( O_{ij} - {\rm fIoU}_{ij} \right)^2 \Biggl) \end{aligned} \\
    + & \sum_{i=0}^{N_g}\sum_{j=0}^{N_b} \mathds{1}_{ij}^{\textrm{void}} \lambda_{\textrm{void}}^{j} 
    \begin{aligned}[t] & \Biggl( \lambda_{\textrm{prob}} \left ( P_{ij} - 0 \right)^2 + \lambda_{\textrm{obj}} \left ( O_{ij} - 0 \right)^2 \Biggl). \end{aligned} \\
\end{split}
\label{eq:method_loss} 
\end{align}

In this equation, in addition to already defined parameters, the first sum runs over all the elements of the output grid for a single image $N_g$, and the second sum runs over all the detection units in a grid cell $N_b$. All the values with a hat \textcolor{black}{represent the targets for the corresponding predicted values, and the $\mathds{1}_{ij}^{X}$ are masks of predictions that pass the quality limit of each subpart. This loss is written for a classification based on a soft-max activation and a cross-entropy error classification, but it can be modified for a sigmoid activation with a sum-of-square error.}

\subsection{Prediction filtering and non-maximum suppression}
\label{sec:filtering_and_nms}

\textcolor{black}{We expect a properly trained detector to order its predictions by quality based on the objectness score for each detection unit. The raw detector output is always a static list of boxes of size $\left< g_w, g_h, N_b{\times} (6+N_c+N_p)\right>$ regardless of the input content. Consequently, the predicted boxes must be filtered based on their objectness score to remove those unlikely to represent an object. By design, the number of detectable objects in the image should be low compared to the total number of detection units. Therefore, most predicted boxes belong to the background type with a low objectness score.} 

While the continuous objectness score for all the predictions is the best direct representation of the sensitivity of the detector, it is incompatible with some final metric that needs a list of considered ``good'' detections. Visualizing the predicted boxes also requires filtering to preserve only the plausible detection. In such cases, an objectness threshold can be used to remove low-confidence detections. \textcolor{black}{The threshold is usually optimized to maximize a metric score on a validation or test dataset not used for training but for which the targets are known.}

\textcolor{black}{Due to the fully convolutional structure of the network, objectness scores from the same detection unit can be compared over the full grid, meaning that the same threshold can be used. On the contrary, the predicted objectness between two independent detection units is not comparable as it depends on the type and frequency of targets associated with each of them during training. A possible solution is to fit an individual objectness threshold for each detection unit. However, it relies on the assumption that predictions from different detection units are independent, which is not true for most applications.} In practice, it is still a good solution to remove the vast majority of false positives. To achieve the best results, the objectness regimes must be homogenized between the different detection units from the start by adjusting the individual $\lambda_{\textrm{void}}^b$ factors (Sect.~\ref{sec:multiple_boxes}). \textcolor{black}{This is done by balancing the ratio between detection and background cases based on how the objects from the training sample are expected to distribute over the detection units.}

With most false positives removed, there can still be multiple high-objectness predictions that represent the same underlying object in the image. To preserve only the best-detected box for each object, we use a classical post-processing step called non-maximum suppression \citep[NMS,][]{paper:det_2012_nms, paper:rcnn}. It consists of an iterative search for the box \textcolor{black}{with the highest objectness score in the image that is then used to remove all the overlapping predicted boxes. To consider that there is an overlap, the two boxes must verify ${\rm fIoU} > L^{{\rm fIoU}}_{\rm NMS}$ with the ${\rm fIoU}$ being computed between the two predicted boxes. The best box is then stored in a static list of selected detection.} This process is repeated until no boxes are left in the raw-prediction list. It is illustrated in Fig.~\ref{fig:nms_filtering}. 

The NMS is done regardless of what detection unit generated the predicted boxes, demonstrating that they are not independent as they can remove each other based on their respective objectness. This is one of the main reasons we force all detection units to have similar objectness distributions. The detection quality can only be evaluated after the NMS, so searching for the best $\lambda_{\textrm{void}}^b$ factors is dependent on the $L^{{\rm fIoU}}_{\rm NMS}$ and respectively.

\section{Dataset description and network training}
\label{sec:dataset_and_training}

In this section, we present the main properties of the SDC1 data along with the expected products and the associated metrics. A complete description of the SDC1 challenge can be found in \citet{paper:sdc1}, while the underlying T-RECS simulation is detailed in \citet{paper:t-recs}. We also present the preprocessing of the data to construct our training sample. From this, we describe our best-performing network architecture and specify the corresponding setup and hyperparameters for our detector.

\subsection{Subchallenge definition}
\label{sec:sub_challenge}

The SDC1 is a source detection and characterization task in simulated SKA-like data products (Sect.~\ref{sec:introduction}) that comprises nine 4GB images (three frequencies, with three integration times each) of the same field. The SDC1 is only modestly challenging regarding data volume, especially compared to the SDC2 with its 1TB data cube. Still, it represents significant challenges for detection methods in many other aspects. All the images have an identical pixel size of 32768x32768. As the frequency increases, the angular resolution improves while the field of view reduces. Therefore, images at different frequencies only partially overlap, meaning that the problem to solve varies with the position in the field. In addition, the number of detectable sources varies significantly with the integration time and frequency \citep[Table 2 in][]{paper:sdc1}. All the images are considered noise-limited, even at the highest 1000h integration time. As a benchmark for our YOLO-CIANNA method, we used only the 560 MHz - 1000h image due to its wider field of view, higher total number of sources, and higher source density per square pixel. \textcolor{black}{This image is a fair example of a typical astronomical detection context with a high dynamic range and a high density of small sources with occasional blending.} While our method could technically work for the other SDC1 images, it would not be more informative regarding its capabilities. The consequences and limits of this choice are further discussed in Sect.~\ref{sec:methode_improvement_disc:sdc1_lim}.

In the original challenge, participating teams were provided a True catalog for a small fraction of the image. This was supposed to facilitate method development and tuning but also allow training of supervised ML approaches. At the time, the full underlying True catalog was unavailable, and the teams had to submit their result to a remote scoring service. After the challenge, the organizers released the full True catalog, the scoring code in \href{https://gitlab.com/ska-telescope/sdc/ska-sdc}{open source}\footnote{https://gitlab.com/ska-telescope/sdc/ska-sdc} \citep{paper:sdc_scoring_code}, and the submitted source catalogs from \textcolor{black}{the participating teams. In this study, we use only the training catalog to constrain our detector so our results can be compared to those published in the context of the challenge.} We then use the full True catalog only to present an in-depth analysis of our trained detector performances. Since the scorer allows individual image scoring (for all frequencies and exposure), we can produce a detection catalog for the 560 MHz - 1000h image that is directly comparable to other team submissions during and after the challenge.

\subsection{Image and source catalog description}
\label{sec:sdc1_data}

The 560 MHz - 1000h image has a field of view of $5.5$ square degrees, which contains the primary beam out to the first null. It has a full size of 32768 square pixels, a pixel size of $0.6$ arcsec, and an imaging resolution in the Gaussian approximation of $1.5$ arcsec full width at half-maximum. \textcolor{black}{A subfield of 512 square pixels from this image is presented in Fig.~\ref{fig:example_field}. The simulated image contains no systematic instrumental effects like calibration, pointing, or deconvolution errors, making it unrealistically clean (Sect.~\ref{sec:real_data_pred}). The image is not primary-beam corrected, but the corresponding primary beam image is available as ancillary data. With this setup, the instrument sensitivity decreases on the edges, but it preserves a uniform noise level over the full field. The thermal noise level reported for this image in \citet{paper:sdc1} is $2.55{\times}10^{-7}$ $\mbox{Jy}\,\mbox{beam}^{-1}$; however, a direct estimate of the noise using the provided image is about $3.45{\times}10^{-7}$ $\mbox{Jy}\,\mbox{beam}^{-1}$. We use this second value for the rest of the paper when necessary.} While we could convert the image to a beam-corrected one, it is better for the detector to work on the constant noise image and to detect sources from their apparent brightness (Sect.~\ref{sec:sdc1_training_area}). The apparent flux of a source can be obtained by multiplying the flux from the True catalog by the interpolated primary beam value at the central position of the source.

The SDC1 challenge task is to detect and characterize the sources. The expected parameters for each source are the central coordinates (RA, Dec), the integrated source flux $f$, the core fraction $c_f$ if different from zero, the major and minor axis ($\textrm{Bmaj}, \textrm{Bmin}$), the major axis position angle $\textrm{PA}$, and a classification $C$ (one of AGN-steep, AGN-flat, or star-forming galaxy). The provided True catalog supplies all these properties for each source, allowing the training of supervised methods. From Sect. 5.3 in \citet{paper:sdc1}, the classification part of the challenge was considered problematic a posterior, as it is only feasible on the tiny fraction of the field where all frequencies are available. For this reason, we focused on getting the best performances on a detection and characterization problem only.

\subsection{SDC1 scoring metric}
\label{sec:sdc1_metric}

The first element of the scorer is a match criterion. Due to source density, relying only on the central positions of the sources for matching predictions with the underlying True catalog is likely to result in many false positives. The SDC1 scorer uses a combination of the position, the size, and the predicted flux accuracies to represent a global matching score defined as
\begin{align}
\label{eq:scorer_match_dist}
E_{\rm tot} &= \sqrt{E^2_{\textrm{pos}} + E^2_{\textrm{size}} + E^2_{\textrm{flux}}}\text{, with}\\
E_{\textrm{pos}}  &\propto \sqrt{(x-\hat{x})^2 + (y-\hat{y})^2}/\hat{S}'\text{,}\\
E_{\textrm{size}} &\propto \left|S - \hat{S}\right|/\hat{S}'\text{, and}\\
E_{\textrm{flux}} &\propto \left|f - \hat{f}\right|/\hat{f},
\end{align}
where the true values from the catalog are indicated with a hat, ($x,y$) is the central coordinates in pixels corresponding to (RA, Dec), $S$ is the average value of $\textrm{Bmaj}$ and $\textrm{Bmin}$, $S'$ is the largest axis convolved with the synthesized-beam, and $f$ is the source intrinsic flux. We note that the difference between $S$ and $S'$ is present in the scorer code but not specified in either \citet{paper:sdc1} or in the challenge documentation. To prevent false detections due to the high source density, the scorer uses a strict match threshold value for $E_{\rm tot}$. Each subpart of the error is normalized to a value considered representative of a $1\sigma$ error. In the scorer, the normalization coefficients are set to $0.93$ for the position, $0.36$ for the flux, and $4.38$ for the size, all using the catalog units defining a global $1\sigma$ error. These individual limits were obtained by fitting the distribution of the corresponding values on the combination of all the submitted catalogs at the challenge time. A match is defined if $E_{\rm tot} < 5\sigma$, which was optimized to reduce the average random association chance of all submitted catalogs below 10\%. We note that the scorer distinguishes false detection into two categories: either ``False'' if there is no target source closer than $1.5\times S'$ or ``bad'' if it passes this distance limit but has a too-high $E_{\rm tot}$.

\begin{figure}
    \centering
    \includegraphics[width=1.0\hsize]{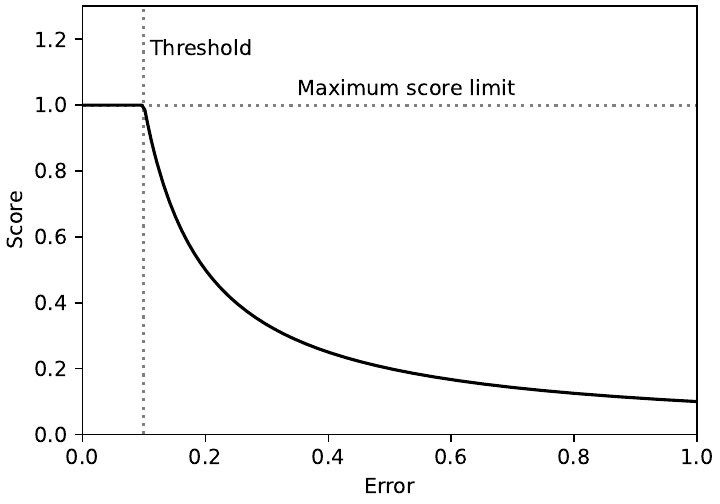}
    \caption{Generic scoring function response for a source parameter with $T=0.1$ as a function of the prediction error.}
    \label{fig:param_score_fct}
\end{figure}

The list of identified matches is then converted to a global score that captures the detection and characterization quality. The SDC1 scorer attributes a score of up to one point for each match depending on its characterization quality and penalizes every false detection with a strict minus one point. The characterization evaluation is decomposed into seven individual subscores that all respect the following scoring rule (based on the scorer code) for a single source
\begin{equation}
s_i^j = \min{\left(1,\frac{T^j}{E_i^j}\right)},
\label{eq:score_response_fct}
\end{equation}
where $j$ represents the subscore part and $i$ a single source from the match list, $T^j$ is a threshold for this subscore part, and finally, $E_i^j$ and $s_i^j$ correspond to the error term and the final subscore part for this source. We list all the subscore parts and their corresponding $E^j$ function and $T^j$ values in Table~\ref{table:sdc1_scorer_sub_scores}, and we illustrate the typical score as a function of the error regarding a given threshold value in Fig.~\ref{fig:param_score_fct}. 

We note that all subpart error functions are based on a relative error, which has an asymmetric behavior. When overestimating the value, the error can rise infinitely, while underestimating a strictly positive prediction will never lower the relative error below -1. The issue is that these errors are associated with symmetric score response functions. Consequently, the score will be higher when underestimating the predictions, which is unlikely to naturally happen on quantities with intrinsic minimum values linked to instrumental limits like the flux, $\textrm{Bmaj}$, and $\textrm{Bmin}$. While the effect is minor for the subpart scoring, we note that for the matching criteria, it will result in excluding properly detected faint sources for which the noise caps the minimum predictable flux (Sects.~\ref{sec:results:field_detection} and~\ref{sec:challenge_and_score_disc}).

The final score for a given source and the average subpart score for all sources are
\begin{equation}
s_i = \frac{1}{7} \sum_j^7 s_i^j \quad \text{ and } \quad \bar{s}^j = \sum_i^{N_{\textrm{match}}} s_i^j.
\end{equation}
To obtain the full SDC1 score, there is a scaling between the different frequencies and integration times. However, since we limit this study to a single image, there is no need for these definitions. The average source score for matching sources is
\begin{equation}
\bar{s} = \frac{1}{N_{\textrm{match}}}\sum_i^{N_{\textrm{match}}}{s_i},
\label{eq:average_score}
\end{equation}
and the final SDC1 score for a single image is then
\begin{equation}
M_s = \sum_i^{N_{\textrm{match}}}{s_i} - N_{\rm false}.
\label{eq:total_score}
\end{equation}
When scoring a submitted catalog, the training area is excluded, so detection performances are evaluated only for examples that were not used to constrain the detector.

\begin{table}
\centering
\caption{\label{table:sdc1_scorer_sub_scores} Error functions and threshold values for all subscores.}
\begin{tabular}{ l c c }
 \hline
 \hline
 Subpart & Error function $E^j$ & Threshold $T^j$ \\
 \hline\\[-2.5ex]
 Position & $\sqrt{(x-\hat{x})^2 +(y-\hat{y})^2}/\hat{S}''$& 0.3\\
 Flux Density & $\left| f - \hat{f}\right| / \hat{f}$ & 0.1 \\
 Major axis & $\left| \textrm{Bmaj} - \hat{\textrm{Bmaj}}\right| / \hat{\textrm{Bmaj}}$ & 0.3\\
 Minor axis & $\left| \textrm{Bmin} - \hat{\textrm{Bmin}}\right| / \hat{\textrm{Bmin}}$& 0.3\\
 Position angle & $\left| \textrm{PA} - \hat{\textrm{PA}}\right|$ & 10.0\\
 Classification & $1 \text{ if } C = \hat{C}\text{, } 0 \text{ otherwise} $& \\
 Core fraction & $\left| cf - \hat{cf}\right| / 0.75$ & 0.05\\
 \hline
\end{tabular}\\
{\raggedright \vspace{0.2cm} Note: $\hat{S}''$ is the target source average size convolved with twice the synthesized beam size, which is different from $\hat{S}'$ \par}
\end{table}

\subsection{Selection function}
\label{sec:sdc1_selection_function}

The full simulation of the SDC1 contains more than five million sources. Due to the added simulated noise, only a fraction of these sources are detectable in the image. In \citet{paper:sdc1}, they estimated that only around 758000 sources are above the noise level by $5 \sigma$ in the 560 MHz - 1000h image. This construction is problematic for supervised ML methods as they are sensitive to wrong labeling during training. For an object detector, it goes two ways: i) if a source is detectable but not labeled as a target box, the network lowers the \textcolor{black}{detection objectness of all predicted boxes that try to detect sources with similar features}; ii) if there is a target box for a source that the network is incapable of detecting, it increases the objectness of nonrepresentative features, likely background noise. To achieve good detection performances, we must construct a training sample that is both complete and pure regarding the detectability of the sources it contains. For this, we defined a selection function based on the source properties of the provided True catalog.

\begin{figure}
    \centering
    \includegraphics[width=1.0\hsize]{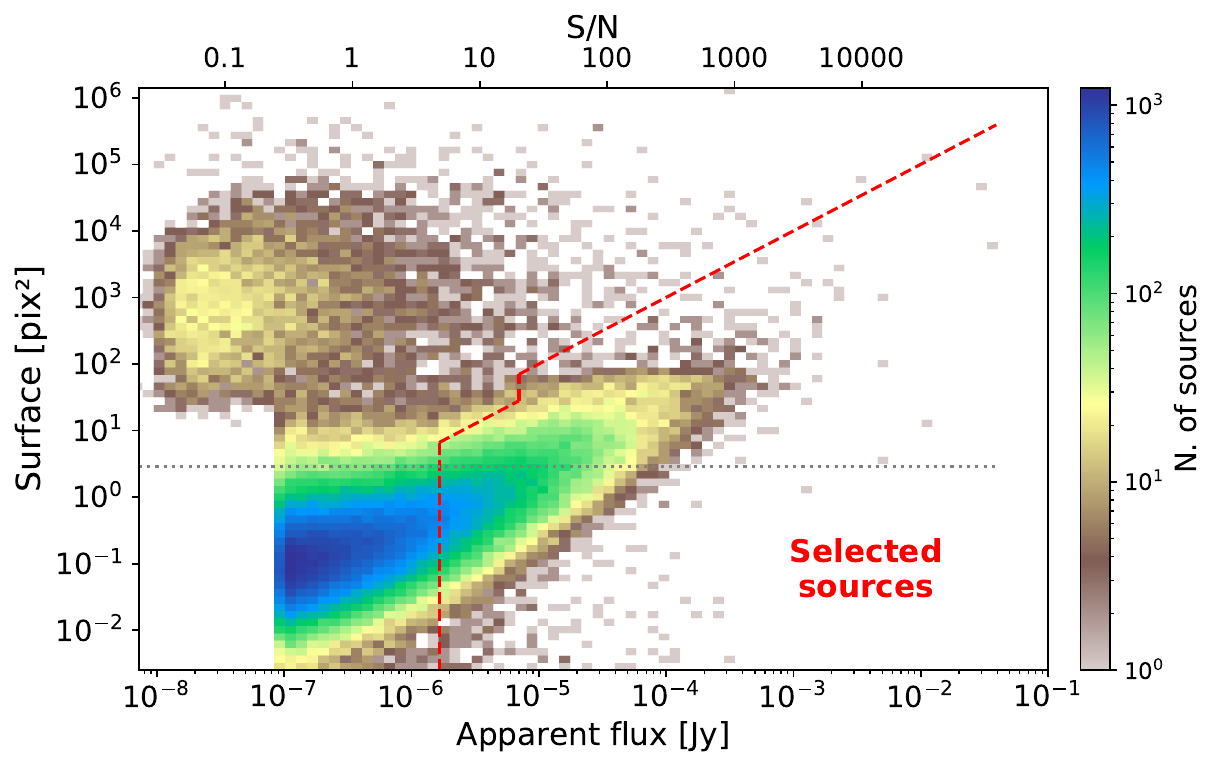}
    \caption{\textcolor{black}{Two-dimensional histogram of the source surfaces as a function of their apparent flux. The dotted gray line indicates the saturated minimum surface with clipping.} The red dashed line represents our selection function, with all the sources on the right of the line being selected.}
    \label{fig:select_function_surface_luminosity}
\end{figure}

Our selection function combines the apparent source flux $f_a$ in $\mbox{Jy}$ (Sect.~\ref{sec:sdc1_data}), and the surface brightness $S_b$ computed as
\begin{align}
S_b &= f_a / (W_{\rm pix} \times H_{\rm pix})\text{, where}\\
W_{\rm pix} &= (2/P_s) \times\max{(1.2,\textrm{Bmaj})} \text{ and }\\
H_{\rm pix} &= (2/P_s) \times\max{(0.6,\textrm{Bmin})}.
\end{align}
The size saturation values are in arcsec and correspond to a two and one-pixel size, respectively, and $P_s$ represents the pixel size in arcsec used to obtain the box size in pixels. The selection function to keep a source in our training sample is then defined as
\begin{equation}
\begin{cases}
    S_b > 1.0\times10^{-7} &\text{if } f_a >= 7.0\times10^{-6} \\ 
    S_b > 2.5\times10^{-7} \text{ \& } f_a > 1.65\times10^{-6} & \text{otherwise}.
\end{cases}
\label{eq:selection_function}
\end{equation}
\textcolor{black}{The $f_a > 1.65\times10^{-6}$ hard cut in apparent flux is equivalent to a signal-to-noise-ratio (S/N) selection with $S/N>4.8$. However, we observed that S/N cuts were insufficient to achieve good results as too-low cuts failed to properly remove extended faint sources, while too-high cuts tended to remove detectable point sources. Consequently, we added the surface brightness cuts that helped to remove undetectable objects. We }represent our selection cuts over a 2D histogram of the source surface against apparent flux for the training area in Fig.~\ref{fig:select_function_surface_luminosity}. \textcolor{black}{In this figure, the top left patchy distribution corresponds to steep spectrum AGN. The flat spectrum AGN and star-forming galaxies distribute similarly and occupy the rest of the distribution, but they are strongly unbalanced with only a few flat spectrum AGN.} In practice, both types of AGNs are mostly undetectable with the 560 MHz - 1000h setup, and are therefore strongly under-represented after our selection function. This also contributes to the identified difficulty of the classification task in the SDC1 result paper \citep[Sect.~5.3 in ][]{paper:sdc1}. The hard cut at the bottom left of the distribution results from a per-class S/N prefiltering done directly by SKAO, characterized by the ``selected'' flag in the training catalog. Due to our size clipping, most of the surface range of this space is collapsed at a minimum surface, represented by the dotted gray line. We observed that lowering the clipping limits below the current values increases the number of nonvisible objects that pass our selection, which should be avoided. 

\begin{figure}
    \centering
    \includegraphics[width=1.0\hsize]{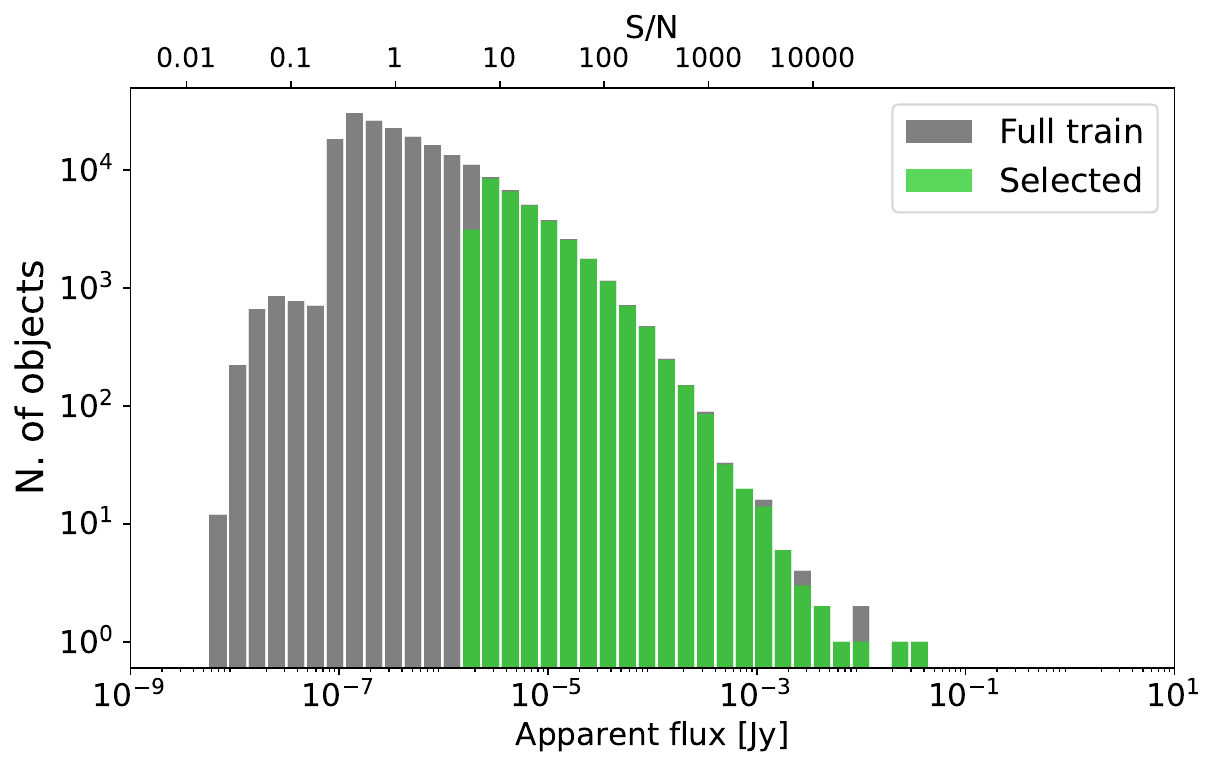}
    \caption{\textcolor{black}{Source apparent flux distribution histogram in log scale and using log-bins for the True catalog and the selected sources}.}
    \label{fig:select_function_dist_flux}
\end{figure}

\begin{figure*}
    \centering
    \includegraphics[width=1.0\hsize]{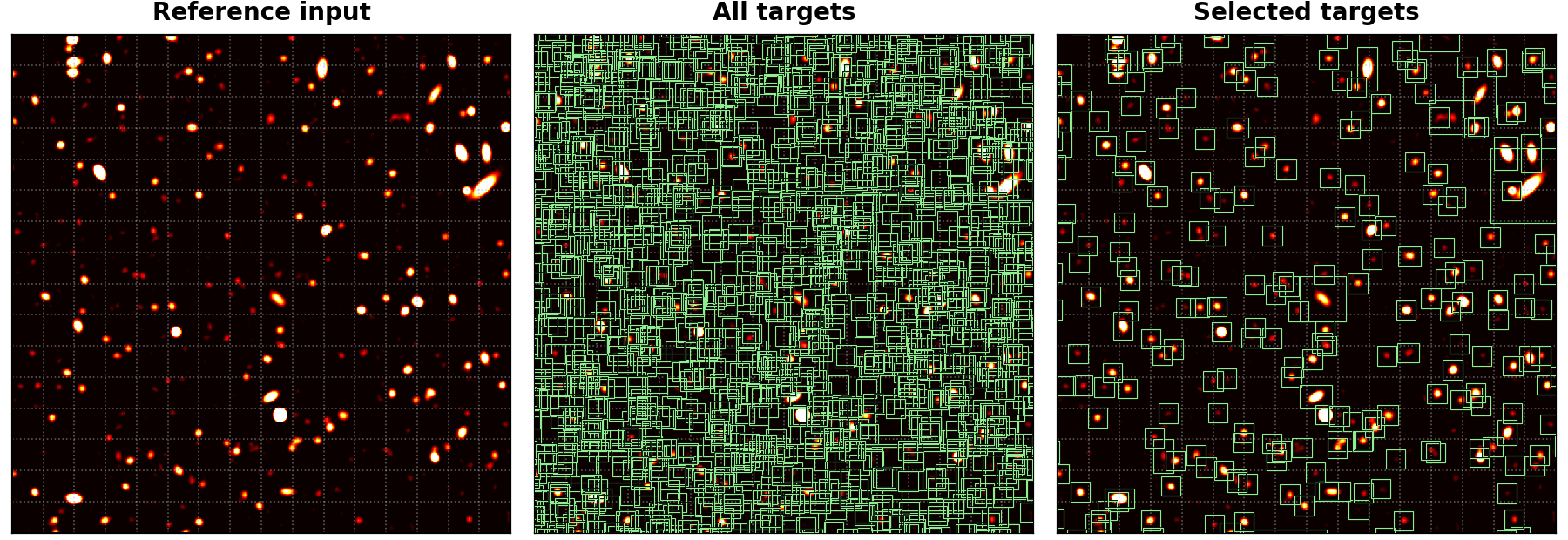}
    \caption{Illustration of the selection function over a typical input field. The background image represents an identical $256{\times}256$ input patch in all frames, centered on $\textrm{RA}=0.1$ deg, $\textrm{Dec} = -30.2$ deg, using the renormalized input intensity but saturated at half the maximum value to increase visual contrast. The displayed grid corresponds to the detector output grid mapping for a network reduction factor of 16. The \textit{left} frame is provided as an image reference, the \textit{middle} frame represents all the target boxes from the True catalog, and the \textit{right} frame represents the remaining target boxes after our selection function.}
    \label{fig:select_function}
\end{figure*}

\textcolor{black}{We represent the effect of our selection function on the source apparent flux distribution in Fig.~\ref{fig:select_function_dist_flux}. We also illustrate our selection on a small field in Fig.~\ref{fig:select_function}.} We see that it misses some apparent compact signals, which can result from several effects: i) the local noise contribution can increase the perceived apparent flux of a faint source, \textcolor{black}{ii) blended faint sources can add their flux at the same location}, and iii) it can be a bright and compact part of an extended faint source below the surface brightness limit. \textcolor{black}{Regardless of their origin, these nonlabeled compact signals will likely confuse the detector training. We tried to adapt the selection process and our threshold values, but the current formulation produced the best results.} We also tried to define our surface brightness using a more common astronomical size definition by convolving the $\textrm{Bmaj}$ and $\textrm{Bmin}$ with the synthesized beam. Still, it resulted in lower detector scores after training.

\textcolor{black}{We emphasize that this hand-made selection function relies on parameters from the True source catalog. Therefore, applying the method to observed data instead of simulated ones will require either using estimates from another approach to define the target or constructing a selection function that does not rely on these source properties (Sect.~\ref{sec:real_data_pred}).} In Appendix~\ref{sec:appendix:training_area:larger_and_bootstrap}, we discuss an alternative way to construct a selection function iteratively using the prediction of a naively trained detector to evaluate the detectability of the sources. 

\subsection{Training area}
\label{sec:sdc1_training_area}

\begin{figure}
    \centering
    \includegraphics[width=1.0\hsize]{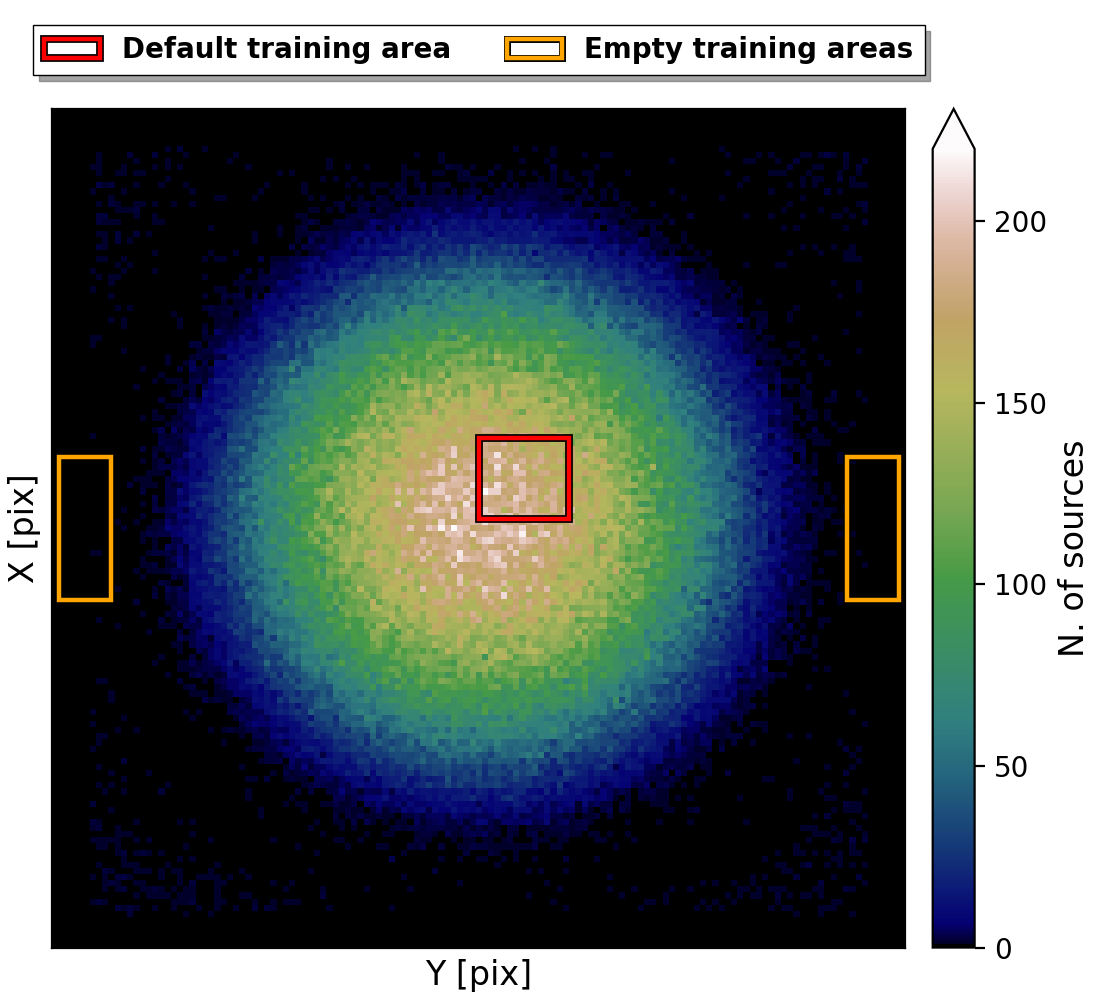}
    \caption{Two-dimensional histogram of the central coordinates of the sources from the full True catalog that pass our selection function. The red box indicates the default SDC1 training area, and the orange boxes indicate our additional noise-only regions.}
    \label{fig:sdc1_training_sample}
\end{figure}

Using the challenge setup, the provided training area only spans a small part of the image center, corresponding to about $0.308$ square degree. The associated training catalog contains 190552 sources (using the predefined ``selected'' flag), corresponding to roughly $3.6\%$ of the sources from the full True catalog. Applying our custom selection function to this region drops the number of sources to 33813. Due to the effect of the primary beam sensitivity over the image field, the central area is not a good representation of other parts of the image. The learned features and context awareness of a detector trained on this region are unlikely to generalize properly to other parts of the image. \textcolor{black}{We represent the distribution of sources that pass our selection function for the whole image field using the full True cat in Fig.~\ref{fig:sdc1_training_sample}}. The footprint of the primary beam sensitivity is clearly visible. The provided training area corresponds to the red box. A more suited training area definition could have been a narrow band over a complete beam radius of the image field, which we explore in Appendix~\ref{sec:appendix:training_area}. To mitigate the generalization issue while still following the original challenge definition, the detector could be constrained to either be flux agnostic and only perform morphological detection or to reject all sources outside a given radius from the image center. 

Instead, we propose another approach that uses other regions of the image without adding any target sources. We observed that detecting an object above a given radius from the image center becomes very unlikely. From this, we added two ``noise only'' regions to our training sample that are sufficiently far from the image center to be considered devoid of any detectable source. We selected two rectangular regions of identical width and height of 2000 and 5600 pixels, respectively, that are both vertically centered in the image but on opposite sides horizontally with a margin to the image edge of 250 pixels (Fig.~\ref{fig:sdc1_training_sample}). These regions lie between the main lobe and the first sidelobe of the primary beam. \textcolor{black}{We note that the regions are, in fact, not fully empty and that it will result in a few unlabeled visible sources. Still, this approach remains vastly beneficial regarding global detector performances. During training, examples are drawn randomly} in the default training area but also at a low $5\%$ rate in one of the two noise-only regions. This forces the detector to understand the input intensity dynamic of parts of the image where it is not expected to detect anything. We observed that detectors trained with this process can interpolate between the two regimes and provide much better results for the full image without manually excluding difficult regions. \textcolor{black}{Other things being equal, the best achievable challenge score is increased by more than 10\% with this approach compared to training only with the default region.}

\subsection{Network backbone architecture}
\label{sec:cnn_backbone}

As stated in Sect.~\ref{sec:bounding_boxes}, our method requires a fully convolutional neural network backbone to create a mapping from a 2D input image to a regular output grid. Efficient neural network architectures can be built by stacking convolutional layers while progressively reducing the spatial dimension to filter the relevant information and construct higher-level representations of the input content \citep{paper:deep_cnn, paper:deep_learning}. With a fully convolutional structure, the last layer is responsible for encoding the output grid of the detector. Its spatial dimension represents the output grid, and the filters encode all the output vector elements for each grid cell. It also results in the output grid size being controlled by the reduction factor of the backbone network, which corresponds to the total spatial reduction of the network from its input to the output. The properties of a fully convolutional architecture and its impact on our method design are further detailed in Appendix~\ref{sec:appendix:conv_net_yolo}.

With a fully convolutional architecture, the input size does not impact the model structure and number of parameters. In this section, we consider the input size to be $256{\times}256$ pixels (Sect.~\ref{sec:detector_training}). Regarding the reduction factor, we found that a value of 16 resulted in the best detection accuracy, which means that the training output grid is composed of $16{\times}16$ cells. \textcolor{black}{Considering the source density of the 560 MHz - 1000h image, each grid cell has to detect multiple sources, which was the motivation for using a prediction-aware association process (Appendix~\ref{sec:appendix:association_function})}.

\begin{figure}
    \centering
    \includegraphics[width=1.0\hsize]{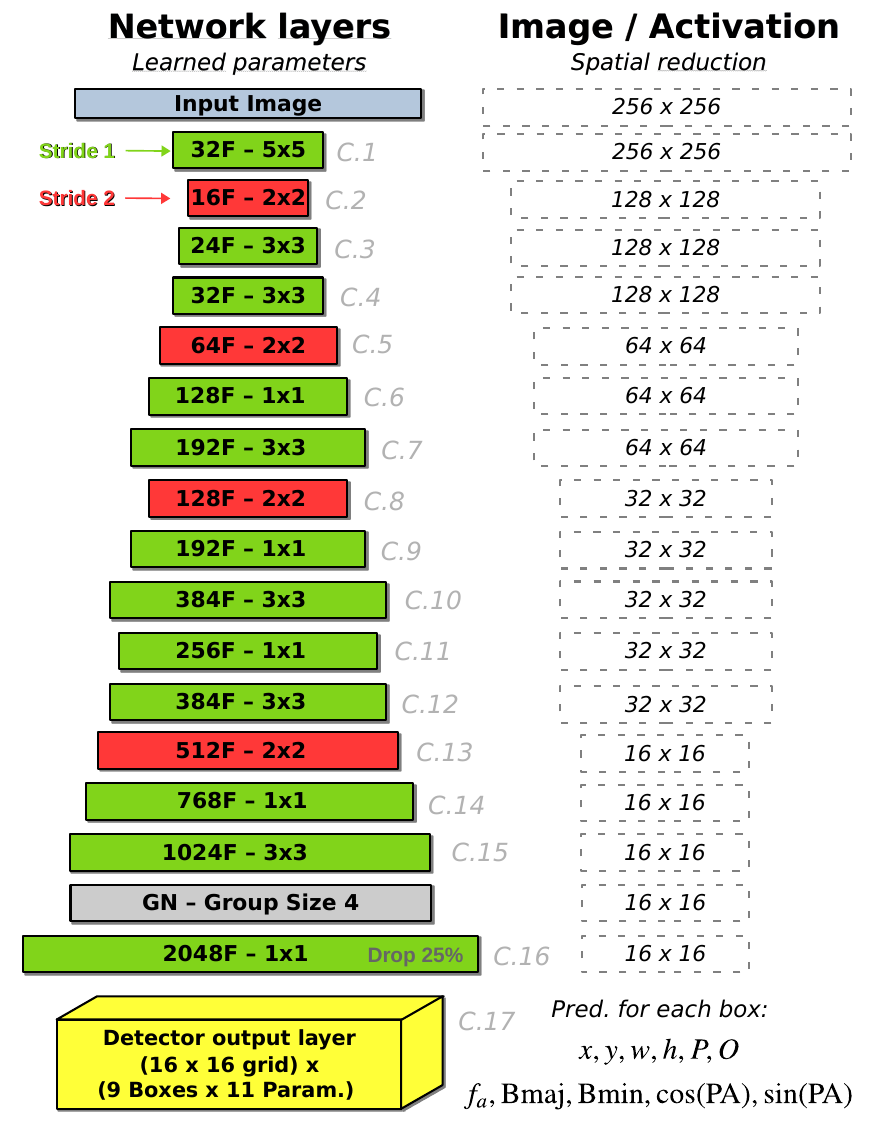}
    \caption{Illustration of our final CNN backbone architecture. The \textit{left} column provides the layer structural properties, while the \textit{right} column indicates the spatial output dimension for each layer starting from a $256{\times}256$ input size. The input image is on the top, and layers are stacked in order vertically. The width of a layer represents its number of filters. The green color indicates a layer that preserves the spatial dimension, while the red indicates a reduction. The output grid dimension and the list of the predicted parameters for each box are also indicated.}
    \label{fig:network_architecture}
\end{figure}

\textcolor{black}{At this point, it would be tempting to adopt the classical darknet-19 backbone network introduced in YOLO-V2 \citep{paper:yolo_v2}. We notably used it successfully with our custom association process for other contexts  (Appendix~\ref{sec:appendix:yolo_cianna_benchmarks})}. We discuss this possibility and explain why it would lead to poor performances in Appendix~\ref{sec:appendix:yolo_v2_arch}. Instead, we meticulously explored increasingly complex custom backbone architectures. We specifically looked for an architecture that is both computationally efficient and capable of high detection accuracy. An illustration of our final architecture is presented in Fig.~\ref{fig:network_architecture}.

\textcolor{black}{Our final architecture was based on a few educated guesses, but it also required exploration through score optimization. We note that the spatial dimension is always reduced by convolution operations instead of pooling operations, which helps preserve the apparent flux information and better represents continuous objects with no sharp edges.} The first layer has larger filters to extract continuous luminosity profiles better. The second layer performs a local compression (both spatially and in the number of filters), mainly acting as a local noise filter. Then, for a few layers, we progressively increase the number of filters while decreasing the spatial dimension at a rate that maximizes computing efficiency. \textcolor{black}{Starting with the seventh layer, we begin alternating large layers with $3{\times}3$ filters and smaller layers with $1{\times}1$ filters, which is typical of the YOLO darknet architecture.} This structure alternates searches for local spatial coherency with representation compressions. It improves compute performance and reduces the global number of parameters compared to a more classical stacking of identically sized $3{\times}3$ layers. \textcolor{black}{We note that our third spatial-dimension reduction layer (eighth layer in global) is also considered a compression layer in this scheme.} The second last layer has a 25\% dropout rate, which is used for regularization \citep{paper:dropout}. 

\textcolor{black}{We tested adding group normalization \citep{paper:group_norm} at various places in the network, but it almost always degraded the best achievable score by a few percent (Appendix~\ref{sec:appendix:yolo_v2_arch}).} This is likely because in-network normalization tends to lose the absolute values of the input pixels, making it more difficult to predict the flux accurately. \textcolor{black}{Since group normalization works at the scale of the whole spatial dimension, it might also affect the input dynamic in a way that makes faint sources more challenging to detect in the presence of bright sources in the image. The only place where it produces a beneficial effect is near the end of the network, after the last spatial correlation, where the flux value is likely fully re-encoded in the high-level features.} This specific normalization layer has several beneficial effects, including a speed up and stabilization of the training process and a slight improvement of the best achievable score of about $2\%$.

The complete architecture contains 17 convolution layers for around 12.62 million weights, but 76\% of these weights are concentrated at the end of the network in the connections between layers 14 to 16. With this architecture, the receptive field for a given grid cell at the output layer is about 100 pixels. It limits both the maximum size of the sources that can be detected with this architecture and the context windows accessible to each detection unit. We discuss the limits of the current network structure and potential architecture improvements in Sect.~\ref{sec:methode_improvement_disc}.

\subsection{YOLO-CIANNA configuration for the SDC1}
\label{sec:sdc1_settings}

\subsubsection{Input normalization}
\label{sec:sdc1_settings:input_norm}

Input normalization is critical to obtain good detection performances in our specific context. \textcolor{black}{This aspect is key for astronomical images due to their very high dynamic range. Everyday-life images are usually encoded using three 8-bit integers, resulting in 256 possible values for each color. Converting an astronomical image to a similar format often induces a significant information loss. We note that some off-the-shelves deep learning detectors or frameworks implicitly convert images to this format without warning, sometimes explaining poor performances.} We found that 16-bit floating-point quantization was sufficient for the SDC1 image after other input transformations, but some astronomical datasets are likely to require 32-bit floating-point encoding. It is also useful to offset the raw dynamic to better represent the low flux regime. In our case, we redefine our minimum and maximum values as $\min_p = 4{\times}10^{-7}\,\mbox{Jy}\,\mbox{beam}^{-1}$ and $\max_p = 4{\times}10^{-5}\,\mbox{Jy}\,\mbox{beam}^{-1}$, and apply a scaled hyperbolic tangent for all pixel values, which can be summarized as 
\begin{equation}
p_i' = \tanh{\left(3\times \frac{p_i - \min_p}{\max_p - \min_p}\right)},
\end{equation}
where $p_i$ is the raw pixel value in $\mbox{Jy}\,\mbox{beam}^{-1}$ clipped with the two limits, and $p_i'$ is the pixel value as it is presented to the detector. \textcolor{black}{This normalization remaps all input data in the 0 to 1 range and grants most of this range to low signal values using an almost linear regime to help distinguish faint sources from the noise.} The counterpart is a flattening of the dynamic for high fluxes, but bright sources require less accuracy on the pixel values to obtain a good relative flux estimate.

\subsubsection{Detection units settings}
\label{sec:sdc1_settings:box_prior_selection}

To configure our detection units, we must first define our target boxes. The True catalog does not contain the necessary information to define them in the classical computer vision way. Instead, we define our boxes as centered on the source central coordinates and with a size that is scaled on its major and minor axes. For each target source, we define a rectangular box with $\hat{w}_{\textrm{PA}=0} = 2 \times \textrm{Bmaj}$ and $\hat{h}_{\textrm{PA}=0} = 2 \times \textrm{Bmin}$ corresponding to $\textrm{PA}=0$. This box is then rotated to correspond to the actual $\hat{\textrm{PA}}$ of the source, and we search for the smallest square box that contains the four rotated vertex. The resulting dimensions are clipped in the 5 to 64 pixels range to obtain the final $\hat{w}$ and $\hat{h}$ dimensions that are used to define our target box for the corresponding source (Fig.~\ref{fig:select_function}). The minimum clipping implies that all unresolved point sources get the same minimum size\textcolor{black}{, which was calibrated to optimize the association process of our detection layer. We stress that box sizes do not have to be very accurate in our specific context. Firstly, they do not constrain the receptive field in any way, meaning that the detector can use information outside the box to detect or characterize the source}. Secondly, they are only used during association so the detector can be trained and are not used in the scorer matching criteria and characterization score. The important $\textrm{Bmaj}$ and $\textrm{Bmin}$ values are instead predicted as additional parameters (Sect.~\ref{sec:sdc1_settings:box_properties}).

\begin{figure}
    \centering
    \includegraphics[width=1.0\hsize]{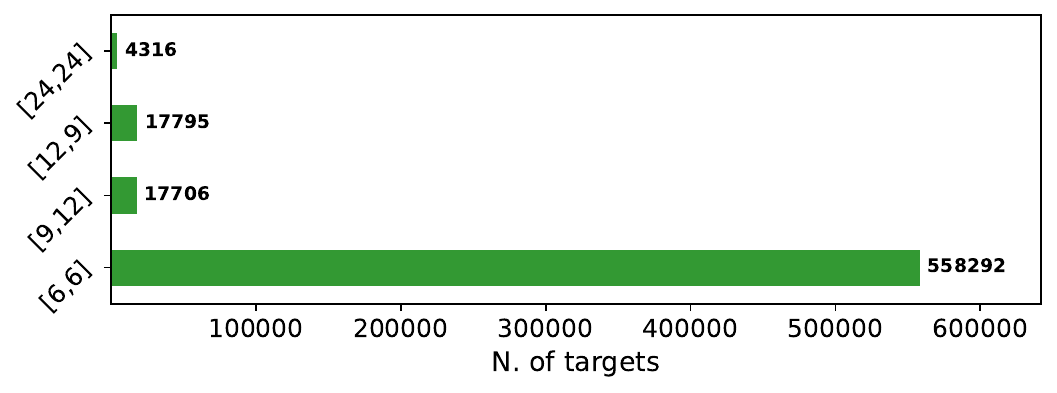}
    \caption{Target distribution over the closest box size prior. The association uses the Euclidean distance in the 2D box size space.}
    \label{fig:targ_prior_size_dist}
\end{figure}

We define the number of detection units and their size priors based on the target source density and the box-size distribution. \textcolor{black}{We chose to have three size regimes: i) a small regime that is composed of several identical units of $6{\times}6$ size prior, ii) an intermediate regime with two units of identical surfaces but two aspect ratios with size priors of $9{\times}12$ and $12{\times}9$ respectively, and iii) a large regime with a single unit of $24{\times}24$ size prior.} We illustrate how the target sources distribute over these size-priors based on the smallest Euclidian distance with their respective size in Fig.~\ref{fig:targ_prior_size_dist}. This indicates that most sources would theoretically be associated with the smallest size regime. \textcolor{black}{We tried an alternative setup with all our detection units in the small regime, but it resulted in lower detection performances for all size regimes.} This confirms that the source size remains an appropriate first-order criterion for distributing the network expressivity over the detection units, even with such a massive target size regime imbalance (Appendix~\ref{sec:appendix:association_function}).

\textcolor{black}{To detect multiple small sources in the same grid element, we must populate the small-size regime with several identical detection units. We observed good results for all models trained using four to height small units with optimum results for six.} Too few detection units limit the number of detectable objects. It can also force some units to encapsulate multiple contexts, preventing them from being detected simultaneously at prediction time. Conversely, too many boxes dilute the context diversity and increase the training difficulty. Combined with the three detection units from the two other size regimes, our detector predicts nine independent boxes in the same grid cell. \textcolor{black}{In practice, all these detection units are never used simultaneously. We discuss how the actual predictions are distributed among them in Sect.~\ref{sec:results:full_catalog}.}

\subsubsection{Source properties to predict}
\label{sec:sdc1_settings:box_properties}

\textcolor{black}{With our method, source characterization is handled directly by the detector in the form of a single-stage network. We discuss the possible impact of this design choice in Sect.~\ref{sec:challenge_and_score_disc:alternative_metrics} and Appendix~\ref{sec:appendix:characterization_impact}. In practice, each detection unit must always predict the box center coordinates ($x,y$), the box size ($w,h$), the probability $P$, and the Objectness $O$. For all the remaining parameters requested by the SDC1 task, namely source flux $f$, major axis size $\textrm{Bmaj}$, minor axis size $\textrm{Bmin}$, and position angle $\textrm{PA}$, we can use the extra-parameter prediction capability of YOLO-CIANNA (Sect.~\ref{sec:add_pred})}. As stated in Sect.~\ref{sec:sub_challenge}, we ignore source classification and the associated core fraction and do not predict these properties. Still, the scorer requires values for these parameters to compute the per-source score. We adopted the simple approach of setting a constant value for these parameters for all sources. We found that setting $cf = 0.0375$ and $C = 2$ (corresponding to SFG) resulted in the best average subscores over the full catalog with $\bar{s}^{cf}= 0.9865$ and $\bar{s}^C = 0.9734$.

The output activation for all extra parameters is linear (Fig.~\ref{fig:yolo_out_vect}), so we have to normalize in a similar range. For the flux, \textcolor{black}{we convert it to apparent flux as described in Sect.~\ref{sec:sdc1_data}. Then we apply minimum and maximum clipping limits so $f_a$ is in the $[1.9{\times}10^{-6},2{\times}10^{-3}]$ Jy range, which is then passed through a $\textrm{log10}$ function. The resulting distribution is then linearly rescaled in the 0 to 1 range using the clipping limits as interval edges. For scoring, the predicted apparent flux can be converted back to intrinsic flux by inverting the operations. $\textrm{Bmaj}$ and $\textrm{Bmin}$ follow the same normalization scheme, with $\textrm{Bmaj}$ clipped in the $[0.9,60.0]$ arcsec range and $\textrm{Bmin}$ clipped in the $[0.3,30.0]$ arcsec range. They then go through a $\textrm{log10}$ function before being rescaled in the 0 to 1 range.} For $\textrm{PA}$, it is crucial to consider whether the source is resolved. Trying to predict the position angle of an unresolved source would result in training noise. We first search for all sources with $\textrm{Bmaj} \leq 1.8$ arcsec and attribute them a target of $\hat{\textrm{PA}} = 0$. This filtering only applies when defining the target, while nothing prevents the detector from predicting an angle for smaller sources at prediction time. We define two predicted parameters, $\cos{(\textrm{PA})}$ and $\sin{(\textrm{PA})}$. Considering that the target $\textrm{PA}$ is in the $[-90,90]$ degree range, we linearly rescale $\sin{(\textrm{PA})}$ to the 0 to 1 range. \textcolor{black}{Due to angular symmetries and degeneracies of \textrm{PA}, we obtained better results by predicting both $\cos{(\textrm{PA})}$ and $\sin{(\textrm{PA})}$ as output extra parameters, then reconstructing the predicted $\textrm{PA}$, than with a direct angle prediction.} In summary, each box predicts the following vector for each source $\left<x, y, w, h, P, O, f_a, \textrm{Bmaj}, \textrm{Bmin}, \cos{(\textrm{PA})}, \sin{(\textrm{PA})}\right>$. As discussed in Sect.~\ref{sec:add_pred}, we can set independent $\gamma^p$ scaling values for all parameters to balance their respective importance in the loss. We list the $\gamma^p$ we used for all parameters in Table~\ref{table:hyper_param_list}.

\subsubsection{Remaining hyperparameters}
\label{sec:sdc1_settings:hyper_parameters}

\textcolor{black}{For reproducibility purposes, we list all the hyperparameters for our application, which include those described in the appendix Appendix~\ref{sec:appendix:association_function} with the advanced description of the association process of YOLO-CIANNA. Understanding the specific behavior of each parameter should not be necessary to interpret the obtained result, and we refer to the Appendix when necessary.}

First, we use the ${\rm DIoU}$ as our matching metric (Appendix~\ref{sec:appendix:association_function:match_metric}), which implies that the $L^{{\rm fIoU}}$ limits can take values in the -1 to 1 range. Following our nine detection units definition with three scales, we set $\lambda_{\textrm{void}} = 0.15$ for the six small detection units and $\lambda_{\textrm{void}} = 0.01$ for the three larger ones. The $\lambda$ scaling factors, along with the preactivation scaling and limit values, are given in Table~\ref{table:scaling_param_list}. The various $L^{{\rm fIoU}}$ and some other remaining hyperparameters are given in Table~\ref{table:hyper_param_list}. For a more exhaustive view of all the detector hyperparameters, we recommend reading the provided example SDC1 scripts archived with CIANNA V-1.0.

\begin{table}[t]
\centering
\caption{\label{table:scaling_param_list} Scaling factors and limits related to the loss subparts.}
\begin{tabular}{ l c c c c c}
\hline
\hline
& Pos. & Size & Prob. & Obj. & Param. \\
\hline
$\lambda$ & 36.0 & 0.2 & 0.5 & 2.0 & 5.0 \\
Pre-activ. scaling & 0.5 & 0.5 & 0.2 & 0.5 & 0.5 \\
Pre-activ. max & 6.0 & 1.2 & 6.0 & 6.0 & 1.5 \\
Pre-activ. min & -6.0 & -1.2 & -6.0 & - 6.0 & -0.2 \\
\hline
\end{tabular}
\end{table}

\begin{table}[t]
\centering
\caption{\label{table:hyper_param_list} Detection layer configuration hyperparameters.}
\begin{tabular}{ l c c c c c}
\hline
\hline
 Parameter & $\gamma^{f_a}$ & $\gamma^{\textrm{Bmaj}}$ & $\gamma^{\textrm{Bmin}}$ & $\gamma^{\cos{(\textrm{PA})}}$ & $\gamma^{\sin{(\textrm{PA})}}$ \\
 Value & 2.0 & 2.0 & 1.0 & 0.5 & 0.5 \\
 \hline\\[-2.5ex]
 Parameter & $L^{{\rm fIoU}}_{\textrm{GBNB}}$ & $L^{{\rm fIoU}}_{\rm low}$ & $L^{{\rm fIoU}}_{P}$ & $L^{{\rm fIoU}}_{O}$ & $L^{{\rm fIoU}}_{p}$\\
 Value & 0.5 & -0.1 & -0.3 & -0.3 & -0.1 \\
 \hline
 Parameter & $S_{ar}$ & $\alpha_{\rm small}$ & $N_{\rm rand}$ & $\alpha_{\rm best}$ & $\alpha_{\rm rand}$\\
 Value & 0 & 0.0 & 16000 & 0.90 & 0.02 \\
\hline
\end{tabular}
\end{table}

\begin{table}[t]
\centering
\caption{\label{table:first_nms_limits} First NMS rejection limit pairs.}
\begin{tabular}{ l c c c c}
\hline
\hline
 $L^{{\rm fIoU}}_{\rm NMS}$ & 0.05 & -0.1 & -0.3 & -0.5 \\
 $L^{\rm obj}_{\rm NMS}$ & 1.0  & 0.7 &  0.5 &  0.3\\
 \hline
\end{tabular}
\end{table}

\subsection{Network training}
\label{sec:detector_training}

\begin{figure*}
    \centering
    \includegraphics[width=0.98\hsize]{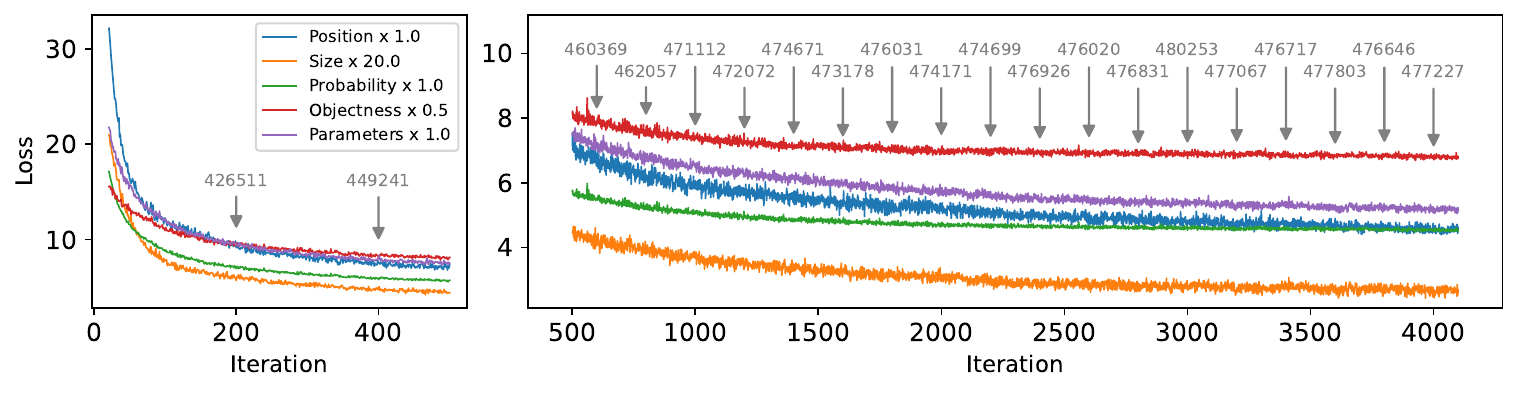}
    \caption{Evolution of the validation loss subparts during training, using the ``natural'' loss representation (Appendix~\ref{sec:appendix:cascading_loss}). The first 20 iterations are skipped. Optimized scores obtained using the procedure described in Sect.~\ref{sec:detector_inference} are displayed every 200 iterations. The small difference in exact score that can be observed compared to Table~\ref{table:team_scores} is due to rounding of the objectness thresholds, some minor adjustment of the largest prior threshold, and the use of FP16 mixed-precision at inference.}
    \label{fig:loss_curve_with_score}
\end{figure*}

The layer weights are initialized to random values following a Glorot normal initialization \citep{paper:xavier_init}, the batch size is set to $b_s=16$, and the numerical resolution is set to full 32-bit floating point for all network elements. \textcolor{black}{The training images are dynamically generated and augmented from the training area, as described later in the current section. The learning rate starts low at around $3\times10^{-6}$ and is increased linearly over the first 64000 training images up to the default value or $1.5{\times}10^{-4}$. Then, it follows an exponential decay of $3.13{\times}10^{-7}$ as a function of the number of training images up to a minimum value of $5\%$ of the default learning rate.} A constant weight update momentum of 0.8 is also used. No weight decay is used since it always resulted in lower scores. The initial random association is set to $N_{\rm rand} = 16000$ examples (Appendix~\ref{sec:appendix:association_function:random_startup}). 

With our fully convolutional architecture, we could technically use any input size that is a multiple of the network reduction factor (see Appendix~\ref{sec:appendix:conv_net_yolo}). A too-small input size can cause many cutout edge effects, \textcolor{black}{and a too-large input size reduces the number of independent examples we can generate from a finite training area}, increasing the chances of overtraining. We observed that a training input size of $256{\times}256$ pixels resulted in the best achievable score for our specific application. 

\textcolor{black}{To generate a training image, we draw a random pixel position inside the training area, extract a small region corresponding to our network input size, and apply pixel-preserving operations selected randomly from vertical or horizontal flips and $90$ or $-90$ degree rotations. Images from the additional noise region are selected at a low $5\%$ rate and follow the same augmentation rules. With this dynamic augmentation, it becomes impossible to define an epoch traditionally. By cutting out patches of the larger image, sources on the edges of a patch might be cut, which can negatively impact the detector capabilities (Sect.~\ref{sec:sdc1_selection_function}).} We tried various approaches for handling these sources (Appendix~\ref{sec:appendix:association_function:diff_flag}) but obtained the best scores with a simple exclusion of any target box that is not fully contained in the input patch. \textcolor{black}{With this dynamic generation scheme, the classical epoch definition cannot be used. Therefore, to ease the monitoring of the training, we consider 1600 generated images to form a group that we call an iteration and use this quantity instead of the number of seen examples to express the advance of the network training.}

Because the labeled area is small, splitting it into independent training and validation subsets would significantly reduce the context diversity for both subsets. The detection performance would be negatively affected, and we would have no guarantee that the validation set is representative. For this reason, our training examples are randomly drawn from the full labeled area. Our validation dataset is built from the same region and catalog but using a fixed grid of nonoverlapping patches for a total of 100 images. \textcolor{black}{We acknowledge that this validation set would fail to identify overtraining. However, we can rely on the SDC1 scorer to provide an independent metric on a separate dataset to evaluate our detector performances at a regular iteration interval.}

With this setup, the detector reaches a high score after only 400 iterations, but the extended sources are not properly constrained yet, and the global characterization is suboptimal. The detector usually requires around 3000 iterations to converge to its best score. After that, it oscillates for a few hundred iterations and eventually exhibits overtraining. We present the validation loss with scoring stamps every 200 iterations in Fig.~\ref{fig:loss_curve_with_score}. \textcolor{black}{We note that using stricter values for parameters that control the association process like $L^{{\rm fIoU}}_{\textrm{GBNB}}$ or $L^{{\rm fIoU}}_{\rm low}$ would speed up the training. It would result in more stable results, but it also decreases the best achievable score by up to $2\%$.}

On an RTX 4090 GPU and by doing image augmentation on the CPU in parallel, we can reach almost 400 images per second of training performances. Reaching iteration 3000 takes around 3.5 hours. With the current training setup, the GPU memory footprint of the network is around 8GB. A save file representing the model at a given iteration requires 50MB of storage. \textcolor{black}{We tried using mixed-precision training to leverage tensor-cores acceleration, but it always resulted in a score decrease of about $2\%$. Still, it speeds up network training by a factor of up to $1.8$ while preserving the relative performance impact of various hyperparameters}, allowing a more efficient exploration of network architectures and hyperparameter combinations.

\subsection{Prediction pipeline}
\label{sec:detector_inference}

To use our trained detector on the full SDC1 image, we need to decompose it into patches. \textcolor{black}{As stated before, our method allows us to use any input size for the prediction, even if it differs from the one used for training. By decomposing the image into patches, some sources will end up close to the edges of a patch, resulting in poor detection or characterization. To overcome this issue at prediction time, we use overlapping patches with an overlap size of 32 pixels. This value was chosen to be half the maximum size for our target boxes. It is also a multiple of the reduction factor, making it an integer number of complete grid cells.} With this setup, each source will likely be well-represented by at least one patch. In the case of multiple detections of the same source by two overlapping patches, having perfectly overlapping grid cells ensures that objectness scores are comparable. Thus, these multiple detections can be filtered through an inter-patch NMS process. From there, we observed that larger patches produce better results by reducing the proportion of sources close to an edge. We settled for a prediction input size of $512{\times}512$ pixels, which results in a good balance between prediction quality and compute efficiency. By setting a small offset on the edges of the full image, we can obtain a map composed of $67{\times}67$ partially overlapping patches. 

For each prediction patch, we first apply a per-detection-unit objectness threshold filtering. The threshold values are obtained through score optimization by identifying the objectness interval contributing positively to the score (Sect.~\ref{sec:sdc1_metric} and Eq.~\ref{eq:total_score}). \textcolor{black}{This optimization is only a refinement, and we would only lower our best score by a few percent by using naive guesses on threshold values instead of searching for the best ones. We also add a rejection criterion based on the average pixel flux inside the predicted box area as a function of the objectness score to exclude small false detections induced by very extended and bright sources.} Multiple detections are then filtered by the NMS process for each patch (Sect.~\ref{sec:filtering_and_nms}), using multiple $L^{{\rm fIoU}}_{\rm NMS}$ and $L^{\rm obj}_{\rm NMS}$ thresholds pairs that are given in Table~\ref{table:first_nms_limits}. Following the NMS process, considering the best current score box, any other box that respects one of the condition pairs ${\rm DIoU} > L^{{\rm fIoU}}_{\rm NMS}$ and $O < L^{\rm obj}_{\rm NMS}$ is removed. With this setup, the rejection chance is based on a combination of the detection confidence and the distance to other confident detections. Multiple detections from overlapping patches are filtered using a secondary inter-patch NMS process. It works as the first NMS but with a constant $L^{{\rm fIoU}}_{\rm NMS}=-0.15$. \textcolor{black}{From the filtered box list, we construct a catalog in the SDC1 scorer format by inverting our normalizations on the predicted parameters.}

On an RTX 4090 GPU, the raw prediction compute performance at the largest available batch size is around 300 images per second for an input size of $512{\times}512$ pixels, or almost 80 million pixels per second. The raw processing time for the full 4GB image of 32768 square pixels after the network loading is around 15 seconds. To this, we need to add the post-processing time of the prediction pipeline, which strongly depends on the minimum objectness threshold values for the different detection units. Using an end-to-end Python pipeline that produces the converted detection catalog from the raw detection catalog, the post-processing time is about 40s on a single CPU core (Ryzen 9 5900X). Interestingly, the detection performances are fully preserved when doing the prediction in mixed-precision using 16-bit numerical computations with a 32-bit accumulator when necessary, allowing the use of Nvidia AI-dedicated tensor-cores compute units available on modern GPU architectures. With this, we can reach around 500 images of $512{\times}512$ pixels per second, or around 130 million pixels per second, of raw computing performance. At this point, \textcolor{black}{the time required to save the results} on our high-end SSD storage becomes the limiting factor, and the network loading time becomes comparable to the prediction time over the full image. This confirms that our method is suited for application over large surveys and real-time detection (Sect.~\ref{sec:real_data_pred}).

\section{Results and Analysis}
\label{sec:results}

In this section, \textcolor{black}{we present the results of our best model over the full field of the selected SDC1 image.} We stress that there are small variations of results from retraining a network with the same training setup, with typical variations of the best achievable score around $\pm 0.5\%$. In practice, the detailed analysis of this specific training holds perfectly with any training we have done using the same setup. The results in this section are ordered following the post-processing pipeline from the raw predicted quantities up to the predicted source catalog using the challenge format. The validation loss curve as a function of the training iteration is presented in Fig.~\ref{fig:loss_curve_with_score}. This figure also indicates the optimized score at a regular interval of 200 iterations. The following results are presented for \textcolor{black}{iteration 3000,} which achieved the best score for this specific training.

\subsection{Patch-divided prediction results}
\label{sec:results:patched_results}

\begin{figure}
    \centering
    \includegraphics[width=1.0\hsize]{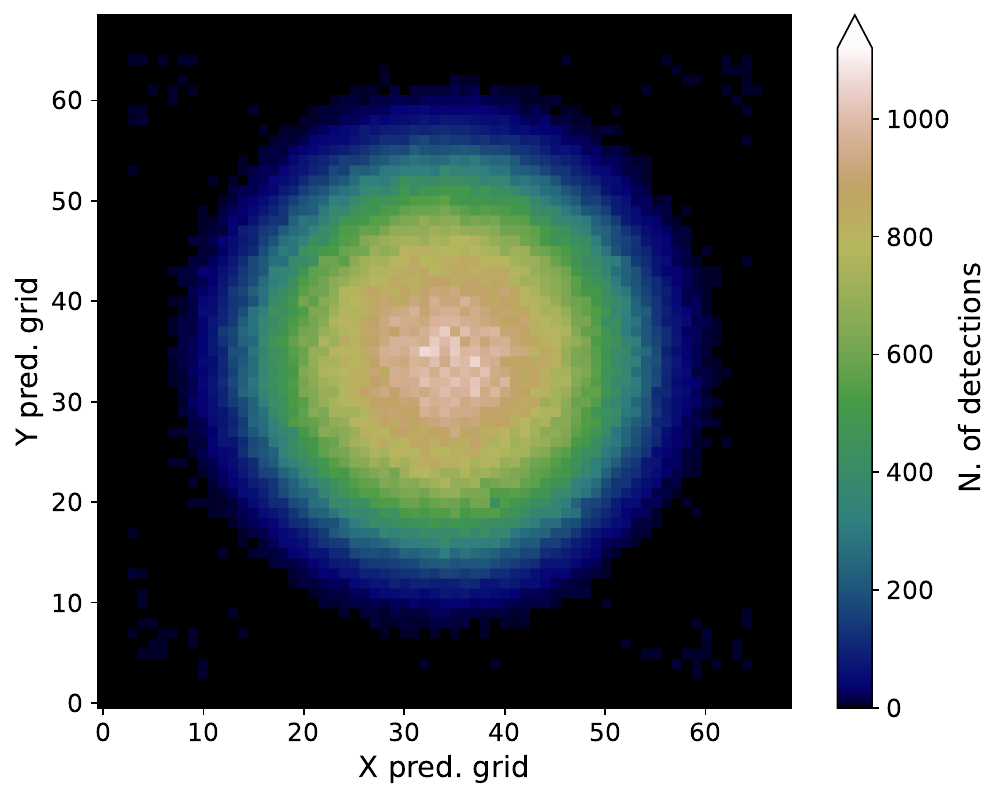}
    \caption{2D representation of the full SDC1 image field decomposed into overlapping patches. Each pixel corresponds to a specific patch, and the color represents the number of detected sources after objectness filtering and in-patch NMS. The central coordinates of the field are $\textrm{RA}=0 $ deg, $\textrm{Dec}=-30$ deg.}
    \label{fig:pred_grid_density_map}
\end{figure}

From the raw model prediction, after applying the first objectness filtering and the in-patch NMS, we can obtain a list of all the detected sources per patch. We represent the respective number of sources for all patches over the whole image field in Fig.~\ref{fig:pred_grid_density_map}. We note that some sources contribute to multiple bins in this representation because the inter-patch NMS has not yet been applied. We see that the detector sensitivity follows the primary beam imprints over the full image. We do not observe a clear overdensity of detected sources in the training area region, which tends to indicate the absence of overtraining.

\begin{figure}
    \centering
    \includegraphics[width=1.0\hsize]{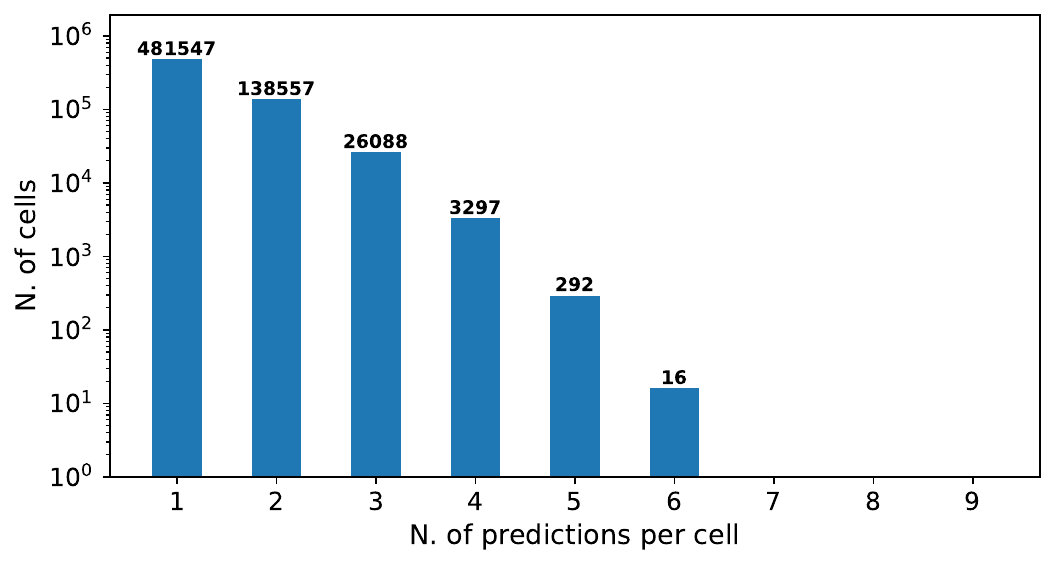}
    \caption{Distribution of the number of detections in a single cell (nonempty only) after objectness filtering and in-patch NMS.}
    \label{fig:pred_per_reg_dist}
\end{figure}

\begin{figure}
    \centering
    \includegraphics[width=1.0\hsize]{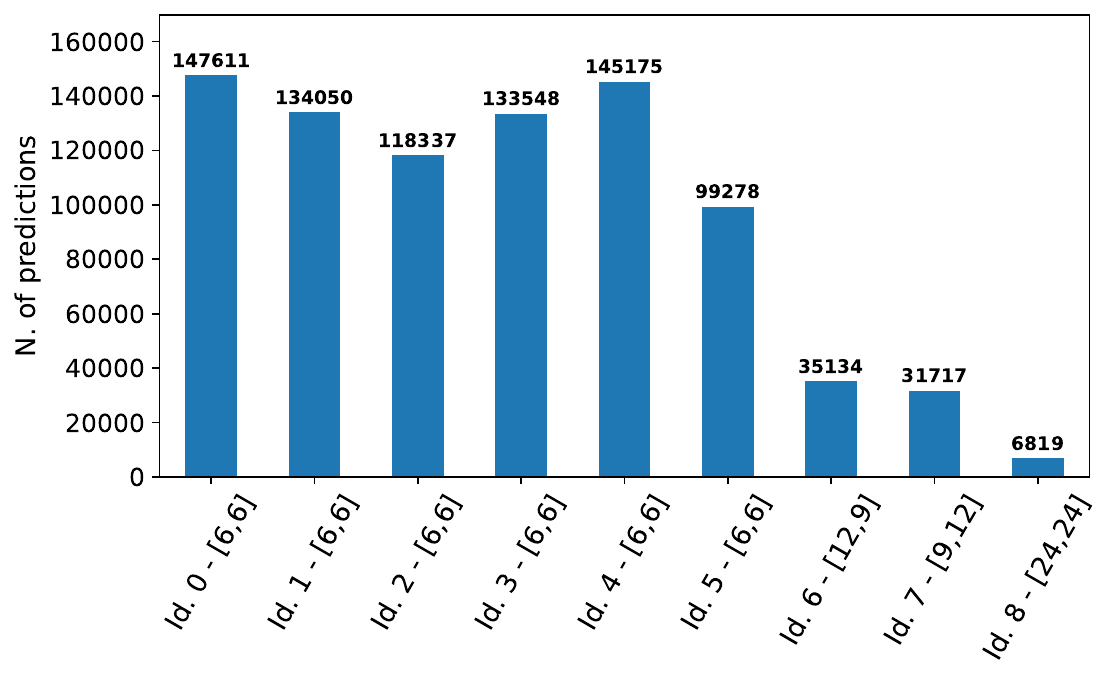}
    \caption{Distribution of the detections regarding the detection unit that produced them, after objectness filtering and in-patch NMS. The size prior is indicated for each detection unit.}
    \label{fig:pred_prior_size_dist}
\end{figure}

To analyze in detail the behavior of the individual detection units from our detector output (Sect.~\ref{sec:multiple_boxes}), we represent a histogram of the number of found-useful detections per grid cell for all patches in Fig.~\ref{fig:pred_per_reg_dist}. \textcolor{black}{Here, empty grid cells are excluded, but they represent the most common case.} We observe that cases with six or more detections are very few, four or five detections are not common but significant, and most nonempty grid cells contain less than three detections. Further analysis indicates that only around $26\%$ of the nonempty grid cells have more than one detection. This proportion drops to $4.5\%$ for more than two detections. In practice, the in-patch NMS process removes $78\%$ of the raw good detections. This indicates that many detection units are \textcolor{black}{in use simultaneously and capable of detecting real sources,} but they often end up detecting the same sources. Still, we observed that lowering the number of detection units lowers the score, indicating that they are useful for some specific contexts or that having more detection units better distributes the information. To further verify this idea, we represent in Fig.~\ref{fig:pred_prior_size_dist} a histogram of the \textcolor{black}{number of predictions per detection unit for the whole image after the in-patch NMS.} The distribution over the different detection units from the small-size regime is relatively homogeneous, confirming that all detection units are useful. We also observe that the relative proportions of each size regime are well preserved regarding the theoretical target distribution from Fig.~\ref{fig:targ_prior_size_dist}, which confirms that our association process properly distributes the information over the available size regimes.

\begin{figure*}
    \includegraphics[width=\textwidth]{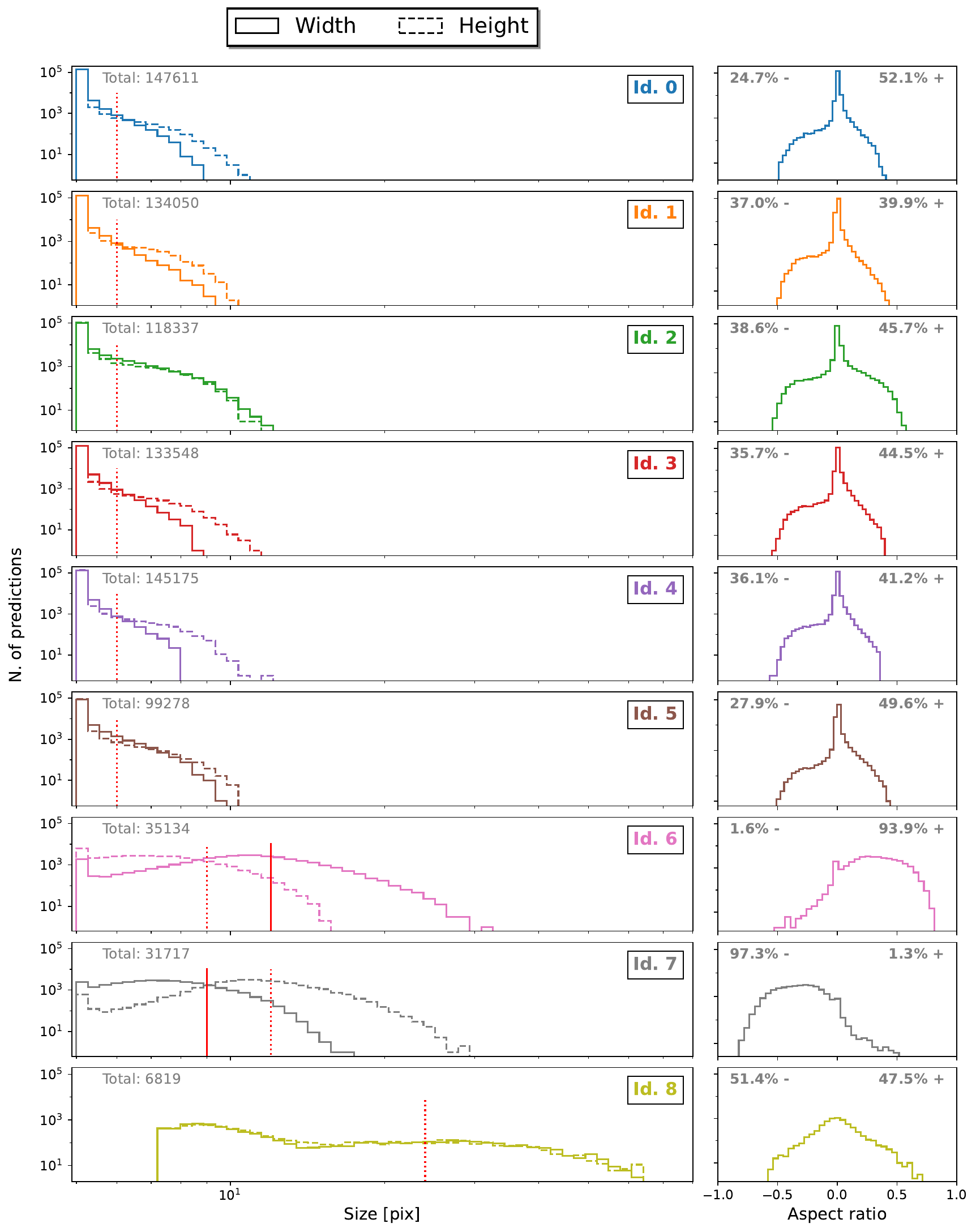}
    \caption{Histograms of the predicted heights, widths, and aspect ratios for every detection unit. \textcolor{black}{The object count and size axes are in log scale.} The aspect ratio is computed as \mbox{$(w - h)/\max{(w,h)}$}. The number of predictions and the proportion of strictly positive and negative aspect ratios are also given for each detection unit. The red vertical lines indicate the width and height of the size priors.}
    \label{fig:pred_dist_size_per_prior}
\end{figure*}

To better understand the representation regime of individual detection units, we can look at their respective distribution of predicted sizes and aspect ratios in Fig.~\ref{fig:pred_dist_size_per_prior}. We observe that some boxes are more specialized regarding aspect ratios, which is natural for the intermediate size regime due to the two different ratios of size priors. \textcolor{black}{The behavior of the small-size regime detection units is more surprising, as they all appear to prefer a positive aspect ratio when predicting small boxes and a negative aspect ratio when predicting large boxes. The cause of this behavior remains unidentified for the moment, but it is reproducible in all of the backbone and hyperparameter configurations we tried. Finally, the detection unit from the large-size regime and usually one or two from the small-size regime also predict sizes representative of the intermediate regime. Due to the difference in aspect ratio, only one of the intermediate-size regime units is likely to be used in a given grid cell. When two or more sources of intermediate size have to be detected simultaneously, a small or large box with a square prior might be better than using the wrong aspect ratio from the appropriate size regime.}

\subsection{Full catalog scoring}
\label{sec:results:full_catalog}

After the inter-patch NMS filtering and conversion of all the predicted source parameters, we obtain a source catalog for the full image that can be put through the SDC1 scorer (Sect.~\ref{sec:sdc1_metric}). \textcolor{black}{It automatically excludes the training area and produces a detection score for the rest of the image. Our score is presented in Table~\ref{table:team_scores} under the name MINERVA (YOLO-CIANNA), along with scores from other methods. In addition to the challenge score, we also indicate the number of sources retrieved, the precision of the submitted catalog, and the average source characterization score. For a fair comparison, we recomputed the scores from the available detection catalogs for each method using the same scoring code. The top three results from the original challenge leaderboard on the same single 560 MHz - 1000h image are indicated for reference.} The Engage-SKA team used the ProFound package \citep{paper:profound}, a collection of astronomical image processing and analysis tools that produce segmentation maps. The Shanghai team worked with a combination of three classical astronomical source-detector packages, namely AEGEAN \citep{paper:aegean}, DUCHAMP \citep{paper:duchamp}, and SExtractor \citep{paper:sextractor}, but ended up using only the last one to construct its catalog. Finally, the ICRAR team used the deep learning CLARAN method \citep{paper:claran}, with a region-based method similar to R-CNN \citep{paper:rcnn}. We note the participation of another team (not in our table) that used a very shallow convolutional network structure for segmentation, achieving a much lower score with the ConvoSource method \citep{paper:convo_source}. We also include another post-challenge score from \citet{paper:jlrat2}, which is a refinement of the JLRAT team that participated in the original challenge but with a lower submitted score, also with a deep learning-based method using a combination of a region-based approach with a feature pyramid network. Their new catalog is publicly available and is provided in the scorer format, allowing us to include their result on the same image in the table. \textcolor{black}{Our SDC1 source catalog is also made publicly available \citep{data:minerva_sdc1_catalogs}\footnote{Our reference SDC1 detection source catalog along with other interesting catalogs described in the present paper have been archived at \href{https://doi.org/10.5281/zenodo.13141772}{10.5281/zenodo.13141772} in the scorer format} with the present paper so the full score table can be reproduced.}

\begin{figure*}
    \centering
    \begin{subfigure}{0.49\hsize}
    \caption*{\textbf{Score-focus selection}}
    \includegraphics[width=1.0\hsize]{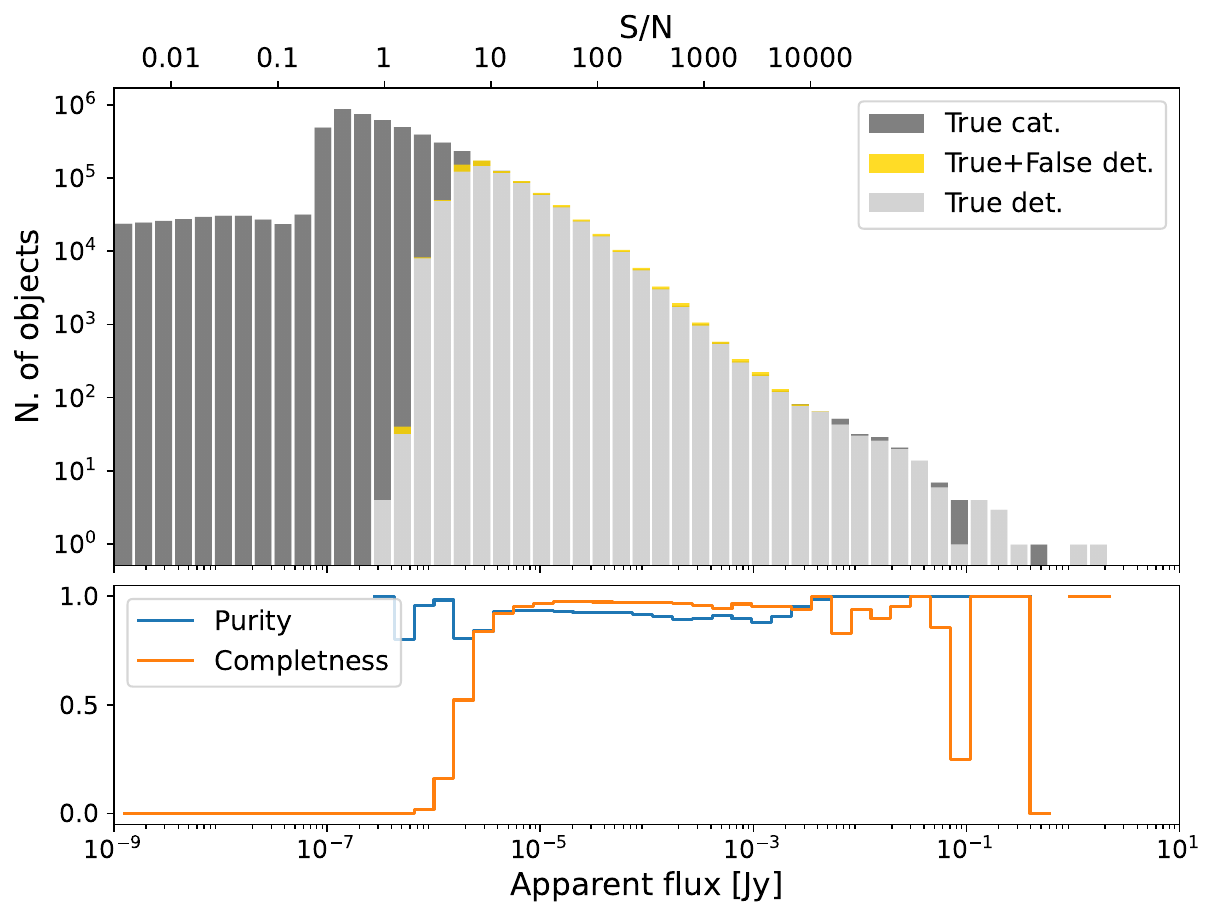}
    \end{subfigure}
    \begin{subfigure}{0.49\hsize}
    \caption*{\textbf{Purity-focus selection}}
    \includegraphics[width=1.0\hsize]{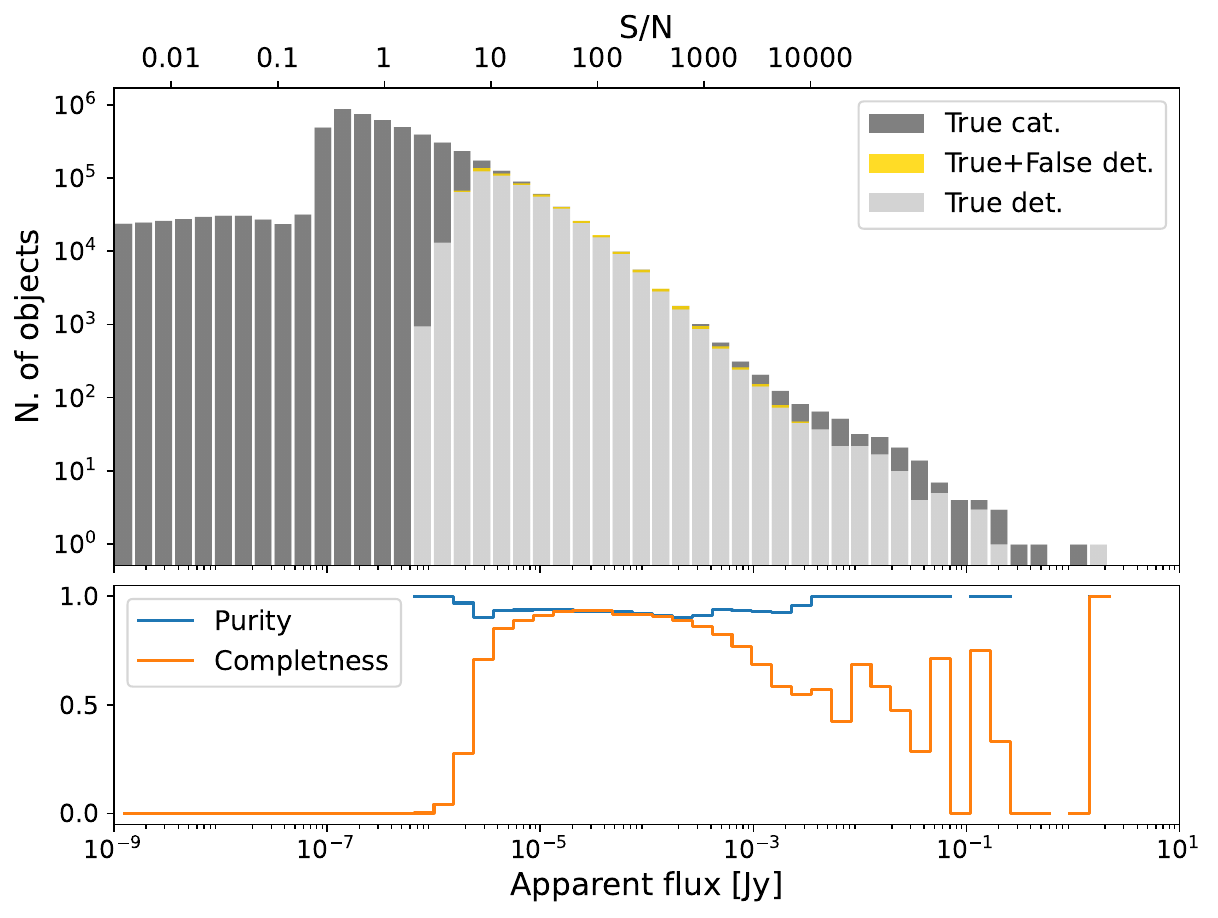}
    \end{subfigure}\\
    \begin{subfigure}{0.49\hsize}
    \includegraphics[width=1.0\hsize]{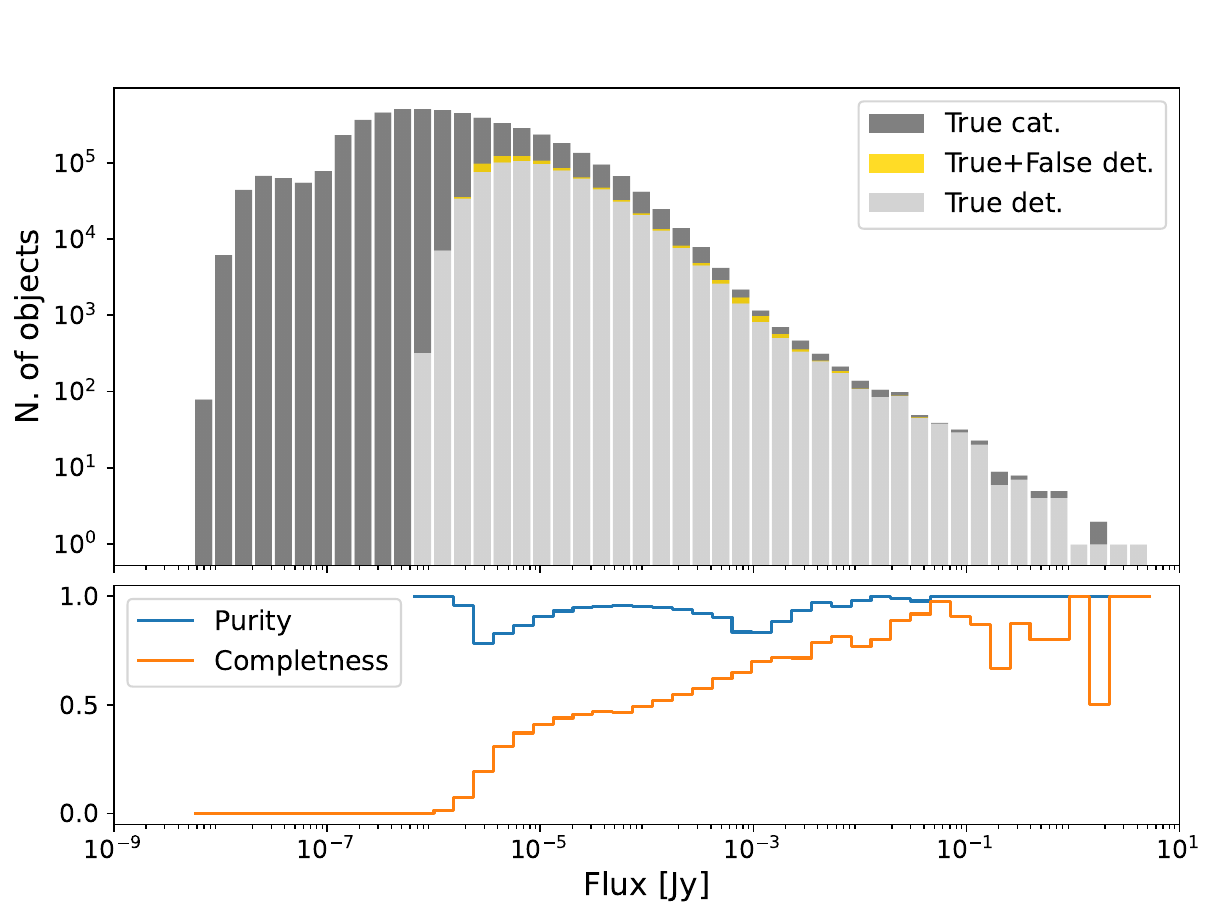}
    \end{subfigure}
    \begin{subfigure}{0.49\hsize}
    \includegraphics[width=1.0\hsize]{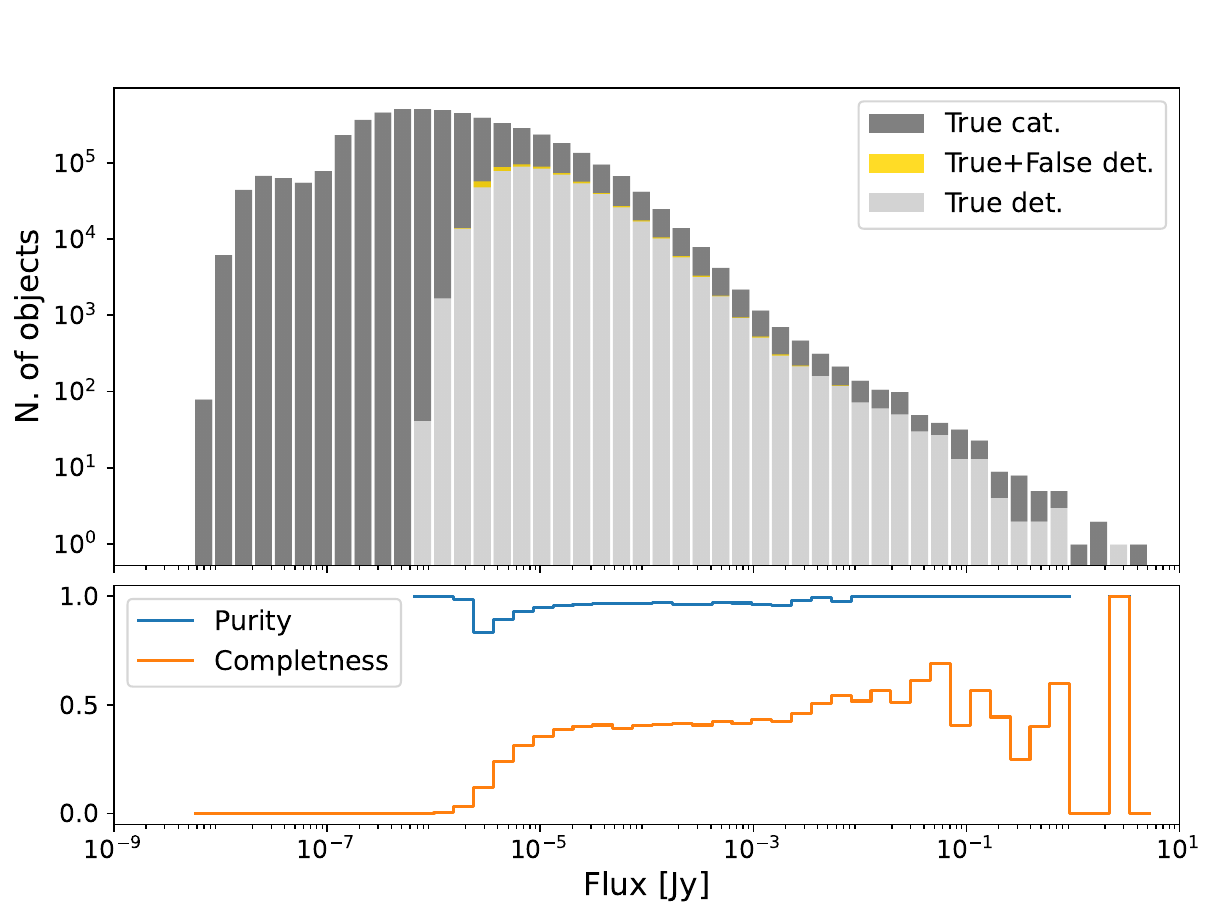}
    \end{subfigure}
    \caption{Histograms of the sources as a function of their flux using logarithmic bins for the underlying True catalog and the predicted sources in the testing area. The true source flux is used for matches, while the predicted flux is used for false positives. The bottom part of each frame represents the purity and completeness of each bin of the histograms (defined in Sect.~\ref{sec:results:full_catalog}). The top frames use the apparent flux after multiplication with the primary beam, while the bottom frames use the integrated flux. The columns represent the two possible post-process objectness-thresholds search objectives over the same trained YOLO-CIANNA model.}
    \label{fig:flux_distribution_full_false_true}
\end{figure*}

The source catalog obtained with our YOLO-CIANNA detector improves the best-submitted score of the originally participating teams by +139\% and the score of the only other published post-challenge score by +61\%, which already outperformed what was submitted during the challenge. \textcolor{black}{We observe that most teams focused on catalog purity to achieve high scores.} This is likely to be due to their relatively low average source characterization score, which is discussed in Sect.~\ref{sec:results:source_characterization}. As the apparent flux should be an appropriate first-order proxy to represent the detection difficulty, we present the apparent flux distribution of our detection list in comparison to the True catalog in Fig.~\ref{fig:flux_distribution_full_false_true}. \textcolor{black}{We also present the corresponding completeness and purity for each flux bin.} The completeness represents the proportion of total sources we successfully detected, and the purity represents the proportion of the proposed detection that are real sources. These quantities are computed based on the full True catalog and SDC1 matching criteria. Comparing this flux distribution with the one from our selected training sample (Fig.~\ref{fig:select_function_dist_flux}), we observe that our detector can identify sources outside its training flux range close to the noise limit. This is likely to correspond to sources for which the perceived apparent flux was increased by the local noise or blended faint sources that add their flux, leading the network to detect them as a single brighter source (Sect.~\ref{sec:sdc1_selection_function}). These two cases will result in sources with a target apparent flux close to or below the detection limit to be detected. This out-of-range detection capability is also visible in the example fields from Fig.~\ref{fig:detection_fields}. The completeness is high in the intermediate apparent flux range but drops progressively at lower flux values. While it can be explained by sources getting under the detection noise limits, it can also come from sources that are too blended to be detected individually. In such a context, the prediction will be associated with the brightest target, removing mostly low apparent flux sources. This is also visible in the example fields from Fig.~\ref{fig:detection_fields}. \textcolor{black}{On the opposite side of the flux range, our detector misses a few bright sources, which can be explained by the absence of similar examples in our training sample.} While they are too few to influence the scoring by their absence, they can still have a strong effect by adding false positives, which is discussed in Sects.~\ref{sec:detector_inference} and~\ref{sec:results:patched_results}.

\begin{table*}
\centering
\caption{\label{table:team_scores}. SDC1 scores and related properties for source catalogs from different teams and methods.}
\begin{tabular}{ l c c c c c c c}
 \hline
 \hline
 Team (method) & $M_s$ (Score) & $N_{\rm det}$ & $N_{\textrm{match}}$ & $N_{\rm false}$ & $N_{\rm bad} \in N_{\rm false}$ & Purity & $\bar{s}$ \\
 \hline
 \textit{Post-challenge results} & & & & & & &\\
 \hline
 \textcolor{black}{MINERVA (YOLO-CIANNA)}         & \textbf{480450} & 724480 & 680000 & 44480 & 16839 &         93.86\%  & 0.7719\\
 \textcolor{black}{\textit{\quad $\hookrightarrow$ purity-focus thresholds}} &         418434  & 541542 & 536412 &  5130 &  2506 & \textbf{99.06\%} & 0.7896\\
 JLRAT2 (JSFM2)                                                                          & \textbf{298201} & 502146 & 484212 & 17934 &  2274 &         96.43\%  & 0.6529\\[2ex]
 \hline
 \textit{Original challenge results} & & & & & & &\\
 \hline
 Engage-SKA (PROFOUND)                      & \textbf{200939} & 421992 & 418384 &  3608 &  2677 &         99.15\%  & 0.4889\\
 Shanghai (multiple methods)                & \textbf{158841} & 292646 & 291553 &  1093 &   698 &         99.63\%  & 0.5486\\
 ICRAR (CLARAN)                             & \textbf{142784} & 279898 & 259806 & 20092 &  6875 &         92.82\%  & 0.6269\\
 ... & ... & ... & ... & ... & ... & ... & ... \\
 \hline
\end{tabular}
\caption*{\vspace{-0.2cm}\\ Note. The bold element represents the target metric that is optimized for each result.}
\end{table*}

\subsection{Alternative purity focused detection}
\label{sec:results:purity_focus_results}

\textcolor{black}{By trying to optimize for the best possible SDC1 score, we are forced toward a predefined catalog purity. Consequently, our per-source characterization score defines the ratio of true-to-false detections above which new detections add score points (Sect.~\ref{sec:sdc1_metric}). This relation is nonlinear, as objects that are difficult to detect will likely be challenging to characterize. This is why we optimize the score using the ordered detection in objectness, which reflects the detector confidence (Sect.~\ref{sec:detector_inference}). If catalog purity is the main concern, we can adjust the objectness threshold selection to preserve only more confident detections using the exact same model (no retraining is necessary). With this approach, we can evaluate the performance of a given trained detector in different confidence regimes. To illustrate this capability, we present the result obtained with the same trained YOLO-CIANNA model but with a post-process search for objectness thresholds that enforce a 99\% purity. The score for this purity-focus catalog is presented in Table~\ref{table:team_scores} as a subproduct of our model. As expected, keeping only the most confident sources reduces the score by around 13\% compared to our score-optimized catalog, as it removes sources that were contributing a positive score. Still, this score remains much higher than the other methods with a similar purity. We observe that the average source score also increases with this selection, which confirms that the predicted objectness captures information about both the detection and characterization difficulty of the sources.}

We represent the apparent flux distribution of this purity focus catalog in Fig.~\ref{fig:flux_distribution_full_false_true}. Comparing this distribution with our default result shows that the detections at low and high apparent fluxes get removed first. While the link between low flux and difficulty is straightforward, it is less clear why bright objects get removed more than intermediate apparent flux regimes. \textcolor{black}{It is likely due to the small number of bright source examples in the training sample that leads to poor characterization, resulting in lower detection confidence. It can also be related to the fact that extended sources often exhibit bright-compact regions over much fainter extended structures, resulting in target bounding boxes that do not correlate well with the detectable part of the objects. This naturally prevents the detector from predicting the appropriate box sizes and lowers the geometrical confidence reflected in the objectness of extended sources.}

As discussed in Sect.~\ref{sec:probability_objectness}, having a predicted confidence score mapped by a continuous function implies that the trained detector can sample all its sensitivity regimes simultaneously at prediction time. \textcolor{black}{By optimizing for the best SDC1 score or for a given purity, we only select specific sensitivity thresholds.} From a user perspective, a given trained model can be adapted to various applications with either completeness or purity requirements. We discuss how this principle could be used to design an alternative detection metric in Sect.~\ref{sec:methode_improvement_disc}.

\subsection{Source characterization}
\label{sec:results:source_characterization}

\begin{figure*}
    \centering
    \includegraphics[height=0.90\textheight]{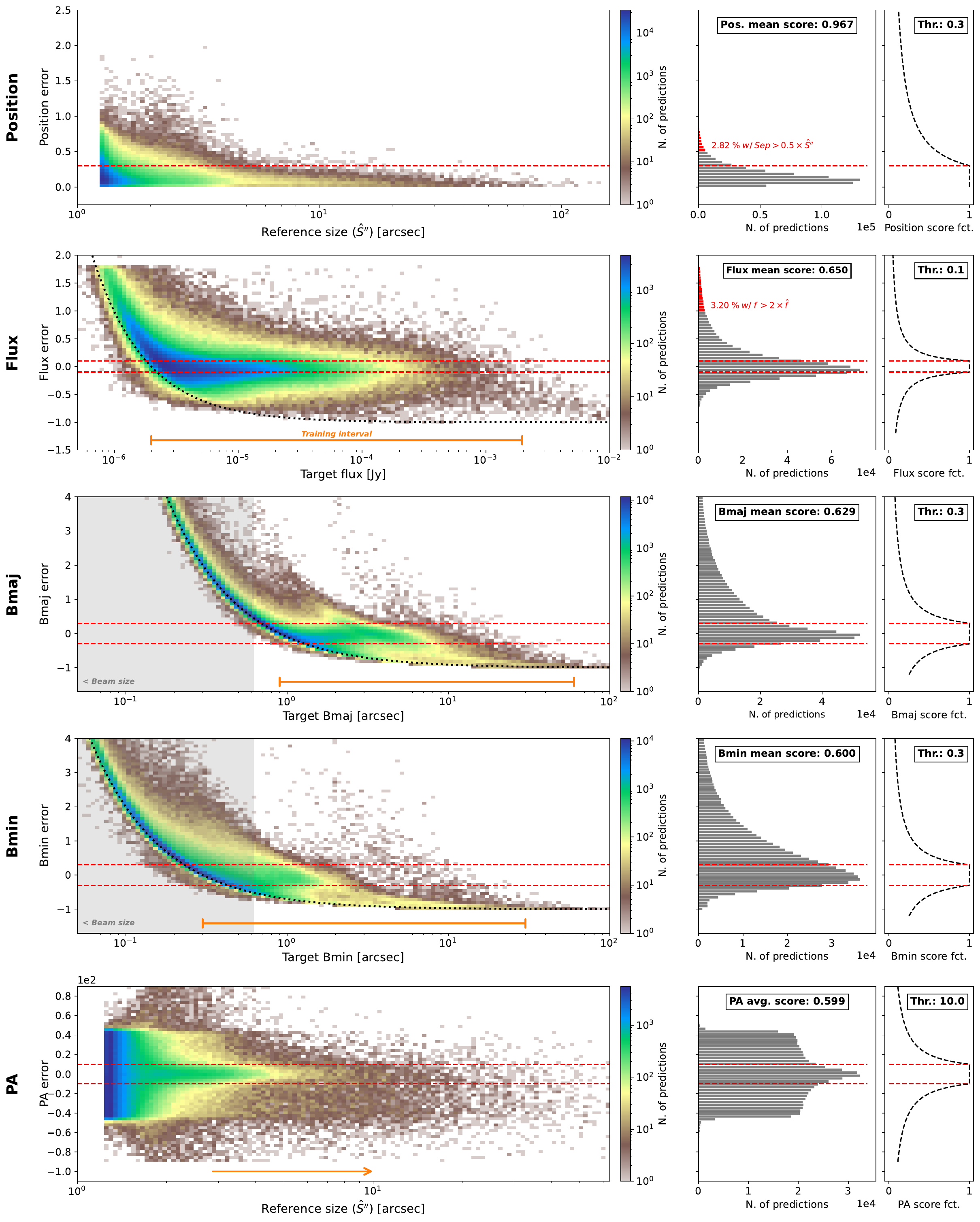}
    \caption{Two-dimensional histograms of the different parameter errors defined by the scorer as a function of relevant difficulty proxies. The \textit{left} column represents the two-dimensional histograms. The right column represents a 1D histogram of the relative error projected over the same axis as the \textit{left} column. To the right of this projection and using the same axis projection, we indicate the specific response of the score function for each parameter following Eq.~\ref{eq:score_response_fct} and Fig.~\ref{fig:param_score_fct}. The red dashed line indicates the edges of the area for which the parameter score is saturated at its maximum value. \textcolor{black}{The orange lines and arrow indicate the value range available in the training sample.} The gray overlay represents the size regime for which the sources are smaller than the beam size.}
    \label{fig:all_param_pred}
\end{figure*}

\textcolor{black}{In addition to the global score, the scorer code can produce detailed results, which include a list of matches along with the predicted and target properties for each.} In the following sections, we use these products to perform a detailed analysis of the characterization capabilities of our YOLO-CIANNA detector.

A primordial aspect of our produced source catalog is its high source characterization performance with $\bar{s} = 0.7703$. Before analyzing it in detail, it is important to highlight that parameter prediction accuracy is impossible to decorelate from the detection capability in the SDC1 scorer. The average parameter score ignores the fact that the characterization is not equally difficult for all sources. For example, we could choose to detect only the ten sources that are the easiest to characterize and obtain a very high characterization score. In contrast, a method that detects many sources, including some that are difficult to characterize, would get a lower average characterization score even if it is better at characterizing the specific ten easiest sources. This effect is well illustrated by the increase in average source score for our alternative purity-focused result compared to our default score result in Table~\ref{table:team_scores}. Overall, this average source score metric penalizes more efficient detectors that properly identify more difficult sources. To overcome this limitation, it would be possible to order or bin the sources by difficulty before evaluating the characterization capability of different methods on each bin. However, defining a characterization or detection difficulty proxy that would be method agnostic is likely impossible. The best proxies we have are source brightness or signal-to-noise ratio (S/N) measurements. A better solution to compare method characterization capabilities would be to use subsets of sources detected by the different methods. In our opinion, this is a significant limitation of the analysis from both SDC1 and SDC2 summary papers. 

Our approach to overcome this difficulty is to use 2D histograms of \textcolor{black}{the parameter errors from sources matched by the scorer as a function of a difficulty proxy,} presented in Fig.~\ref{fig:all_param_pred}. This figure also comprises histograms of the sources regarding only the parameter errors, which are directly aligned with their respective score response functions to ease the comparison. The following subsections detail the prediction capability for each type of predicted parameter.

We achieve a very high positioning score of $\bar{s}^{\textrm{pos}} = 0.964$, which is, in fact, a requirement for reaching a high detection recall and precision regarding the scorer match criteria. For point sources to be considered a match, we must reach subpixel positioning accuracy (Sect.~\ref{sec:sdc1_metric}). This typically explains why we configured our detector with a strong focus on the position loss in Table~\ref{table:scaling_param_list}. From this point of view, the 0.3 error threshold for the position seems relatively permissive and is not very useful in comparing methods regarding their positioning accuracy.

For the flux prediction, the distribution mostly centers on the score function response, granting a good average flux score of $\bar{s}^{\textrm{flux}} = 0.646$. As for the position, sources with a bad flux prediction are more likely to be rejected based on the scorer match condition (Sect.~\ref{sec:sdc1_metric}), imposing practical limits on the accessible flux dispersion. \textcolor{black}{For faint sources, the detector overestimates the flux, which is typical of a prediction over a noise-limited image. Overall, the detector tends to predict the minimum flux value it was trained on for fainter objects.} Still, the faintest sources are only detectable when the local noise increases their apparent flux. This can result in a selection effect bias that causes the network to overestimate the underlying flux of faint sources. This can explain the spreading of the flux predictions above the minimum training value prediction. \textcolor{black}{For the few bright sources, our detector underestimates the predicted flux, which results from the combination of several factors: i) our maximum value at training time will push the network to underestimate all fluxes above the limit, ii) high flux sources are represented mainly by a rare class of objects (Sect.~\ref{sec:sdc1_selection_function}) that are only detectable at this flux regimes meaning that they are unconstrained, and iii) our hyperbolic tangent transform applied in the input renormalization might slightly reduce the accessible information for high flux sources (Sect.~\ref{sec:sdc1_settings:input_norm})}.

Regarding $\textrm{Bmaj}$ and $\textrm{Bmin}$, the global predictions are convincing and centered on the score response function with $\bar{s}^{\textrm{Bmaj}} = 0.627$ and $\bar{s}^{\textrm{Bmin}} = 0.594$. We also observe the same type of out-limit generalization behaviors. Here, these effects are reinforced by the fact that the True values to be predicted are not convolved with the synthesized beam. In practice, the detector is expected to predict nonavailable information for most detectable point sources. Still, we observe that predictions for $\textrm{Bmin}$ are surprisingly not collapsed toward a single solution below the synthesized beam size. This indicates that the network can extract residual information about the source size under the resolution limit or reconstruct this information from correlated parameters. For $\textrm{Bmaj}$ specifically, we observe that the predictions are distributed over two independent prediction regimes that both follow the same type of error function. This separation is induced by the source distribution over different detection units based on their size. This creates a discontinuity as only detection units with a large size prior are properly constrained to predict large $\textrm{Bmaj}$ values, while small detection units with small size priors are constrained to predict small $\textrm{Bmaj}$ values. \textcolor{black}{This effect is already strongly mitigated by the addition of the random box association regularization in our association function (Appendix~\ref{sec:appendix:association_function:association_tricks} and Fig.~\ref{fig:association_algorithm}) that forces detection units to generalize outside} of their size range. When forcing more detection unit independence, this separation between different size regimes gets even more striking in the figure.

For the position angle, the average score is surprisingly high with $\bar{s}^{\textrm{PA}} = 0.601$, considering that most detected sources are smaller than the beam resolution limit. From the distribution, we observe that the prediction is not too noisy and that small sources distribute evenly over the error range, while the prediction accuracy improves progressively with the increase in source size.

\subsection{Example fields}
\label{sec:results:field_detection}

\begin{figure*}
    \centering
    \includegraphics[width=1.0\textwidth]{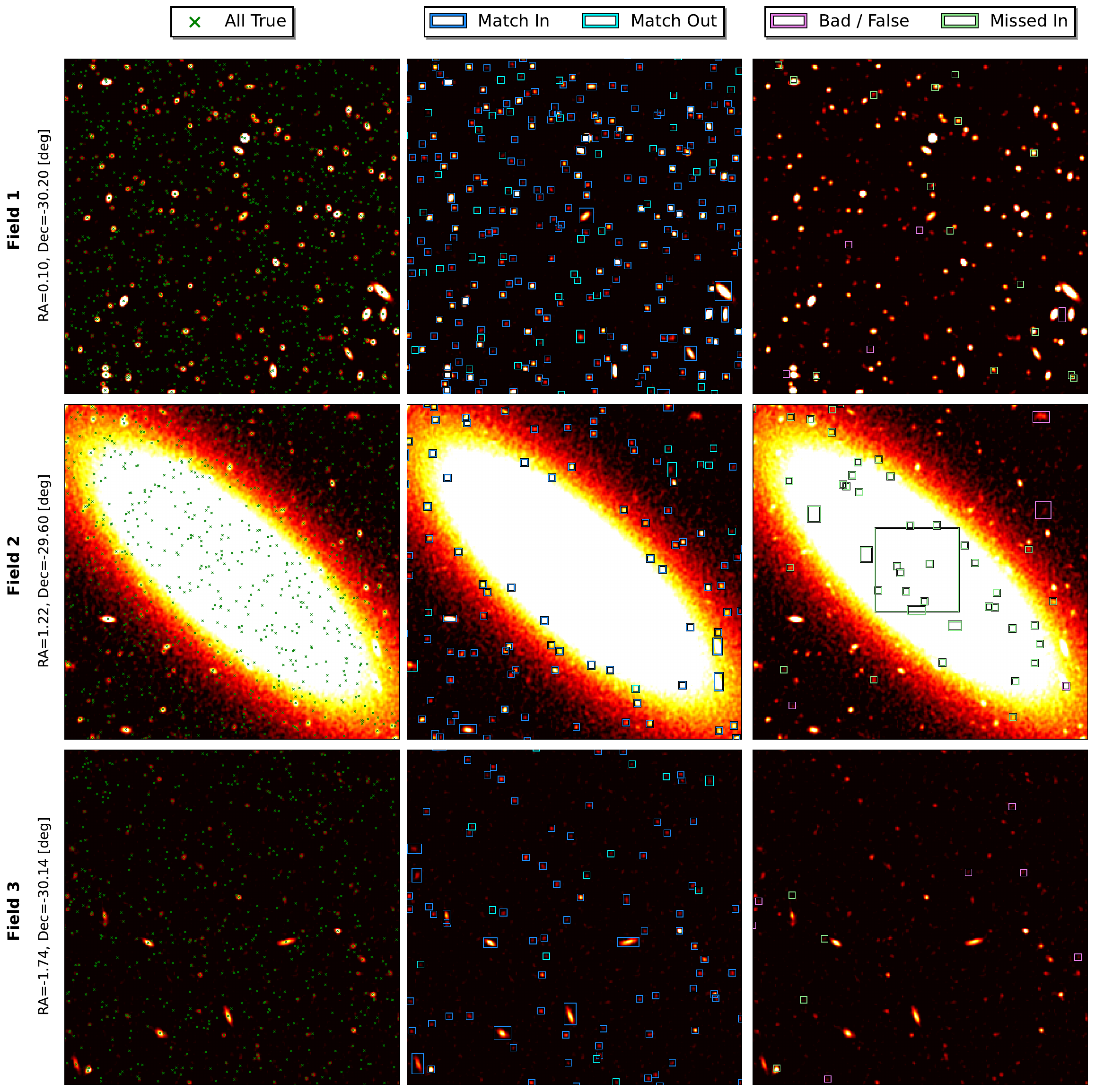}
    \caption{Distribution of sources and predictions for three 256 square pixels example fields. The background images use the same clipping and normalization as the network input (Sect.~\ref{sec:sdc1_settings:input_norm}) but are saturated at $50\%$ of the maximum. Each line represents a different field. \textit{Left:} scatter plot of the central source coordinates from the full True catalog. \textit{Middle:} boxes for predictions that match a true source based on the scorer. The match-out boxes indicate properly detected sources with a target flux below the minimum value in the training sample, so outside our selection function criteria (discussed in Sect.~\ref{sec:results:full_catalog}). \textit{Right:} boxes for true sources that pass our selection function but are not detected, and boxes for predictions that are false positives based on the scorer.}
    \label{fig:detection_fields}
\end{figure*}

\textcolor{black}{We illustrate the detected boxes from our model over a few example fields outside the training area in Fig.~\ref{fig:detection_fields}. We use the same field size as our training input window, namely $256{\times}256$ pixels. The first field represents the central high-sensitivity region where our detector reaches peak detection performance. This field illustrates the capability of our detector to identify faint sources outside its training interval} (Sect.~\ref{sec:results:full_catalog}). A few sources are considered missed, but most of them are blended with sources already correctly detected. The few false detections can be separated into three cases: i) it is near a properly detected source, but our NMS process did not remove it, ii) it is near a source, but the matching criterion failed, certainly due to the target flux being lower than the minimum accessible flux, or iii) it is near a compact signal that does not correspond to a source. The first and second cases are the most common.

The second field illustrates the extreme case of an extended resolved galaxy. As previously stated, there is no similar source in the training area that the detector could have learned to identify. Therefore, the large source itself was expected to be missed. However, this field illustrates our capability to prevent false detections in such a context. Only a few detections are considered false, and most of them seem to represent a real underlying source, indicating that the matching criteria likely rejected it because of the predicted flux or size. Several blended sources are also properly detected on the edges of the extended source. \textcolor{black}{The detector manages to identify blended sources in bright regions of the extended central source that appear saturated. The background images in this figure are, in fact, oversaturated to increase contrast over faint sources. Using the actual input image provided to the network, these detected sources are also obvious to the eye. There are many missed sources in this context, which are completely hidden by the bright source and are therefore undetectable. These missed undetectable sources highlight a typical possible improvement of our training selection function, which could be modified in order to remove them.}

\begin{figure*}
    \centering
    \includegraphics[width=1.0\hsize]{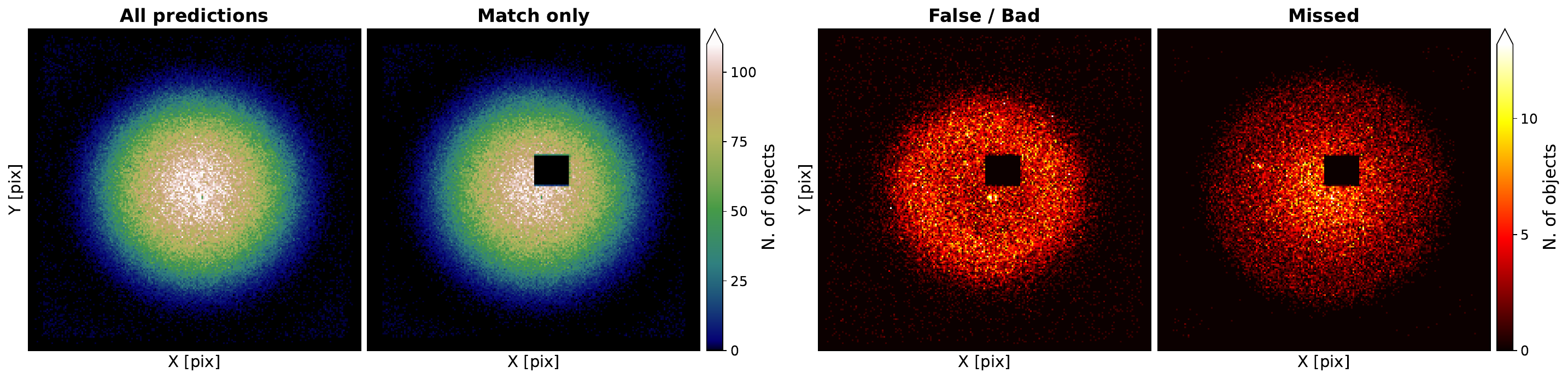}
    \caption{Two-dimensional histograms of various kinds of ``objects'' representing their distribution over the full SDC1 field  \textcolor{black}{for the YOLO-CIANNA model. The training area is masked when necessary. All densities are binned using the same 200x200 grid.} The match and false detections are based on the scorer, while the missed ones are based on our selection function. The central coordinates of the field are $\textrm{RA}=0$ deg, $\textrm{Dec}=-30$ deg.}
    \label{fig:pred_match_false_field_map}
\end{figure*}

The last field illustrates a region far from the image center but still in a primary beam regime where several sources can be detected. This is typical of the intermediate regime between the default training area and our noise-only areas. With this field, we can verify the capability of our detector to interpolate between the two training contexts. \textcolor{black}{We observe that most sources are properly detected. Only a few are considered missed, and their origin is likely the same as in the two other example fields. This field is typical of the regions of the whole image for which we identify a small overdensity of false detection in Fig~\ref{fig:pred_match_false_field_map}.}

\subsection{Whole image field distribution}

Using the match criterion from the scorer, we can look at the match distribution over the whole field and the distribution of false positives and missed sources, which are all illustrated in Fig.~\ref{fig:pred_match_false_field_map}. All distributions follow the primary beam imprint over the whole field. \textcolor{black}{The striking feature of this figure is that the false positives appear more concentrated in a ring that spans from the farthest point of the training area up to the edge of the primary beam.} This is a direct consequence of the training area definition, and this specific distribution is more pronounced when we do not include the two additional noise-only regions in our training sample (Sect.~\ref{sec:sdc1_training_area}). The distribution of missed sources based on our selection function mostly follows the primary beam footprint, illustrating that our detector completeness is uniform over the whole image field. The distribution of false and missed both exhibit compact over-densities of a few pixels, corresponding to the position of the few extended and bright sources.

\section{Discussion}
\label{sec:discussion}

\subsection{Challenge definition and scorer biases}
\label{sec:challenge_and_score_disc}

The SDC1 successfully gathered research teams to develop, adapt, and evaluate detection methods on a controlled dataset with a fixed evaluation metric. Even now that the challenge has ended, it remains a valuable dataset in its challenge form, even more so now that they have provided the full True catalog. \textcolor{black}{However, as stated in Sect.~\ref{sec:sdc1_data}, with only 4GB images, the SDC1 does not prepare the community for handling data-intensive workloads. The following SDC2 and SDC3 editions are more challenging in this regard.} In addition, our study highlighted a few limitations that are the result of some design choices of the challenge. In this section, we discuss these limits and propose a few directions that could help design future challenges.

\subsubsection{Training area}
\label{sec:challenge_and_score_disc:training_area}

In \citet{paper:sdc1}, the authors highlight that the source classification aspect is challenging due to the small number of detectable sources for which all frequencies are available. As the frequency increases, the size of the training area is also reduced, leading to a very small training sample, especially for rare classes. \textcolor{black}{More generally, they do not illustrate or discuss the distribution of the matched sources and false positives of the submitted catalogs over the whole image field to highlight the potential effects of the training area selection.} We extensively discussed how having only access to a central part of the image for training affects the generalization capability of our detector Sect.~\ref{sec:sdc1_training_area} and proposed a workaround by adding the two extra noise-only training regions. In Appendix.~\ref{sec:appendix:training_area}, we explore an alternative training region that spans an entire beam radius to encapsulate all the sensitivity regimes from the primary beam. We demonstrate that this simple change solved most of the related issues. Overall, the SDC1 falls under the classical supervised learning hiccup of having a training sample that does not represent the final task to perform. Interestingly, this aspect was corrected for the SDC2 \citep{paper:sdc2}, for which the training sample was provided as an independent data cube representing another realization of the same underlying simulation.

\subsubsection{True catalog properties}
\label{sec:challenge_and_score_disc:true_catalog}

In Sect.~\ref{sec:sdc1_selection_function}, we discussed how the construction of the training sample matters, especially regarding the selection of detectable sources. While sampling the transition between detectable and nondetectable sources with a sufficient margin is a good thing, keeping the vast majority of un-detectable objects in the reference True catalog causes several issues. At testing time, it increases the chance of random association. To counterbalance this effect, restrictive matching criteria have been used in the SDC1 scorer, notably on the flux, which then causes the rejection of actually detected sources too often (Sect.~\ref{sec:challenge_and_score_disc:scorer}). We note that in \citet{paper:sdc1}, the authors indicate that background faint sources contribute to the realism of the image noise. Still, most of them could have been ignored during scoring to reduce the random matches over undetectable sources. In all cases, a selection function must be defined to select a fixed point in the detectable to undetectable transition. In this regard, the provided True catalog could have contained more information directly from the simulation instead of just the source properties to predict. \textcolor{black}{For example, the underlying simulation produces pixel masks for all sources, which would help recognize blending or identify compact detectable regions in extended resolved sources.} Such extra information provided only for the True catalog would have allowed the construction of a more advanced selection function while not changing the amount of information in the image from which the detection has to be made.

Some other design choices do not directly impact the detector performances but could have been made differently to serve different purposes. \textcolor{black}{For example, the True catalog contains a unique true value for all the parameters of every source.} An alternative approach would have been to provide a single noised realization for each target parameter and accompany them with uncertainty measurements. It would have improved the realism of the produced catalog, considering how target sources can be acquired from real observations. Consequently, uncertainty predictions could have been added to the scorer.

\subsubsection{Scorer metric}
\label{sec:challenge_and_score_disc:scorer}

The main impactful choices are naturally in the evaluation metric itself in the form of the scorer (Sect.~\ref{sec:sdc1_metric}). Firstly, several error thresholds are permissive. Therefore, multiple detection methods with different characterization capabilities will have similar scores for the corresponding source parameters. Secondly, most parameters use an asymmetrical relative error with a symmetrical score response function. This implies that underestimated predictions tend to achieve better scores. \textcolor{black}{Finally, the error functions do not consider some structural or observational limits of the predicted parameters. For example, the limited instrumental resolution constrains the accessible values for the $\textrm{Bmaj}$, $\textrm{Bmin}$, and $\textrm{PA}$ parameters, while the noise limit constrains the accessible flux values. Having target values lower than the observational limits results in an uncapped positive relative error that does not reflect the intrinsic detector capability}. This is striking for the Flux, $\textrm{Bmaj}$, and $\textrm{Bmin}$ error distribution illustrated in Fig.~\ref{fig:all_param_pred}. Since the score of a properly detected source depends on its characterization, successfully detecting difficult sources is poorly rewarded compared to detecting obvious ones. \textcolor{black}{Another issue specific to the flux is that the target value is the intrinsic source flux instead of the apparent flux. As the sensitivity varies with the primary beam, the flux score estimate and matching criteria are inhomogeneous over the image. This can also contribute to explaining the faint outer ring of false detection visible in Fig.~\ref{fig:pred_match_false_field_map}.} As discussed in Sect.~\ref{sec:results:source_characterization}, characterization scores are averaged over all the detected sources regardless of their intrinsic difficulty. Consequently, a good detector with a good characterization capability will have a lower average characterization score than a less complete detector with poor characterization capability. \textcolor{black}{A better approach would be to use subsets of sources identified by the two detectors.} In this regard, we strongly encourage any future method application to the SDC1 dataset to include figures that represent the distributed parameter errors as a way to evaluate their characterization capability (e.g., Fig.~\ref{fig:all_param_pred}).

\textcolor{black}{To maximize its score, a detector has to identify as many objects as possible while preserving a given purity dictated by its average source characterization score. This challenge is strongly detection-focused as all identified sources} always contribute a similar score, while a false positive always contributes -1. It is equivalent to having set a fixed purity requirement and tasked teams to achieve the best completion at this purity level. \textcolor{black}{Still, the set of match criteria includes predicted characteristics like the flux and the size of the sources, which are used to prevent random association with the many nondetectable objects}. As for the source characterization, the match criterion uses asymmetrical relative errors on quantities with instrumental limitations. When defining a match, the effect of the synthesized beam on the size and position has been accounted for, but nothing was done regarding the predicted flux. Consequently, perfectly detected faint sources regarding the instrument limits can be considered false positives if their True catalog flux is slightly overestimated. In addition, since these sources get lower characterization scores, the detection purity in this regime has to be higher to add a positive score, biasing the detector to remove them more than necessary. \textcolor{black}{This results in underestimating the capability of strong detectors for low S/N, which is exactly the regime in which the comparison of method capabilities would be interesting. While using the flux in the matching criteria is a good idea to reduce the chance of random matches, it would have been better to incorporate the matching score in the global scoring instead of rejecting sources based on an arbitrary threshold. Also, the target apparent flux correlates well with the detection difficulty and could have been used as a conditional threshold for the match criteria. Another approach could have been to directly scale the contribution of each target to the total score based on their target flux.}

\subsubsection{Alternative custom AP metrics}
\label{sec:challenge_and_score_disc:alternative_metrics}

\begin{figure*}
    \centering
    \begin{subfigure}{0.49\hsize}
    \caption*{\textbf{SDC1 scorer matching}}
    \includegraphics[width=\hsize]{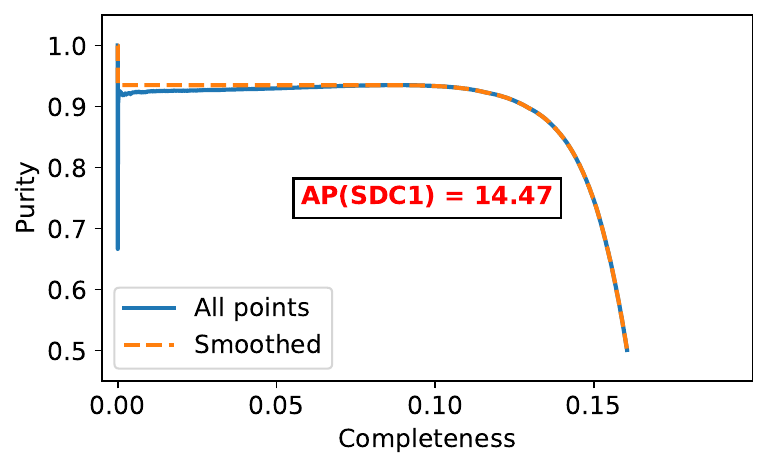}
    \end{subfigure}
    \begin{subfigure}{0.49\hsize}
    \caption*{\textbf{DIoU-based matching}}
    \includegraphics[width=\hsize]{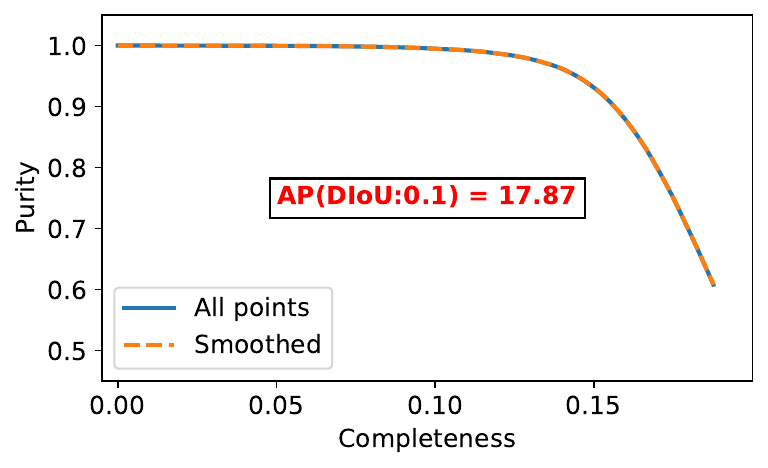}
    \end{subfigure}\\
    \begin{subfigure}{0.49\hsize}
    \includegraphics[width=\hsize]{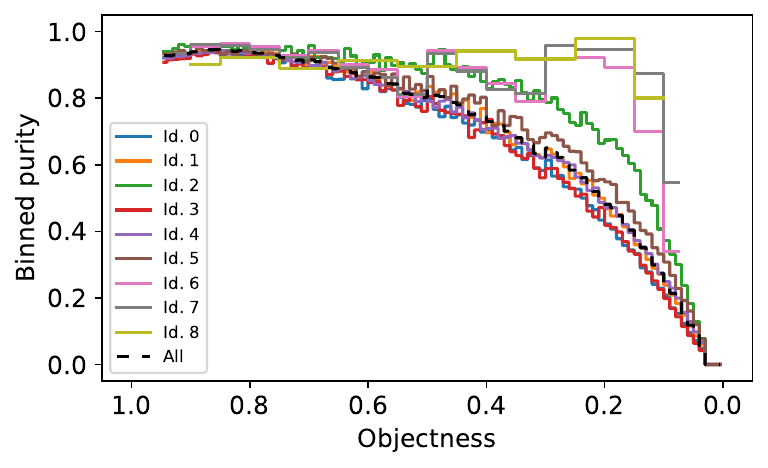}
    \end{subfigure}
    \begin{subfigure}{0.49\hsize}
    \includegraphics[width=\hsize]{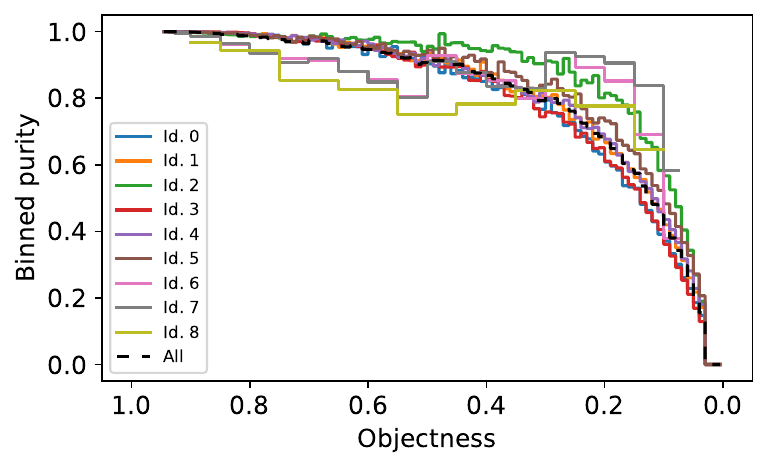}
    \end{subfigure}
    \caption{Detector sensitivity curves for the YOLO-CIANNA model. The \textit{top} frames represent purity-completeness curves built from running purity and precision scores on an objectness-sorted prediction list. The smoothed version considers that the purity at any completeness point is the maximum purity at any superior completeness. The \textit{bottom} frames represent the binned purity as a function of the objectness for each detection unit. The bin size is 0.01, 0.05, and 0.1 for the small, intermediate, and large size-prior regimes, respectively. The \textit{left} and \textit{right} columns present curves produced with the SDC1 scorer matching rule and a purely geometrical DIoU-based with a 0.1 threshold matching rule, respectively. All curves are produced from the same candidate catalog where all detections with an objectness less than 0.03 have been removed.}
    \label{fig:sdc1_ap}
\end{figure*}

Another drawback of this scoring approach is that it expects a list of considered true detections, which does not exploit the capability of some methods to produce a continuous confidence score. In classical computer vision challenges, the predicted score is used to construct a completeness-purity curve based on a relatively permissive match criterion. The integral of this curve is then used to \textcolor{black}{evaluate the detector capability over its full sensitivity range (Appendix.~\ref{sec:appendix:yolo_cianna_benchmarks:ap_metric}). For some challenges, they also build a metric that is the average of sensitivity curve integrals using multiple matching thresholds. Interestingly, since the SDC1 scorer represents a matching criterion, it can be used to sample the completeness-purity curve and build a custom AP metric.} We present the raw and smoothed completeness-purity curve for our YOLO-CIANNA model over the test region in Fig.~\ref{fig:sdc1_ap}, along with the binned purity as a function of the objectness for each detection unit. \textcolor{black}{To speed up the scorer cross-match, we rejected all predicted sources with an objectness score below $0.03$, which cuts the tail of the curve in the high-recall law precision regime, slightly reducing the AP score as well.} We see that the purity drops quickly to a $0.9$ plateau up to $0.1$ completeness. This indicates that some confidently detected sources are rejected by the SDC1 scorer match criteria, which is likely the result of the scorer limitations discussed in Sect.~\ref{sec:challenge_and_score_disc:scorer}. As a reference, we achieve ${\rm AP_{SDC1}}=14.47$. Since most of the objects in the simulation are nondetectable, the resulting AP is low, but it should remain sensitive to variations in detection capability. While this is an interesting alternative detection metric, it only accounts for the characterization of the sources through the match criteria.

\textcolor{black}{While the source characterization is a core part of the SDC1 challenge and is likely to be required as well for observational data, we can try to evaluate how our method performs with a purely geometric detection-only metric. For this, we use the classical AP definition with a DIoU as our matching criteria. This allows us to focus mostly on source positioning while still accounting for the box size to some extent. Since our target boxes are small, this metric should only be lightly affected by random matches. We build the completeness-purity curve over the full image, which should not cause issues as the training area represents only a small fraction of the image, and we do not observe any sign of over-training. We also rejected all predicted sources with an objectness score below $0.03$. We present the sensitivity curve of this alternative metric in Fig.~\ref{fig:sdc1_ap}, along with the binned purity as a function of the objectness for each detection unit. As a reference, we achieve ${\rm AP_{DIoU:0.1}}=17.87$. We note that a DIoU limit of $0.1$ is already quite demanding regarding position accuracy, especially for small objects, as illustrated in Fig.~\ref{fig:iou_illust_cases}. We see that the purity is near the maximum value up to a completeness of $0.1$, showing that at this DIoU threshold, all confident detections are considered to match a real source. However, for higher completeness values, the purity drops significantly, indicating that the low objectness source remains less reliable.}

\textcolor{black}{At this point, one can wonder if our choice of merging the source detection and characterization in a single network model was the best approach. As detection-related characteristics and extra-parameters contribute to a common loss (eq.~\ref{eq:method_loss}), it would be possible that the network struggles to find a balance, reducing its predictive capability for both. However, in our experience, trying to optimize correlated parameters simultaneously usually produces better results than when optimizing them independently. As stated before, we consider the flux and source size to correlate with source detectability. Thanks to the detection-only AP metric we just defined, we can verify that a network trained to do the detection and characterization simultaneously performs better in terms of detection-only performances than a network trained on the detection task only. This demonstration is made in Appendix~\ref{sec:appendix:characterization_impact}.}

\textcolor{black}{Finally, we note that the obtained detection-only AP cannot be directly compared with the SDC1 equivalent. However, it shows that our method produces a high-quality detection and allows for future comparisons with other methods. For now, it is impossible to compute this metric using catalogs from other teams listed in Sect.~\ref{sec:results:full_catalog}, as the SDC1 file format does not ask for the required per-source detection score. These catalogs are also cut to optimize the SDC1 score and therefore do not sample the full sensitivity range of the associated detectors. We do not provide our full sensitivity catalog as an end product, but it can easily be reconstructed from the provided trained network models and associated scripts.}

\subsection{Method weaknesses and possible improvements}
\label{sec:methode_improvement_disc}

\subsubsection{SDC1 specific limitations}
\label{sec:methode_improvement_disc:sdc1_lim}

\textcolor{black}{In this study, we tackled only a subproblem compared to the original challenge description. We applied our method to the single 560 MHz - 1000h image and excluded the classification and core fraction estimate. We could have used a combination of images at different frequencies as independent input channels, but we would have been limited to their overlapping area.} The combination of the 560 MHz and 1400 MHz images would be an interesting case study for our method as the highest frequency would help reduce source confusion and even allow a simple attempt of classification. Still, our application to the single 560 MHz - 1000h image was sufficient to demonstrate the capabilities of our YOLO-CIANNA source-detector in a context with a high density of small sources and occasional blending. \textcolor{black}{It also demonstrated the characterization capability of our method.}

\textcolor{black}{Regarding the specific case of very extended and bright sources} that are not detected, we discussed that it results from the lack of similar examples in the training area content and is not an intrinsic limitation of our method. One approach to recover these objects in the current setup would have been to downsample the image to reduce the apparent size of all objects and apply the same trained detector to recover the extended sources. As discussed in Sect.~\ref{sec:detector_training} and Appendix~\ref{sec:appendix:conv_net_yolo}, changing the apparent resolution of the objects has a significant impact on what is considered to be the instrumental limits, hence the absence of rescaling augmentation in our training. However, using rescaled input with a detector trained to recover only sources that are far from the resolution limit is a viable solution. Regarding the predicted parameters, simple scaling rules could be used to reconstruct the original quantities. We did not include this approach in this study as it would increase the complexity of the prediction for only a marginal number of recoverable sources.

We exposed in Sect.~\ref{sec:challenge_and_score_disc:true_catalog} that adding noise and uncertainty to the parameters to predict in the training catalog would have been more realistic. Following this idea, we tried adding random noise realizations to the target parameters in our augmentation during the training process as a regularization. While it stabilizes the training, it also reduces the best achievable score. Still, it demonstrates that accounting for the uncertainty is technically straightforward. If predicted uncertainty were included in the scoring procedure, our method would have to predict uncertainty as well. By adding dropout in our network, our YOLO-CIANNA method can already use MC-dropout to predict posterior distributions for all the predicted parameters, which scales with the uncertainty \citep{paper:dropout, paper:dropout_uncertainty}.

\subsubsection{Method limitations}
\label{sec:methode_improvement_disc:method_lim}

\textcolor{black}{Currently, the depth of our network backbone architecture is limited due to vanishing-gradient. We identified that group normalization and other variants known to mitigate this issue are detrimental to our detection and characterization performances with our current setup. The best solution would be to use residual or skip connections \citep{paper:residual_connect}, but those are not yet available in our generic CIANNA framework. The absence of these elements also prevents us from adding an efficient multi-scale detection in our YOLO-CIANNA method, as is the case in most modern object detectors.} Even considering our limited accessible architectural space, and despite exploring hundreds of variations, we might have missed some specific architectures that would outperform our current YOLO-CIANNA model. It goes the same way regarding our selection of hyperparameter values, especially considering their number and strong interdependence.

Another obvious limitation of our YOLO-CIANNA method is that it is fully supervised. In this regard, we already discussed how the training area and catalog representativity affect the capability of our method. We also presented the importance of the purity and completeness of the training sample \textcolor{black}{when designing our selection function (Sect.~\ref{sec:sdc1_selection_function}), which could still be improved.} Consequently, the trained model is constrained to a specific instrumental setup, and generalization toward other datasets is likely to be difficult (Sect.~\ref{sec:real_data_pred}). There are ways to force the detector to be more independent of the exact training dataset. For example, it can be incentivized to base its decision on relative contrasts and morphologies rather than the absolute signal. Still, the absolute source flux would have to be retrieved differently. We could also have a single network trained on images from various instruments and configurations and provide the instrumental setup as an external input. This way, the shared features between similar images would be mutualized into a single model.

Then, we have limitations linked to the construction of the method itself. The most striking is that we predict bounding boxes to detect mostly elliptical objects that typically fade with the distance to the center. \textcolor{black}{In practice, the detector has to fit luminosity profiles, with a significant part that can be below the noise level, to identify the objects correctly.} This task is inherently different than searching for the box that contains all the visible pixels of a given object, \textcolor{black}{which justifies several design choices in the method itself and the network backbone architecture.} This specific property of astronomical source detection explains why most source detection methods are based on segmentation. The main issue with segmentation methods is that they are computationally intensive. In this regard, bounding-box-based detection methods are more efficient in discriminating what part of the input signal is relevant to the specific task. From this, we consider that the best approach would be to use segmentation methods on small regions around sources detected with a method like ours.

\textcolor{black}{Another structural limitation is that we must know in advance the density of sources we expect to detect and at which typical size and resolution. Several modern deep-learning object detectors implement multi-scale detection capabilities. This is usually done by placing detection heads periodically in the network between spatial dimension reduction operations or adding skip connect at similar places toward the single final detection head. This way, the network can access information from small scales before they are compressed into larger-scale patterns \citep[e.g.,][]{paper:fpn_det, paper:yolo_v3}}. In addition, grid-based detection can bias the estimated confidence of objects near grid element edges and construct an artificial discretization of the detection space. A possible solution would be to have overlapping grid cells, which would complexify the association process and the post-process filtering. As a long-term objective, we would like to modify our method to recursively extract boxes of different sizes until nothing is left in the detection area, which should be feasible using transformers \citep[e.g.,][]{paper:detr, paper:vit_yolo, paper:yolo_sequence}. It means we could remove the need to define the number of detection units.

\textcolor{black}{By using a prediction-aware association process, our method exploits information about the current detector confidence to guide the training.} However, we limited our matching score metric to geometrical properties. Another approach would be to allow some of the additional predicted properties to constrain the association. For example, we could include the predicted flux and position angle of a source to select the best current detection in the association process, which would be more similar to classical astronomical cross-matching criteria. \textcolor{black}{As we observed that predicting correlated extra-parameters improves the detection-only score, we are confident that this addition would accelerate and stabilize the training process of our detector and likely result in higher detection scores. We note that this would only affect the association phase and that the resulting trained model could still be applied to a detection-only task with no prediction of the source flux or other extra parameters.}

\subsection{Generalization to real observational data}
\label{sec:real_data_pred}

In this paper, we evaluated and discussed the performances of YOLO-CIANNA on a simulated dataset. We still have to evaluate its capability over real observational data. The only strict requirement for the method is to have a training sample representative of the dataset to which it will be applied. Several approaches could be employed to generate a confident target dataset compatible with an application over observational data. 

One approach would be simulations since they provide complete and unambiguous target catalogs. The difficulty is that it must be as realistic as possible regarding source shapes and properties. \textcolor{black}{But more importantly, it must be realistic regarding instrumental properties and background signals. This includes possible diffuse structures, characteristic instrumental noise, or artifacts of various kinds and origins.} Overall, the instrumental model is often more critical regarding supervised detector performances than the target catalog, which is an identified weakness of the SDC1 dataset (Sect.~\ref{sec:sdc1_data}). Any lack of realism in the training sample would result in missed interesting objects, false detections, or unpredictable behaviors due to unconstrained contexts. While the SDC1 dataset could be used to train supervised methods before applying them to similar radio astronomical data, there is much more value in the underlying T-RECS simulation that was used to construct it \citep{paper:t-recs}. Such simulation could generate mock observations for various radio telescopes on which dedicated detectors can be trained. We also note that much effort has been put into the realism of the instrumental data pipeline for the simulated data used in the latest SDC3. These improvements could be used to construct a new simulated continuum dataset. In practice, even if a detector trained solely on simulated data could be directly applied to observational data, \textcolor{black}{it is often better to consider the simulated images as pretraining data and have a fine-tuning step over some real data, which is a case of transfer learning.}

Another approach would be to use solely observed data to define a training sample for a given instrument. While realism is assured, it is more difficult to define the target list. One classical approach in computer vision is to rely on human labeling through crowd science. While it was already extensively used for astronomical datasets and notably for galaxy identification \citep[e.g.,][]{paper:galaxy_zoo, paper:radio_galay_zoo}, it presents several difficulties. \textcolor{black}{First, analyzing astronomical datasets requires more expert knowledge than tagging everyday-life images.} Then, due to the survey size, source density, diversity of morphologies, and variety of contexts, a large crowd-science campaign would be required to obtain exploitable results. This limitation explains why some studies only call for visual inspection of litigious cases after automated detection. \textcolor{black}{The more robust and classical alternative would be to use strong observational confirmations for a limited number of examples. A typical example would be to observe a small region of a large survey using the same instrument but with a much longer integration time. This way, a classical detection method can be employed to construct a robust training catalog for a deep learning detector that takes input images at the typical survey lower integration time. Such a detector is expected to produce better results than a classical method at the application integration time and can then be used over the whole survey. However, instrument time is expensive and difficult to obtain. A possible solution would be to get strong confirmations only for a limited number of source candidates obtained with a detector trained from a less confident source catalog (see Appendix~\ref{sec:appendix:training_area:larger_and_bootstrap}).}

\textcolor{black}{We tried to apply our SDC1-trained model to surveys from SKA precursor instruments.} We notably explored the respective closest frequency from the LOFAR (LOw-Frequency ARray) Two-metre Sky Survey \citep[LoTSS,][]{paper:lotss_dr2}, and the Rapid ASKAP (Australian Square Kilometre Array Pathfinder) Continuum Survey \citep[RACS,][]{paper:racs}. \textcolor{black}{Performing a direct prediction over images from instruments with a different resolution and sensitivity requires adapting the image dynamic range and the apparent size of the objects to be similar to what was seen during training.} Then, the detector sensitivity can be adjusted through the objectness thresholds to achieve the desired completeness or purity. Our preliminary results with this approach are encouraging for both surveys, the main drawback being that there is a high amount of false detection around bright sources that produce radial artifacts. Still, we are confident that our method will be capable of understanding such context through complementary training using observational examples.

As exposed in Sect.~\ref{sec:detector_inference}, our method can be used to produce a computationally efficient source-detector, \textcolor{black}{making it well suited for real-time object detection.} It could be used to perform low-level analysis and flagging inside data-heavy pipelines or as an integrated module for added-value data visualization. We are currently exploring how to deploy our method as a standardized service using virtual observatory tools to interact with image viewers through an API. It would allow users to perform detection and characterization on the fly from a list of pretrained models with the possibility of adjusting several prediction-related hyperparameters. We are developing experimental support for these functionalities in the YaFITS viewer \citep{soft:yafits}.

\subsection{Numerical environmental footprint}
\label{sec:carbon_footprint}

We want to finish this paper by evaluating the computational numerical footprint of the present study. Most of it was performed on a single workstation equipped with an AMD Ryzen 5900X and a high-end NVIDIA GPU (RTX 3090 at the start, then replaced by an RTX4090). Our typical load is mostly GPU-dominated. We measured the power consumption of our workstation during training at approximately $0.5$ kW. Between the first step of this study and the publication of this paper, three years have passed, during which we estimate the total number of computing hours invested to be around 13000 hours. \textcolor{black}{To convert this into a ${\rm CO_2\text{-e}/kWh}$ estimation, we need to estimate the carbon intensity (CI) of the consumed electricity. We chose to use the average energetic mix value estimated for France in 2022 by the \href{https://base-empreinte.ademe.fr/}{ADEME} of 52 g of ${\rm CO_2\text{-e}/kWh}$. The total impact for this study would then be around 338 kg of ${\rm CO_2\text{-e}}$. Interestingly, we can evaluate the impact of a single training of our current architecture at 156 g of ${\rm CO_2\text{-e}}$. For prediction, the measurement would change depending on the CI of the user localization. Using the same value for CI, every gram of ${\rm CO_2\text{-e}}$ allows us to process around 69000 images when doing the prediction at a $512{\times}512$ input size using mixed-precision.} 

We acknowledge that a ${\rm CO_2\text{-e}}$ metric cannot fully represent the environmental impact. We also only accounted for the power consumption, while the full impact should include hardware production cost and other aspects of the research process. We note that the numerical environmental footprint of our data analysis can be considered negligible compared to the estimated impact of astronomical facilities \citep{paper:jurgen}. This is especially true for the simulated SDC1 image we used, which would correspond to 1000h of integration time with the future SKA. Still, we think that our estimation provides a valuable order of magnitude regarding the impact of deep learning methods when actively trying to produce computationally efficient models.


\section{Conclusion}
\label{sec:conclusion}

In this paper, \textcolor{black}{we present a new deep-learning source detection and characterization method called YOLO-CIANNA, which takes inspiration from the widely adopted YOLO regression-based object detector.} We use the SKAO SDC1 dataset as a benchmark for evaluating the capability of our method when applied to simulated continuum radio images that should resemble upcoming SKA data products.

We highlight that astronomical data have specific properties that challenge the classical design of deep-learning detection methods. Typical astronomical images representing wide fields with high sensitivity have a huge dynamic range and are crowded with small objects of a few pixels, which can be blended. We present several low-level functionalities and discuss method design choices introduced to handle these specificities, like using a conditional prediction-aware association function during training and putting the focus on the central position prediction. This results in a cascading loss process that progressively guides the detector expressivity toward a more complex and complete problem as it gets better at solving the task. Our method is also capable of predicting an arbitrary number of additional parameters, allowing it to characterize the detected astronomical sources while preserving a single prediction stage. In the dedicated Appendix~\ref{sec:appendix:association_function}, we provide more details about our custom association process and highlight the main differences from the classical YOLO method. We also show that our method achieves the same level of performance as YOLO-V2 for classical computer vision datasets using a given network backbone.

Some specificities of the SDC1 challenge design require special attention. We present a selection function that extracts the detectable sources from the complete catalog. In its default configuration, the SDC1 training area does not represent the full image because of the shape of the primary beam response. We present an approach that uses the accessible full image to enrich the training sample with noise-only fields. We demonstrate that this approach allows our detector to interpolate between these two regimes and to achieve good results at the scale of the full image. In the dedicated Appendix~\ref{sec:appendix:training_area}, we propose an alternative training area that represents a beam radius over the image field, and allows us to solve the representativity issue.

We introduce a custom network architecture backbone for our method that we have optimized regarding the specificities of astronomical images. In the dedicated Appendix~\ref{sec:appendix:yolo_v2_arch}, we evaluate the performance of a classical darknet-19 architecture backbone on the SDC1 task using our adapted detection formalism. This test shows that architectures designed for everyday-life images are not suited to astronomical images and that our custom architecture significantly outperforms it. We also disprove the classical assumptions that low-level features trained on everyday-life images are good candidates for domain adaptation toward astronomical images.

To perform the prediction over the large $32768{\times}32768$ input image of the SDC1, we establish a prediction pipeline that decomposes it into overlapping patches used as individual input for our detector. We present a complete post-processing approach that removes multiple detections while preserving blended sources that are confidently detected and that account for the overlapping areas between patches.

In its current setup, our YOLO-CIANNA detector strongly outperforms other methods applied to the SDC1 dataset. We achieved a +139\% score improvement in the SDC1 scorer metric compared to the team that reached first place during the actual SDC1 challenge. Even in comparison with another post-challenge result, we still achieve a score that is higher by 61\%. Our catalog has a detection purity of more than 94\% while detecting 40 to 60\% more sources than the other top-score methods. We also demonstrated that it is possible to constrain the purity regime of a given trained model in post-processing. This capability was used to produce a catalog with 99\% purity that still detects 10 to 30\% more sources than the other top-score methods. \textcolor{black}{Our detector is especially efficient in low-S/N regimes, and we are likely to have reached a structural limit on this aspect. Regarding source characterization, our method also significantly outperforms all the other submissions. The corresponding source catalogs are archived at \href{https://doi.org/10.5281/zenodo.13141772}{10.5281/zenodo.13141772}.}

We are working on applying our YOLO-CIANNA detector to observational data from the LOFAR and RACS instruments and expect it to be a great tool for analyzing upcoming SKA science data products. Our method is also computationally efficient, allowing us to reach real-time detection capability. As a result, it could be deployed as a service and integrated into astronomical image viewers, especially in the context of the future SKA science regional centers.

For reproducibility purposes, the training and prediction scripts that allowed us to construct our YOLO-CIANNA result are provided alongside this paper in the form of an example directory dedicated to the SDC1 in our \href{https://github.com/Deyht/CIANNA}{CIANNA} repository. We also provide simplified notebooks for visualizing the model results and to make it easier to explore the method itself. The exact scripts are archived with the CIANNA V-1.0 release (\href{https://doi.org/10.5281/zenodo.12806325}{10.5281/zenodo.12806325}), while our reference YOLO-CIANNA model for the SDC1 is archived at \href{https://doi.org/10.5281/zenodo.12801421}{10.5281/zenodo.12801421}.

\begin{acknowledgements}

We acknowledge the support of the MINERVA project from the Paris Observatory, including its material and financial support. We acknowledge the Paris Observatory DIO shared computing resources and their handling of the MINERVA dedicated compute resources. We acknowledge the computing resources from the GENCI (Jean-Zay GPU partition) provided to team MINERVA in the context of the SDC2. DC acknowledges the MINERVA project for funding two years of post-doctoral contract. DC acknowledges its PSL fellowship as part of the MCA-IA program. We highlight the great work done by SKAO in organizing all the SDCs and thank them for allowing post-challenge access to the corresponding datasets.

\end{acknowledgements}

%
%

\bibliographystyle{aa}
\bibliography{yolo_sdc1_paper.bib}

\appendix

\section{Method details and comparison to the classical YOLO}
\label{sec:appendix:in_depth_method}

\textcolor{black}{Classical object detection methods have usually been designed explicitly to detect objects in everyday life images.} These respect some implicit properties like objects with sharp features, objects obscuring each other, images being not too noisy and with a limited dynamic range, images being relatively not crowded, and detectable objects being relatively large in pixel size. \textcolor{black}{In contrast, as stated in Sect.~\ref{sec:introduction}, astronomical images can have a dynamic range of several orders of magnitude, be noisy, have complex instrumental artifacts, be crowded with objects that might be as small as a few pixels, have blending between objects, and have object types or classes that are very degenerate. Some of these difficulties were highlighted at multiple places in Sects.~\ref{sec:method} and ~\ref{sec:dataset_and_training}. In this appendix section, we expose some specificities of our YOLO-CIANNA method that contrast with the classical YOLO method, making it more suitable for astronomical source detection. We first discuss how to achieve simultaneous detection of multiple objects per grid cell and the cascading loss principle we implemented to balance the network expressivity. Finally, to verify that our method performs well for classical object detection, we apply it to some standard computer vision datasets and compare our results to the classical YOLO-V2 using an almost identical CNN backbone.}

\subsection{Fully convolutional network specificities}
\label{sec:appendix:conv_net_yolo}

In Sects.~\ref{sec:bounding_boxes} and~\ref{sec:cnn_backbone}, we stated that we use a fully convolutional to create a mapping from a 2D input image to a regular output grid corresponding to independent regions of the input image. \textcolor{black}{This is the choice of most classical object detection methods due to its translation equivariance property. It presents two main advantages. Firstly, each grid cell is the direct spatial reduction of an input region of the size of the network reduction factor. For example, a network that contains four $2{\times}2$ pooling layers has a reduction factor of $2^4=16$ in both dimensions. This structure would take an input image of size $128{\times}128$ to produce a $8{\times}8$ output grid, with each element directly mapping a $16{\times}16$ input region. Adding convolution layers with overlapping filters between the spatial reductions increases the input receptive field. This way, each output grid cell can search for objects in a much larger area but remains ``structurally'' centered on the appropriate input region. Secondly, a fully convolutional architecture preserves translational equivariance for input shift values that are a multiple of the network reduction factor. With a convolutional layer as output, the grid cells share the same weight vector. Therefore, the output layer defines a single ``detector'' that is applied independently to subparts of the image to construct the grided output. It ensures that the objectness score has the same meaning for a given detection unit over all the grid cells for post-prediction filtering (Sect.~\ref{sec:filtering_and_nms}). It also helps the training process since all objects contribute to training the same position-agnostic set of weights regardless of their associated grid cells.}

An interesting side effect of having a fully convolutional structure is that we can use any input size that is a multiple of the reduction factor. Doing so will only change the size of the output grid but not the behavior of the detector at the level of a detection unit. \textcolor{black}{In classical computer vision applications, the detections must be invariant to the apparent scale of an object and its resolutions. For this, the bounding boxes must be expressed in a resolution-invariant format, for example, by predicting the box sizes as fractions of the image size. The network can then be trained using various resolutions.} In the classical YOLO method (starting with V2), this principle is mainly used to adjust the computational cost and detection accuracy of the model by adjusting the input resolution at prediction time. 

\textcolor{black}{For astronomical applications, having a resolution invariance model is not always beneficial. In our datasets, the pixel size is often valuable information regarding the instrument limitations.} Due to the synthesized beam effect acting as a resolution limit, downscaling an image by a factor of two will not result in sources that look similar to sources of half the size. It goes the same way for the flux value that cannot be trivially preserved when changing the scaling of an astronomical image. For these reasons, we decided to drop the detector apparent scale invariance apparent in our method. Instead, we consider all the predicted box quantities directly related to a fixed pixel scale as constrained during training. \textcolor{black}{By doing this, our detector will predict the same detection box for a specific source regardless of the input size. We could, in theory, apply our detector to the whole $32768{\times}32768$ input image of the SDC1 in a single prediction path. This approach should produce the same results as our overlapping patches split, or even better results, as it would eliminate edge effects created by the patch decomposition. However, the maximum input size remains limited by hardware constraints, and it is not always computationally efficient to use the largest possible image size for training or prediction.} We use this principle in Sects~\ref{sec:detector_training} and~\ref{sec:detector_inference}, independently optimizing the training and prediction input sizes regarding their respective constraints.

\subsection{Grid resolution and detection unit density}
\label{sec:appendix:grid_res}

With a fully convolutional architecture, the output grid resolution depends on the input size and the chosen network backbone reduction factor. In addition, the number of detection units per grid cell is fixed before training (Sect.~\ref{sec:multiple_boxes}). Therefore, the network structure sets the total number of detectable objects in an image in advance as a function of the input resolution, \textcolor{black}{which is a common limit of most regression-based detectors. It is a substantial issue for models that are supposed to be scale and resolution-invariant, as changing the input size alters the number of output grid cells and therefore the maximum number of objects that can be detected, which contradicts their claimed resolution-invariant property.} With our nonscale-invariant approach, the only quantity that matters is the density of objects at a given resolution. We consider that our pixel size represents a fixed angular scale defined by the instrument resolution limit. \textcolor{black}{Increasing the input size is then equivalent to having a wider field of view, which does not change the pixel size of the objects. From there, the reduction factor and the number of detection units per grid cell can be chosen to be representative of the source density in the field.}

In the classical YOLO method, the multiple detection units per grid cell were not designed to handle simultaneous multiple detections. In everyday-life images, objects are considered to obstruct each other, and, in most cases, only a single object from a given scale regime can be detected by a grid cell simultaneously. \textcolor{black}{In this classical formalism, the multiple detection units per grid cell are not here to allow multiple detections but to distribute the objects by apparent size or aspect ratio.} Due to this construction, the default YOLO association process during the training phase was not designed to detect multiple objects of similar size centered in the same grid cell (Sect.~\ref{sec:appendix:association_function}). 

\textcolor{black}{From this, as stated in Sect.~\ref{sec:multiple_boxes}, classical detection methods answer to a high object density by lowering the reduction factor so each grid cell maps a small enough area unlikely to contain multiple objects.} The issue with this approach is that it forces the network to work at a higher resolution while still reaching the necessary expressivity and having a large enough receptive field. In practice, typical 2D astronomical images with high sensitivity are so crowded that reaching the appropriate reduction factor is unrealistic, as it would result in a network that is computationally costly and very difficult to train. While using a finner output grid resolution is helpful up to some point, we also need to be capable of detecting multiple similar objects in each grid cell in order to reach the target object density. This can be achieved by switching to a prediction-aware association function as exposed in Sects.~\ref{sec:association_function} and~\ref{sec:appendix:association_function}. \textcolor{black}{For our reference network backbone, we chose a reduction factor of 16 (Sect.~\ref{sec:cnn_backbone}), which is lower than the typical reduction factor of 32 of various backbones, including the YOLO-V2 architecture (darknet-19), but still high enough for our network to remain trainable and computationally efficient.}

\subsection{Prediction-aware association}
\label{sec:appendix:association_function}

\begin{figure}
    \centering
    \includegraphics[width=1.0\hsize]{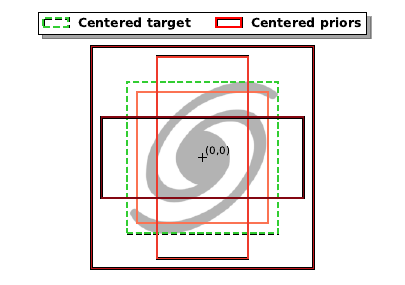}
    \caption{Illustration of a zero-centered comparison between a target in dashed green and different size-priors in shades of red, which represents the classical YOLO association process.}
    \label{fig:theoritical_assoc}
\end{figure}

\textcolor{black}{During training, the association function is responsible for finding the best two-by-two associations from two independent lists of targets and predicted boxes, all in the same grid cell (Sect.~\ref{sec:association_function}). In the classical YOLO method, the association process is based on the best theoretical match using the size priors of each detection unit. For this, it will consider that all target and predicted boxes are centered on a fictive $(0,0)$ coordinate and only compare their respective IoU, as illustrated in Fig.~\ref{fig:theoritical_assoc}. In the classical YOLO, all detection units are implicitly supposed to represent mutually exclusive size categories, and it is considered unlikely for multiple targets of similar size to be present in the same grid cell. When it happens, the second closest prior is used, even if it is badly constrained on the target size regime.}

\begin{figure*}
    \sidecaption
    \centering
    \includegraphics[width=12cm]{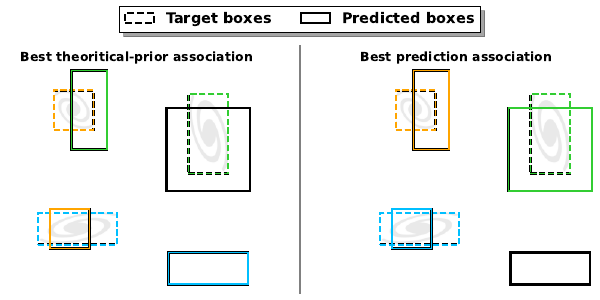}
    \caption{Comparison of the target-prediction association in a grid cell using two different association functions. \textcolor{black}{The size of predicted boxes is considered close to the prior of their corresponding detection unit.} The color indicates which prediction was associated with each target. The association results are illustrated for the default YOLO association function that is based on the best theoretical priors (\textit{left}) and for an association function that uses the best current network prediction (\textit{right}).}
    \label{fig:prior_assoc_comp}
\end{figure*}

\textcolor{black}{To detect multiple small objects in a single grid cell, we first need multiple detection units with comparable size priors. We then use an association function based on the IoU between the targets and the current position and size of the predicted boxes. To illustrate the difference between these two approaches, we present in Fig.~\ref{fig:prior_assoc_comp} a case study with three target boxes and four predictions from independent detection units, all in the same grid cell, and the resulting association based on both the theoretical-prior association and the best current prediction association. We observe that the first one considers that no prediction is placed correctly but that their size is relatively good. In contrast, the second one considers that the network is already good at detecting the objects and that the predicted size needs to be adjusted.}

\subsection{Box matching metric}
\label{sec:appendix:association_function:match_metric}

\begin{figure*}
    \centering
    \includegraphics[width=1.0\hsize]{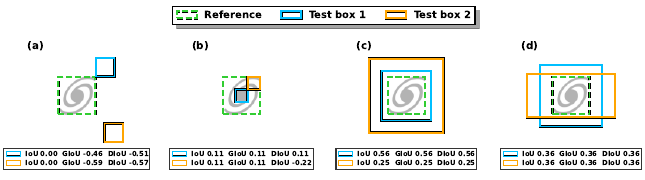}
    \caption{Comparison of three IoU-based metrics in different contexts. For all cases, the reference box is the dashed green box, and two test boxes in blue and orange are compared. The IoU-based metrics are computed between a given case and the reference.}
    \label{fig:iou_illust_cases}
\end{figure*}

A prediction-aware association function has to be capable of selecting which prediction is the best current representation of the target. It means we need a box-matching metric capable of discriminating between all the possible predictions in our specific context. The classical IoU matching criteria defined in Sect.~\ref{sec:bounding_boxes} with Eq.~\ref{eq:iou} present strong limitations. To illustrate them, we compare IoU computations between a reference box and two possible test boxes in different contexts in Fig.~\ref{fig:iou_illust_cases}. We see that the IoU is equal to zero for all couple of boxes that have no intersection regardless of their distance (Fig.~\ref{fig:iou_illust_cases}, frame A). \textcolor{black}{While it is an issue for a prediction-aware association function, the best theoretical association from the classical YOLO is not affected due to the recentering of all boxes to compare.} For our specific astronomical applications where objects can be much smaller than the input size mapped by a grid cell, we need to be capable of attributing a different match score to nonintersecting boxes based on their distance to the reference. For this, we can use the Generalized IoU \citep[GIoU,][]{paper:giou} that makes use of the smallest box C that encloses the two boxes A and B to be compared by refining the default IoU following
\begin{equation}
    {\rm GIoU} = {\rm IoU} - \frac{C - (A\cup B)}{C}.
\label{eq:giou}
\end{equation}
The GIoU can take values from -1 to 1 while still being equal to the IoU in cases where C is equal to one of the two boxes. While leveraging the first limitation, the GIoU remains insufficient to distinguish between test boxes fully contained inside the reference box. In such a context, we would like to attribute a different match score based on the distance between the box centers (Fig.~\ref{fig:iou_illust_cases}, frame B). For this, we can use the Distance IoU \citep[DIoU,][]{paper:diou} that combines the classical IoU with the ratio between the distance of the box centers $A_c$ and $B_c$ and the smallest enclosing box diagonal-length $d$ in the form
\begin{equation}
    {\rm DIoU} = {\rm IoU} - \frac{\rho^2(A_c, B_c)}{d^2},
\label{eq:diou}
\end{equation}
where $\rho$ represents the Euclidean distance between the two center coordinates. The DIoU can take values from -1 to 1 while being equal to the IoU and GIoU in cases where the two boxes have the same center, corresponding to $\rho^2(A_c, B_c) = 0$ (Fig.~\ref{fig:iou_illust_cases}, frame C and D). This last metric is well suited for astronomical applications, but it remains limited in cases where the aspect ratio of the boxes matters. For example, all the metrics we described would give the same results for two test boxes with the same center and identical surface but different aspect ratios (Fig.~\ref{fig:iou_illust_cases}, frame D). This could be included in a more refined matching metric like the Complete IoU \citep[CIoU,][]{paper:diou}.

\textcolor{black}{In our YOLO-CIANNA method, all these IoU variants are only used as match criteria or as a scalar for scaling the objectness. The bounding-box-related loss remains a simple sum-of-square error on the box position and size following what was presented in Sect.~\ref{sec:bounding_boxes} and Eqs.~\ref{eq:loss_pos} and~\ref{eq:loss_size}.} This contrasts with what is described in the papers describing these metrics, where they are presented as alternative loss functions for object detection methods. In practice, all these matching metrics are implemented within YOLO-CIANNA and can be selected regarding the application. The predicted objectness uses the corresponding metric (rescaled between 0 and 1 if necessary) to follow Eqs.~\ref{eq:objectness} and~\ref{eq:loss_obj}, so the predicted score reflects all the related subtleties.

\subsection{Association ordering effect}
\label{sec:appendix:association_ordering}

\begin{figure*}
    \centering
    \includegraphics[width=0.98\hsize]{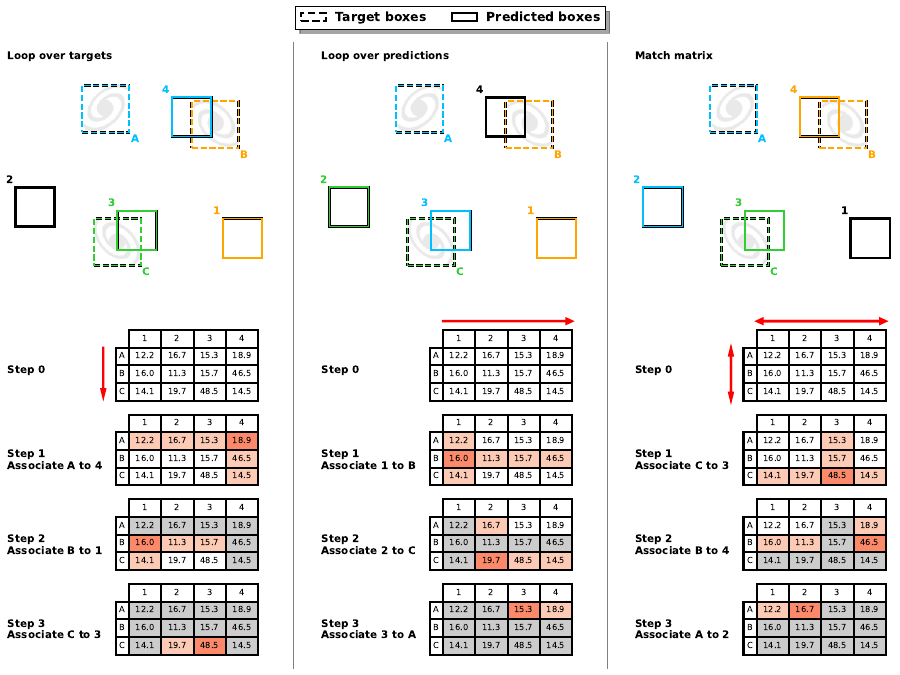}
    \caption{Comparison of the target-prediction association in a grid cell regarding the search order. The color indicates which prediction was associated with each target. All predictions and targets have the same size and shape, respectively. The prediction-target pairs are compared using a scaled matching metric expressed as $({\rm DIoU} + 1){\times} 0.5{\times} 100$. The results for all pairs are displayed in a matching matrix. For each step of the association process, the rows and columns corresponding to the targets and predictions are highlighted in light red, with the intersection being the match score for this pair. For all subsequent steps, the now unavailable elements in the matrix are shown in gray. The three frames compare different association results, along with the corresponding association steps highlighted in the match score matrix, when looping over the targets (\textit{left}), looping over the predictions (\textit{middle}), or searching for the maximum value in the score matrix (\textit{right}).}
    \label{fig:association_ordering}
\end{figure*}

\textcolor{black}{In the classical YOLO, the association is done by looping over the target list for each grid cell, which works well as they are considered unlikely to require the same detection unit. With an association process based on the current network prediction, any detection unit can be the best representation for any target regardless of its size prior. As described in Sect.~\ref{sec:association_function}, we defined our YOLO-CIANNA association function to be a search for the highest value in a matching score matrix that contains the ${\rm fIoU}$ for all the target-prediction pairs}. We illustrate the effect of such an association function compared to a loop over the targets and a loop over the predictions in Fig.~\ref{fig:association_ordering}. \textcolor{black}{The main advantage of our approach is that the well-detected objects will get associated first, so they will not be wrongly forced to adapt toward a different object than the one they were trying to detect. In addition, targets that are not yet well detected are more likely to be associated with a truly idling detection unit.}

\subsection{Association refinements}
\label{sec:appendix:association_function:association_tricks}

Using a prediction-aware association implies that the network is free to distribute the diversity of objects over the detection units independently of their size prior if needed. This added degree of freedom increases training difficulty. Therefore, completely removing the theoretical box-prior association process is usually inefficient. This section describes several refinements we added to our prediction-aware association (Sect.~\ref{sec:association_function}) to constrain the accessible solutions and guide the training process. Regarding our complete loss function (Eq.~\ref{eq:method_loss}), these refinements only affect the content of the $\mathds{1}^{\textrm{match}}$ and $\mathds{1}^{\rm void}$ masks. \textcolor{black}{Most of them occur after a target-prediction couple has been selected in the association search and are verified sequentially in a mutually exclusive way, following the blue B.2 block in Fig.~\ref{fig:association_algorithm}.}

We emphasize that we describe all the refinements available in our method, whether or not they have been employed for the SDC1 application. Listing all the accessible refinements in the same place will allow future reference.

\subsubsection{Strict detection unit association}
\label{sec:appendix:association_function:strict_box_association}

\textcolor{black}{The first refinement is to prevent some detection units from being associated with targets that are too different from their size prior. We do this by defining a ``strict association range'' $S_{ar}$ that represents the number of detection units available to a target object of a given size. For example, with $S_{ar}=3$, each target can only be associated with the three detection units with the closest theoretical size priors. In practice, we do this by masking elements in the matching score matrix before starting the association process (A.3 frame in Fig.~\ref{fig:association_algorithm}). The association is still based on the best current prediction but uses only a subset of detection units depending on each target.}

\textcolor{black}{For this selection to work, we must define which detection units best suit a given target. We implemented three different theoretical proximity measurements. The first uses a zero-centered IoU comparison like the classical YOLO.} The second one uses the Euclidean distance between the target and the prior of each detection unit in the 2D box-size space. The third one searches for the minimum absolute value for the target size by comparing what would be $\hat{o}_w$ and $\hat{o}_h$ for each detection, which is equivalent to measuring a distance in the exponentiated and prior-scaled size space. Our preference goes to the third approach as it is more directly related to the loss expression (Eq.~\ref{eq:loss_size}). \textcolor{black}{The classical zero-centered IoU was mainly kept so our method can emulate a classical YOLO association by setting $S_{ar} = 1$.}

\textcolor{black}{In case identical priors are defined, they are all made available and only count for a single $S_{ar}$ increment. For example, with a list of size-priors $p_w = p_h = \left[6,6,6,6,12,24\right]$ and $S_{ar} = 2$, a square target of size $\hat{w} = \hat{h} = 7$ could be associated with all the detection units except the one with the size prior of 24. With this definition, we can distribute the target in different size categories while allowing the detection of multiple objects in the same grid cell for specific size categories.} Ultimately, we expect detection units with identical size priors to specialize based on other criteria. \textcolor{black}{By construction, we define that a value of $S_{ar} = 0$ turns off this refinement with all detection units available.}

\subsubsection{Startup and forced random association}
\label{sec:appendix:association_function:random_startup}

\textcolor{black}{At the beginning of the training, when all the weights are still close to their initialization values, the network can be exposed to gradient divergence. One solution to mitigate this issue} is to have a low starting learning rate that progressively increases (Sect.~\ref{sec:detector_training}). However, it works best when combined with a second solution that is to deactivate the prediction-aware association for a given number of iterations $N_{\rm rand}$ in favor of a random target-prediction matching (B.2.1 frame in Fig.~\ref{fig:association_algorithm}). \textcolor{black}{This refinement is notably found in specific versions of the classical YOLO. It allows each detection unit to see a variety of examples representative of the full training dataset distribution, constraining them to a permissive range of plausible values while preserving diversity.}

\textcolor{black}{One other problem we can face with a prediction-aware association is solution collapse, with only a subset of detection units trying to represent all the targets. It reduces the capability of the detector to detect multiple objects, resulting in a suboptimum distribution of the network expressivity. While this issue mostly appears at the early stage of training, it is likely to persist indefinitely as unconstrained detection units are less likely to be a good match. To prevent solution collapse, we need to push detection units to specialize in mostly independent categories of objects. However, this might result in discontinuities in the predicted values for some object properties. For example, as detection units distribute over size categories, there will likely be intermediate-size regimes for which none of the available detection is accurate as it is on the edge of the representation space for all of them. Overall, some degree of specialization of the detection units is required, but it must remain permissive enough so each detection unit can sample a wider parameter space.}

\textcolor{black}{Using a strict size association refinement with a permissive range $S_{ar} > 1$ already provides some leverage over the permissivity versus specialization balance.} However, this approach only allows representation overlap and does not guarantee it. As for random startup, \textcolor{black}{forcing random associations at a user-defined rate $\alpha_{\rm rand}$ can force each detection unit to occasionally access parts of the underlying function outside its usual specialized parameter subspace, ensuring better continuity (also B.2.1 frame in Fig.~\ref{fig:association_algorithm}). A very low $\alpha_{\rm rand}$ value around one or two percent is usually enough to prevent mode collapse and to obtain a strong additional regularization effect. However, regarding parameter discontinuity, random association only mitigates the effect that remains measurable in many application cases.}

\subsubsection{Forced specific detection unit association}
\label{sec:appendix:association_function:forced_box_association}

\textcolor{black}{There are some specific cases where more constraints in the association can be beneficial. The first forced association is for small objects, which are often difficult to identify or position due to poorly resolved features. In addition, getting small boxes to overlap requires a higher positioning accuracy, and the chances of random matches in the association process are reduced compared to larger boxes. To overcome these issues, we force the association to the smallest prior if the target surface is smaller than a limit value defined as $\alpha_{\rm small} {\times} s_w {\times} s_h$, with $s_w$ and $s_h$ the width and height of the smallest size-prior, and $\alpha_s$ a user-defined scaling factor (B.2.2 frame in Fig.~\ref{fig:association_algorithm}). If multiple detection units have an identical size-prior that is the smallest, they are all considered viable options. The new target-prediction pair is then selected by searching for the best current prediction in this subset of detection units.}

The second forced association is for targets that no detection units actively try to identify. In such a context, associating the closest prediction results in selecting a random detection unit most of the time. For example, in Fig.~\ref{fig:association_ordering}, target A has no good association, and the two remaining available predictions fall far away in the background. In such a case, instead of using the distance, it might be better to associate the target with the available detection unit that should theoretically be the best suited for it. In our method, this forced association with the best theoretical detection unit is adopted when ${\rm fIoU} \leq L^{{\rm fIoU}}_{\rm low}$ with ${\rm fIoU}$ the score of the current target-prediction pair and $L^{{\rm fIoU}}_{\rm low}$ a user-defined threshold. The best theoretical detection unit is defined as described in Appendix~\ref{sec:appendix:association_function:strict_box_association} accounting for multiple identical priors and excluding already locked boxes (B.2.3 frame in Fig.~\ref{fig:association_algorithm}).

\textcolor{black}{
Finally, we can also use this principle as a regularization approach by defining an $\alpha_{\rm best}$ rate at which the best theoretical detection unit is used regardless of the ${\rm fIoU}$ value for the current target-prediction pair. This is another way to mitigate solution collapse in cases where $S_{ar}$ is high}. We found that $\alpha_{\rm best} = 0.05$ works well as a regularization for datasets based on everyday life images. This last approach can also be used in an alternative way. By setting it to a high value, up to $\alpha_{\rm best} = 0.95$, in association with a high $S_{ar}$, we are emulating a case where $S_{ar} = 1$ but allowing another detection unit for which the current prediction is better to get the association at a low $1 - \alpha_{\rm best}$ rate. This type of setup is especially relevant for contexts with a strongly imbalanced scale distribution in the training sample, and it is also better at improving the continuity of the predicted parameters across the available detection units. We stress that this approach is different from a setup with $S_{ar} = 1$ and a low $\alpha_{\rm rand}$ rate for which the quality of the current prediction would never matter.

\subsubsection{Difficult flagging and quality check}
\label{sec:appendix:association_function:diff_flag}

All target objects are not equally difficult to detect, be it due to specific class features, apparent size, strong context dependency, partial obscuring from another object, or proximity to the edges of an image. Moreover, the detection difficulty is not a discrete property. From this, deciding if an object should be included as a target in the training sample regarding a specific application and set of objectives is complex. \textcolor{black}{There are even cases where both removing or keeping the object can result in poor performance. As stated in Sect.~\ref{sec:sdc1_selection_function}, removing a target that the network would have confidently detected will lower the objectness score of all objects with similar detectable features. In the opposite case, keeping a target that is impossible to detect for the network will add noise to global objectness score training. Several reasons could make an object difficult but not impossible to detect. A typical example arises when doing image augmentation, where objects can be cut on the edge of a zoomed-in image. In such a case, the detectability of the object can depend strongly on what part of the object remains, making criteria based on box size fraction inefficient.}

\textcolor{black}{To help solve this issue, we designed a ``difficult flag'' that can be attributed to any target object and for which we apply a ``positive reinforcement only'' update. Flagged targets can only be matched with predictions for which the network is confident in its prediction.} This is controlled by two user-defined thresholds based on the predicted box quality ${\rm fIoU} \geq L^{{\rm fIoU}}_{\rm diff}$ and current confidence $O \geq L^{\rm obj}_{\rm diff}$ of the network (B.2.4 frame in Fig.~\ref{fig:association_algorithm}). If one of these limits is not fulfilled, we consider there is no match, but the target still counts for defining the GBNB masks. It means that a good match with low confidence is not penalized, allowing its confidence to increase with further training from similar targets. These limits are tested after all the association refinements, so if the quality condition is not fulfilled, the associated detection unit remains free for subsequent associations.

Objects can not only be difficult to detect but also to classify or characterize. For example, predicting the flux of an astronomical \textcolor{black}{source only partially visible on the edge of an image requires the network to learn symmetrical assumptions. To allow more control,} our difficult flag can take different values defining which parameters get updated when the object passes the quality check: ``0'' when the object is not difficult; ``1'' when all the parameters should be updated; ``2'' when the box geometry, objectness, and probability should be updated, but not the class nor the extra parameters; and ``3'' when only the objectness and probability should be updated. \textcolor{black}{All these flagging only affect training and do not impact the prediction process directly but reflect in the predicted objectness score.}

\subsection{YOLO-CIANNA cascading loss}
\label{sec:appendix:cascading_loss}

In sect.~\ref{sec:method_loss}, we introduced the concept of cascading loss, which refers to the loss changing during training as a function of the current state of the network prediction. This is a direct consequence of having a prediction-aware association in addition to various quality checks on the loss subparts and multiple association refinements. \textcolor{black}{In practice, at the beginning of the training, the loss should be dominated by position and size updates toward the most obvious target objects. As the network improves for these objects, their respective loss term will lower, allowing terms from less obvious objects to be more dominant.} Simultaneously, the obvious objects will become sufficiently well detected for other aspects like the objectness, the class, or the additional parameters to be included. From the loss perspective, it results in the search for a balance between \textcolor{black}{better characterizing the most striking objects and starting to detect less obvious ones.}

In summary, the better the network gets, the more complete and complex the loss becomes. \textcolor{black}{The objective is to guide the network expressivity toward the most important aspects at a given point in the training, not bothering to adjust currently irrelevant properties.} Looking at the evolution of the validation loss during training, we usually observe that the loss for the parameters with a quality threshold starts low, then increases before lowering again and reaching a first plateau. This behavior tends to repeat several times during the training with a lower increase and subsequent plateau with a smaller value each time, hence the ``cascading'' name. The same behavior can also appear on objectness and probability independently of the fitting limits due to possible regime change in the object distribution over the detection units or simply due to the ``difficult-flagged'' objects being progressively integrated into the loss. 

\textcolor{black}{Identifying overtraining with a dynamic loss is challenging. To counter this, our implementation can monitor the validation loss with (``complete'' loss) or without (``natural'' loss) the refinements and limits. In the second case, only the base association process is maintained when computing the error on the validation set, and all the loss parameters are computed for every association regardless of their quality.} This usually results in a more classical loss behavior suitable for monitoring overtraining while keeping the cascading loss for the training dataset only. In Fig.~\ref{fig:loss_curve_with_score}, we use the "natural" loss, so the cascading effect is not visible in the monitored validation loss.

\subsection{Classical detection dataset benchmark}
\label{sec:appendix:yolo_cianna_benchmarks}

This section presents a quick summary of the results we obtained on classical object detection datasets using our YOLO-CIANNA method. \textcolor{black}{This aims to validate that our implementation is at least on par with the classical YOLO-V2 performances and that design choices motivated by astronomical image properties have not impaired the capabilities of the method for other applications. We tried to reproduce the darknet-19 backbone from YOLO-V2 as closely as possible, considering the current limitation of the CIANNA framework. The later darknet-53 backbone from YOLO-V3 is currently out of reach with CIANNA. We then present classical benchmark results on the ImageNet, Pascal VOC, and COCO datasets.} We only present the most critical aspects of these benchmarks, as analyzing them is not the scope of this paper. Still, all the corresponding scripts with all the parameters are provided \textcolor{black}{on the CIANNA git repository and archived with the V-1.0 release \citep[\href{https://doi.org/10.5281/zenodo.12806324}{10.5281/zenodo.12806324},][]{soft:cianna}. The different trained models are also available and archived at \href{https://doi.org/10.5281/zenodo.12801421}{10.5281/zenodo.12801421}\citep{data:cianna_models}.}

\subsubsection{Backbone pretraining with ImageNet-2012}
\label{sec:appendix:yolo_cianna_benchmarks:imagenet}

\textcolor{black}{Following the description in \citet{paper:yolo_v2}, we build a network backbone as similar as possible to the darknet-19 architecture, which comprises 19 convolutional layers and a few pooling layers. The main structural difference is that we use group normalization layers \citep{paper:group_norm} instead of batch normalization layers \citep{paper:batch_norm}, which should not significantly impact our result as long as we compensate for the missing batch scale regularization effect with more image augmentation. We also miss the skip connection near the end of the network, which reduces our ability to detect small objects.} 

We first train this backbone on the ImageNet-2012 dataset \citep[ILSVRC-2012,][]{paper:imagenet} composed of roughly 1.2 million single-label \textcolor{black}{images with 1000 possible classes so it can identify generic low-level features that are likely relevant for various applications. We first} trained our network at the typical $224{\times}224$ input resolution up to a loss plateau that achieved a Top-1 accuracy of 70.1\% and a Top-5 accuracy of 89.4\% on the validation dataset. We then continue to train the network at a $448{\times}448$ input resolution, which is closer to the desired input resolution for our object detectors. With access to finer features, the network achieved a Top-1 accuracy of \textcolor{black}{74.7\%} and a Top-5 accuracy of \textcolor{black}{91.7\%}. For comparison, the classical darknet-19 architecture reaches a top-1 accuracy of 76.5\% and a top-5 accuracy of 93.3\%. \textcolor{black}{The remaining difference is likely due to the absence of the regularization effect of batch normalization. When diverging slightly from the original darknet-19 to include more regularization in another form, for example, adding two large layers with 50\% dropout, it is possible to bridge the accuracy gap.}

\subsubsection{Mean average precision metric}
\label{sec:appendix:yolo_cianna_benchmarks:ap_metric}

The classical metric for object detection tasks is the mean average precision (mAP). While mAP can refer to slightly different quantities depending on the challenge, they all usually rely on the computation of the area under a sensitivity curve, representing the precision as a function of recall. The mAP is analog to the receiver operating characteristic curve (ROC) but for object detection tasks. In practice, mAP computation works as follows. All object detectors are expected to produce a score for their detection and sort them for the full dataset. \textcolor{black}{Each prediction is flagged as a true or false detection using a match criterion based on a classical IoU threshold limit (Sect.~\ref{sec:bounding_boxes} and Eq.~\ref{eq:iou}).} From this ordered list of true and false detections, it is possible to construct a running precision and recall quantity. The constructed raw recall-precision curve can present local drops due to some high-confidence objects being flagged as false detections. Therefore, the curve is modified to respect $p(r) = \max_{\tilde{r} > r} p(\tilde{r})$. The average precision (AP) is then obtained by computing the integral over this interpolated precision-recall curve.

Usually, the AP is computed for each class individually. All the AP values are then averaged to construct a more general mAP score. For some challenges, mAP is computed for a single IoU value and expressed as mAP@IoU. For others, multiple IoU thresholds can be used independently or together to construct an averaged mAP over a given IoU range. \textcolor{black}{Overall, this metric has the advantage of sampling the full sensitivity of a given detector, corresponding to its discrimination capability between} the background and the objects to detect (Sect.~\ref{sec:challenge_and_score_disc}).

\subsubsection{PASCAL-VOC object detection}
\label{sec:appendix:yolo_cianna_benchmarks:pascal}

The PASCAL-VOC \citep[PASCAL Visual Object Classes,][]{paper:pascal_voc} dataset was a yearly computer vision challenge that ran from 2005 to 2012, and that included an object detection task. This dataset is still commonly used as a benchmark to compare the performances of different object detectors. \textcolor{black}{One classical approach is to combine the training and validation dataset from the 2007 and 2012 editions onto a single training sample comprising around 16500 images containing multiple bounding boxes belonging to 20 possible classes. The evaluation of the detector performances is then done on the 2007 test dataset of 5000 images. Using our high-resolution pretrained backbone on ImageNet, we can drop the last pooling and output layer and add three new convolutional layers with $3{\times}3$ filters plus group normalization. Finally, we add our detection layer. The input resolution is changed for a $416{\times}416$ input, which results in a $13{\times}13$ output grid. Regarding the setup of our advanced association function, we use the ${\rm DIoU}$ as our matching score, $S_{ar} = 3$, $L^{{\rm fIoU}}_{\rm low} = -0.1$, $L^{{\rm fIoU}}_{C} = -0.1$, $\alpha_{\rm best} = 0.05$, and some difficult object flagging. After training this network for a few hundred epochs, we achieve an ${\rm mAP@0.5}$ of \textcolor{black}{74.98} on the PASCAL 2007 test dataset. At an identical resolution, the original YOLO-V2 model achieves 76.8. Considering the structural limitations of our approximated darknet-19 backbone, this result demonstrates the efficiency of our YOLO-CIANNA method.}

\subsubsection{COCO object detection}
\label{sec:appendix:yolo_cianna_benchmarks:coco}

\begin{table*}[t]
\centering
\caption{\label{table:coco_results} Results comparison for COCO using the dedicated challenge metrics on the test-dev 2017 dataset.}
\begin{tabular}{ c c c c c c c c c c c c c }
\hline
\hline
 & ${\rm AP_{0.5:0.95}}$ & ${\rm AP_{0.5}}$ & ${\rm AP_{0.75}}$ & ${\rm AP_S}$ & ${\rm AP_M}$ & ${\rm AP_L}$ & ${\rm AR_1}$ & ${\rm AR_{10}}$ & ${\rm AR_{100}}$ & ${\rm AR_S}$ & ${\rm AR_M}$ & ${\rm AR_L}$ \\
 \hline
 YOLO-V2 & 21.6 & 44.0 & 19.2 & 5.0 & 22.4 & 35.5 & 20.7 & 31.6 & 33.3 & 9.8 & 36.5 & 54.4\\ 
 \textcolor{black}{YOLO-CIANNA} & 21.9 & 40.1 & 21.8 & 4.1 & 22.8 & 37.9 & 20.7 & 30.2 & 31.2 & 4.5 & 34.6 & 55.0\\
 \hline
\end{tabular}
\end{table*}

The COCO \citep[Common Objects in Context,][]{paper:coco_2014} dataset was also a yearly computer vision challenge that ran from 2014 to 2017, and that included an object detection task. Like PASCAL, COCO is still commonly used as a benchmark to compare the performances of different object detectors. In the 2017 version, the training set comprises roughly 118000 images, the validation set comprises 5000 images, and the hidden test set comprises roughly 40000 images (accessible through a test server). For the detection task, the objects are labeled with over 1000 classes. We use the same network backbone pretrained on ImageNet, with the same set of hyperparameters we used for PASCAL and a $416{\times}416$ input resolution. Our result on the test dataset obtained using the submission server is presented in Table~\ref{table:coco_results} with all the COCO metrics and is compared with the result of the classical YOLO-V2 model. We first observe that our method is competitive with the classical YOLO-V2 detector on this dataset despite our structural limitations. In addition, these metrics indicate that our method retrieves more large objects and better reconstructs the objects it manages to detect. However, the lower ${\rm mAP@0.5}$ score indicates that we recover fewer objects overall. This is likely due to a poorer detection of small objects, confirmed by the small ${\rm AR_S}$ corresponding to recall at small scales. This result was expected, considering that we lack the ending skip connection to allow our method to preserve more resolved features at the detection layer.

Regarding computing performances, we can achieve 690 images per second at this resolution using an RTX 4090 GPU with mixed precision enabled. On a much lighter A2000 mobile GPU, we still reach above 60 images per second, \textcolor{black}{meaning that our implementation allows real-time detection on modern entry-level hardware. We note that this level of performance is similar to the one achieved by the original YOLO-V2 darknet implementation.}

Finally, we highlight that our result is far from the current top scores on the COCO dataset, which require much stronger architectures, but it is not our scope to compete with them. The presented benchmarks validate the method for a given relatively light architecture. Most of the time, the architectures that are truly useful for astronomical datasets are relatively shallow due to much less feature diversity and complexity, the main challenges being placed in other aspects of the method design.

\section{Classical YOLO-V2 backbone for SDC1}
\label{sec:appendix:yolo_v2_arch}

\begin{table*}
\centering
\caption{\label{table:arch_comparison}. SDC1 scores and related properties for source catalogs from different models and architectures.}
\begin{tabular}{ l c c c c c c c}
 \hline
 \hline
 Method & $M_s$ (Score) & $N_{\rm det}$ & $N_{\textrm{match}}$ & $N_{\rm false}$ & $N_{\rm bad} \in N_{\rm false}$ & Purity & $\bar{s}$ \\
 \hline
 \textcolor{black}{darknet-19-regul-scratch}  & 445093 & 688112 & 645626 & 42486 & 15070 & 93.83\% & 0.7552\\
  \textcolor{black}{darknet-19-base-pretrain}  & 423077 & 665304 & 620685 & 44619 & 16003 & 93.29\% & 0.7535\\
  \textcolor{black}{darknet-19-regul-pretrain} & 411541 & 642153 & 600193 & 41960 & 13398 & 93.47\% & 0.7556\\
  \textcolor{black}{darknet-19-base-scratch}   & 386395 & 644371 & 591039 & 53332 & 15020 & 91.72\% & 0.7440\\
 \hline
\end{tabular}
\end{table*}

\textcolor{black}{When exploring the suitability of a given method to a new problem, the classical approach is to look at how this method was applied to a similar problem. This is especially true for network architectures and hyperparameters since exploring these aspects from scratch can be extremely costly in terms of time and computing resources. If the new application is sufficiently close to other well-known applications of the same method, one could even use pretrained models to save some training time or improve the performance of their new model. For example, it is the approach used in \citet{paper:deepl_source_finding} where they test the capability of a YOLO architecture on a source detection problem. They conclude that the results were improved when using a pretrained model. However, achieving apparently good results with known architectures or with pretrained models does not guarantee that the new application will have close to optimal results. In cases where it achieves bad results, it can even lead to the false conclusion that the whole approach is inefficient while only the backbone architecture might be unsuited for the task. To participate in this global discussion, we explore in this section the results we obtained on the SDC1 dataset using the classical YOLO-V2 darknet-19 architecture in a naive way and compare it to our reference result with our custom architecture.}

\subsection{Selection of architectures and training setups}
\label{sec:appendix:yolo_v2_arch:training_setups}

While the classical YOLO-V2 formalism could be emulated with our implementation, \textcolor{black}{doing a thorough and documented comparison of the individual effects of all the differences would be too complicated.} In this section, we preserved all the data preparation, training setup, hyperparameters selection, detection units setup, and prediction pipeline construction that we defined in Sect.~\ref{sec:dataset_and_training} and studied variations in the network backbone and starting weights. \textcolor{black}{To achieve the same reduction factor of 16 as our reference SDC1 model with pretrained darknet-19 backbones, it is necessary to cut the end part of the darknet-19 architecture before it reaches a reduction factor of 32.}

\textcolor{black}{We compare two slightly different architectures. The first one is obtained by taking the first 13 convolution layers of our darknet architecture pretrained on the COCO dataset and adding three new convolutional layers with 1024 filters of size $3{\times}3$, each followed by a group normalization. We then add our detection layer. This first architecture is referred to as darknet-19-base. The second one uses the same pretrained part but then adds layers to recreate the detection head structure of our custom architecture consisting of the last four layers in Fig.~\ref{fig:network_architecture}. This architecture has a more progressive increase in size and includes dropout regularization. This second architecture is referred to as darknet-19-regul. It is essential to note the darknet-19-regul architecture is composed of 14.4 million parameters, which is very similar to our custom architecture. In contrast, the darknet-19-base is almost twice the size with a total of 28.3 million parameters, which is explained by the differences in the last four layers. Regarding computing performances, the darknet-19-regul architecture has almost the same performances as our custom architecture, while the darknet-19-base is around 20\% slower.}

We performed two independent training for each architecture, the first using the pretrained COCO weights for the corresponding layers \textcolor{black}{(-pretrain)} and the second from scratch using classical random weight initialization for all layers \textcolor{black}{(-scratch)}. Therefore, we have a total of four trained models to compare. We use the same prediction pipeline as our reference model for all of them, following the description in Sect.~\ref{sec:detector_inference}.

\subsection{Results and discussion}
\label{sec:appendix:yolo_v2_arch:results_disc}

The best score results for each of the four models are presented in Table~\ref{table:arch_comparison}. \textcolor{black}{We can first observe that the best one has a score $7\%$ lower than our reference model from Table~\ref{table:team_scores}, which reaches better scores after only 400 iterations (Fig.~\ref{fig:loss_curve_with_score}). We stress that the results are reproducible and stable over multiple training with different initial weight values. In addition, the difference in score between models is significant as it is higher than the typical score fluctuation observed over multiple training of the same setup.} From these results, we see that for the darknet-19-base architecture, pretraining improves the detection results, while it degrades it for the lighter darknet-19-regul architecture. The apparent flux distribution for these four models is very similar to our reference model. The main difference between these models is their completeness at a given purity, allowing some of them to detect more sources. We also note that models that achieve the best score have the highest characterization accuracy. This is expected, as the predicted flux accuracy is one of the match criteria.

Our interpretation is that the truncated darknet-19 architecture is not adapted to our application and encapsulates too much expressivity to be properly constrained by our training data. It appears that the pretrained weights do not encapsulate relevant information but still act as a startup regularization by reducing the accessible parameter space, preventing the network from over-fitting. The training of the darknet-19-base-scratch model is very unstable as we observe variations of about 100000 points between successive control steps when continuing the training after the best score iteration, which is a striking proof of overfitting. This model exhibits the poor results of an unconstrained network, while the darknet-19-regul-scratch demonstrates that adding a small amount of structural regularization solves this issue better than pretraining. In our opinion, the most robust demonstration of the irrelevance of the features learned on COCO for our problem is that the darknet-19-regul-scratch model significantly outperforms the darknet-19-regul-pretrain model. In addition, we tested a variety of alternative setups around our darknet-19-regul architecture by changing the number of layers that start from their pretrained weights, \textcolor{black}{the rest being initialized from scratch.} We observed that the best achievable score degrades as the number of pretrained layers increases. In this regard, we consider the common empirical observation about obtaining better results using pretrained models as incomplete and possibly misleading regarding the relevance of low-level features learned from everyday-life images for astronomical datasets \citep[e.g.,][]{paper:deepl_source_finding, paper:claran}. In addition, pretrained models do not necessarily train faster than the others on this specific application, which is another sign of the nonrelevance of the features they are starting from. 

Finally, achieving better results with our custom backbone indicates that architectures that have been proven efficient for everyday-life images are not necessarily suited for astronomical images. A striking difference between the two architectures is the absence of inter-layer normalization. We already identified in Sect.~\ref{sec:cnn_backbone} that using group normalization between most layers in our custom architecture significantly reduces our best achievable score because it alters the absolute value of the input pixels, which contains information about the source flux and the input dynamic. Still, group normalization is not enough to explain the lower result of the darknet-19-regul-scratch model (Sect.~\ref{sec:appendix:yolo_v2_arch}), highlighting that other architectural design choices matter.

\section{Alternative training area for the SDC1}
\label{sec:appendix:training_area}

\begin{table*}
\centering
\caption{\label{table:alt_train_arch_comparison}. Scores and related properties based on the alternative training area for source catalogs from different models.}
\begin{tabular}{ l c c c c c c c}
 \hline
 \hline
 Method & $M_s$ (Score) & $N_{\rm det}$ & $N_{\textrm{match}}$ & $N_{\rm false}$ & $N_{\rm bad} \in N_{\rm false}$ & Purity & $\bar{s}$ \\
 \hline
YOLO-CIANNA-alt-train-large-bootstrap  & 502656 & 723655 & 690894 & 32761 & 12991 & 95.47\% & 0.7750\\
 YOLO-CIANNA-alt-train-large  & 486053 & 709689 & 674404 & 35285 & 14380 & 95.02\% & 0.7730\\
 YOLO-CIANNA-alt-train  & 484145 & 708954 & 672906 & 36048 & 14988 & 94.91\% & 0.7731\\
 \hline
\end{tabular}
\end{table*}

\begin{figure}
    \centering
    \includegraphics[width=0.95\hsize]{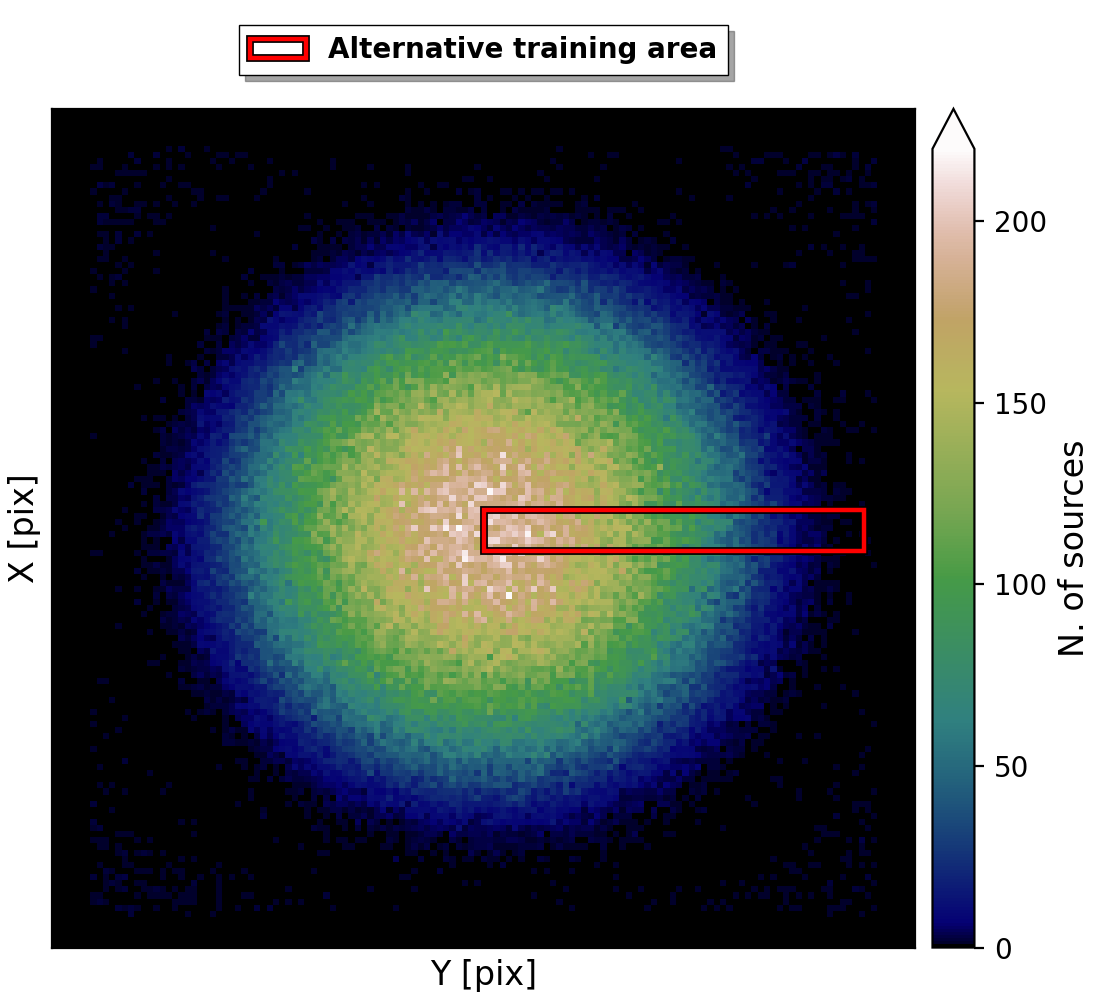}
    \caption{Two-dimensional histogram of the central coordinates of the sources from the full True catalog that pass our selection function. The red box indicates our alternative training area.}
    \label{fig:alt_training_sample}
\end{figure}

\subsection{Reference architecture}
\label{sec:appendix:training_area:same_arch}

\begin{figure}
    \centering
    \begin{subfigure}{\hsize}
    \caption*{\textbf{Alt. training region hand-made selection function}}
    \includegraphics[width=1.0\hsize]{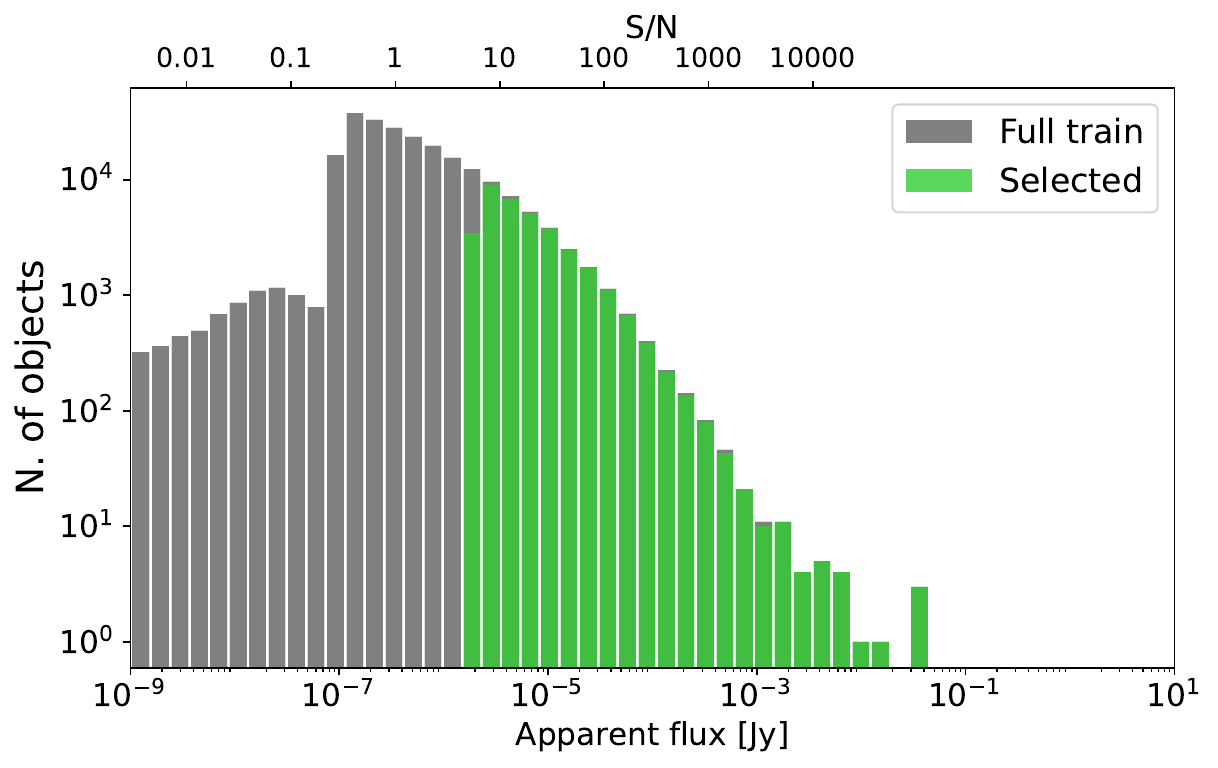}
    \end{subfigure}\\
    \begin{subfigure}{\hsize}
    \caption*{\textbf{Alt. training region bootstrap selection function}}
    \includegraphics[width=1.0\hsize]{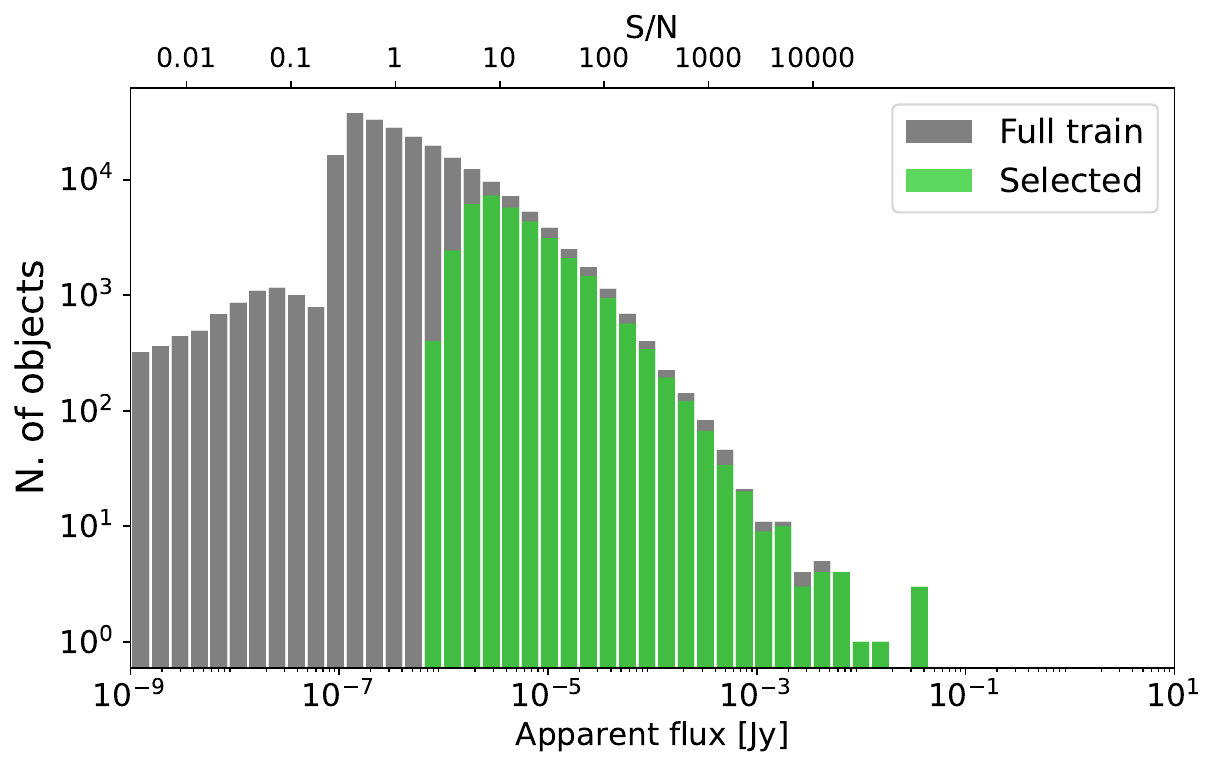}
    \end{subfigure}
    \caption{Source flux distribution histogram in log scale and using log-bins for the True catalog and the selected sources in our alternative training. The \textit{top} frame represents the classical selection function, and the \textit{bottom} frame represents the bootstrap selection function after a single iteration.}
    \label{fig:alt_select_function_dist_flux}
\end{figure}

\textcolor{black}{We exposed in Sect.~\ref{sec:sdc1_training_area} the issues that arise from having access to only a small subpart of the full image to train our detector. This section explores an alternative definition for the training area that is more representative of the image field variations.} We use the post-challenge access to the full True catalog and define a new training area that spans from the center of the image up to the right edge, representing a radius of the primary beam. We do not add extra noise-only regions since they should not be required anymore. \textcolor{black}{To evaluate only the effect of the training region, we adjust the area around the selected radius so the number of selected sources remains similar to our reference case (Sect.~\ref{sec:sdc1_selection_function}). Our alternative training area is then a rectangular region with $-2.80 \leq \textrm{RA} \leq 0.00$ degrees and $-30.12 \leq \textrm{Dec} \leq -29.88$ degrees as illustrated in Fig.~\ref{fig:alt_training_sample}, and that contains 34913 selected sources. Figure~\ref{fig:alt_select_function_dist_flux} represents the apparent flux distribution of the selected sources, which is similar to the distribution we obtained with the default training region.}

\begin{figure*}
    \centering
    \includegraphics[width=0.95\hsize]{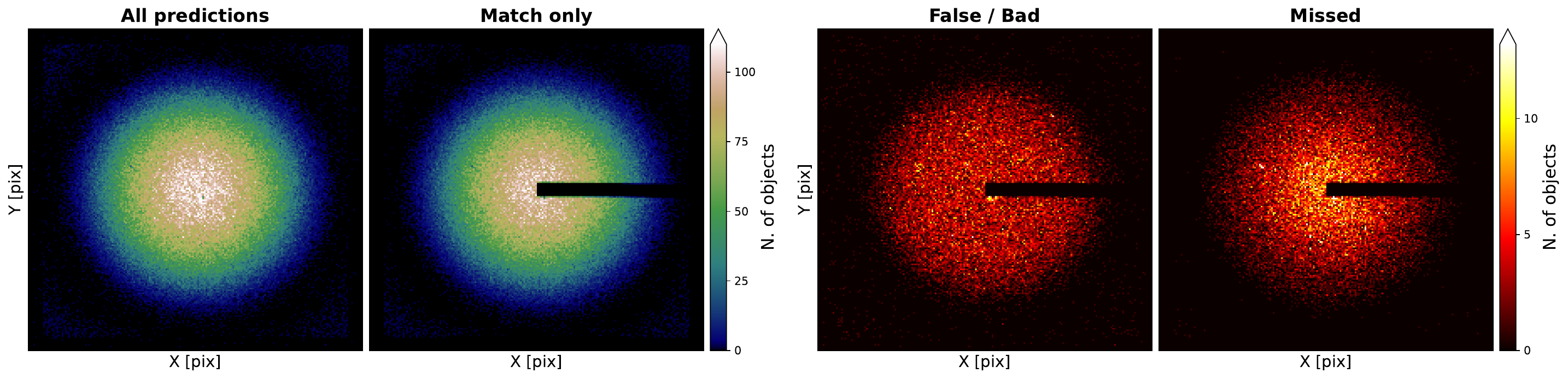}
    \caption{Two-dimensional histograms of various kinds of ``objects'' representing their distribution over the full SDC1 field  \textcolor{black}{for the YOLO-CIANNA-alt-train model. The training area is masked when necessary. All densities are binned using the same 200x200 grid.} The match and false detections are made using the DIoU-based matching metric, while the missed ones are based on our selection function. The central coordinates of the field are $\textrm{RA}=0$ deg, $\textrm{Dec}=-30$ deg.}
    \label{fig:alt_pred_match_false_field_map}
\end{figure*}

\begin{figure}
    \centering
    \begin{subfigure}{\hsize}
     \caption*{\textbf{YOLO-CIANNA-alt-train}}
    \includegraphics[width=1.0\hsize]{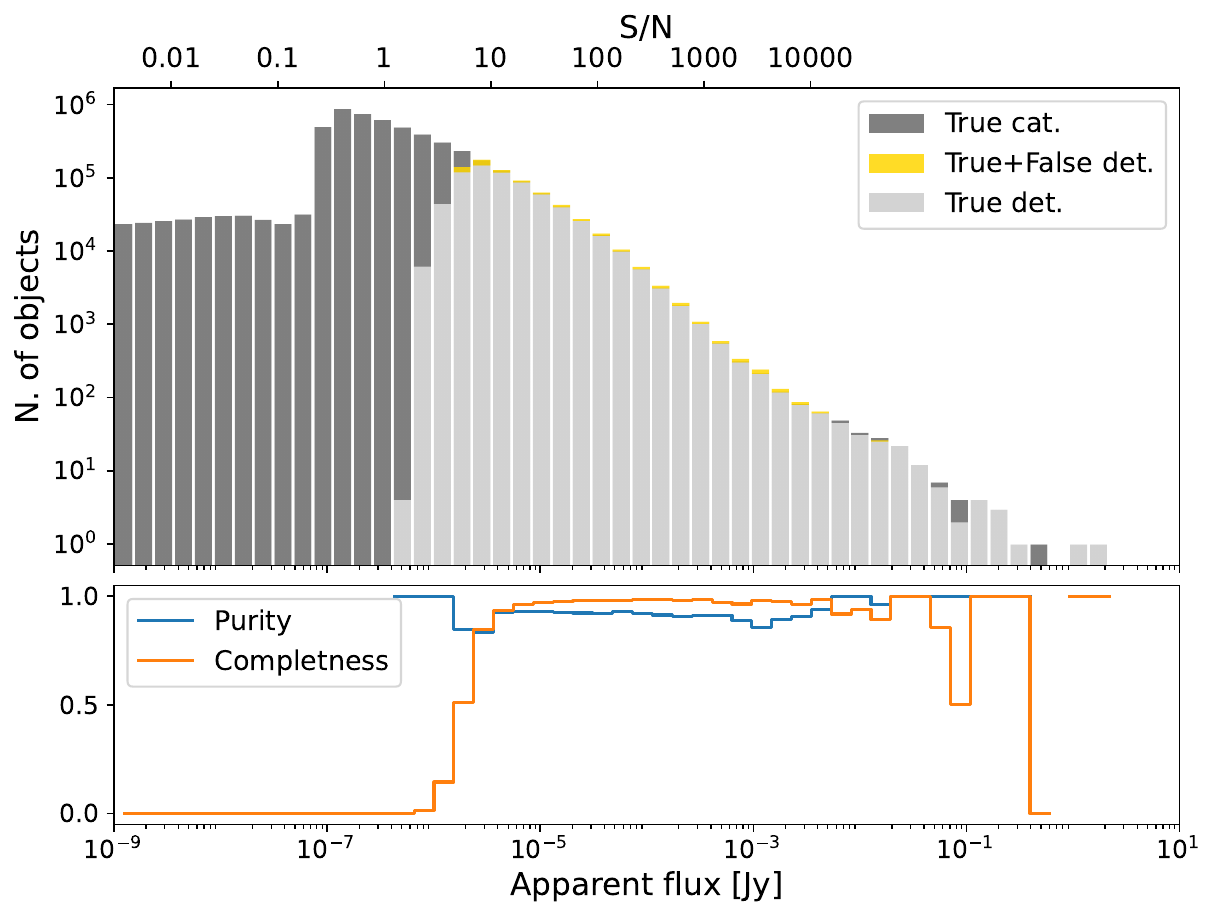}
    \end{subfigure}\\
    \begin{subfigure}{\hsize}
     \caption*{\textbf{YOLO-CIANNA-alt-train-large-bootstrap}}
    \includegraphics[width=1.0\hsize]{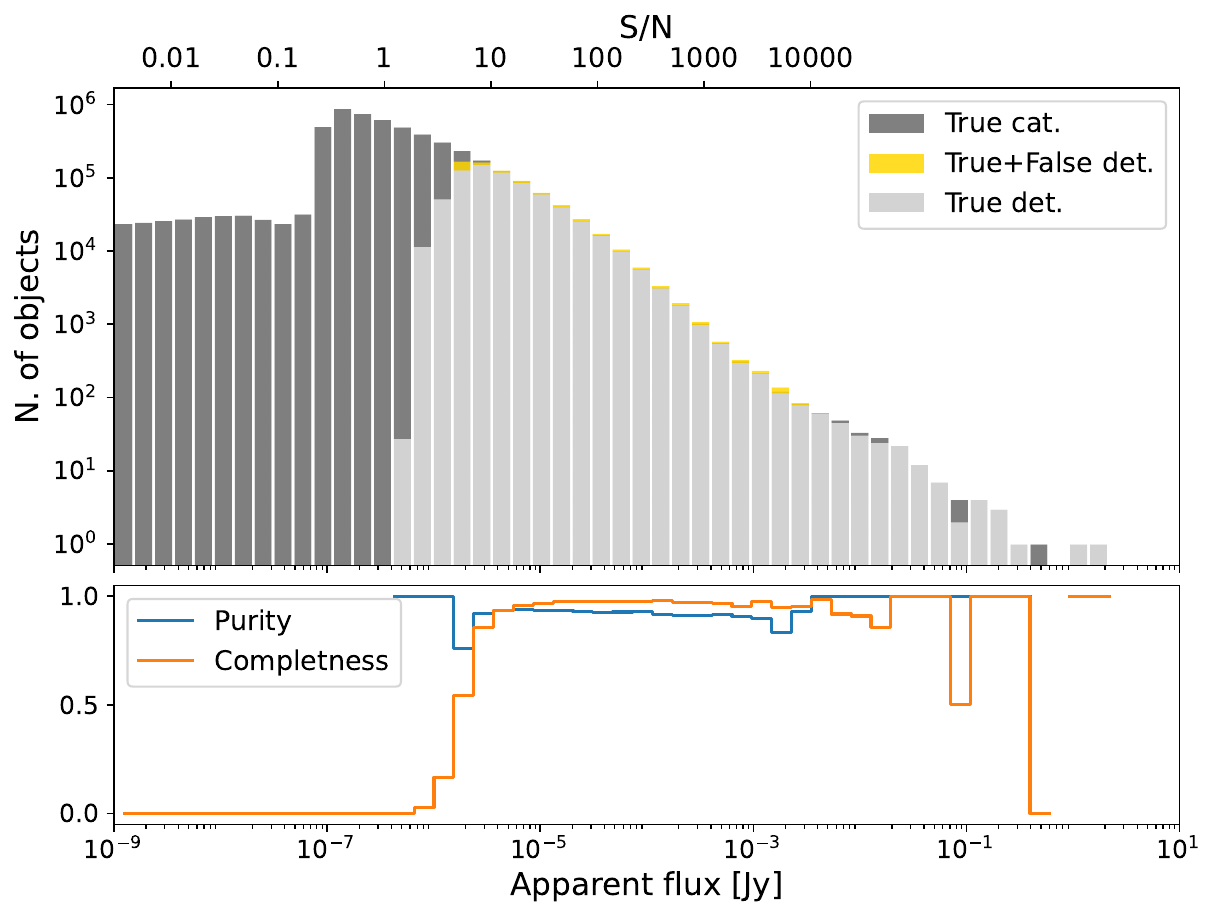}
    \end{subfigure}
    \caption{Histograms of the sources as a function of their apparent flux using logarithmic bins for the underlying True catalog and predicted sources in our modified testing area. The bottom part of each frame represents the purity and completeness of each bin of the histograms. Top and bottom frames represents the alt-train and alt-train-large-bootstrap models respectively.}
    \label{fig:alt_flux_distribution_full_false_true}
\end{figure}

We train our network using the same configuration as our reference model and only change the training sample, which results in the YOLO-CIANNA-alt-train model. The prediction pipeline remained unchanged, but we altered the SDC1 scorer code to modify the definition of the training area. \textcolor{black}{We obtain the score presented in Table~\ref{table:alt_train_arch_comparison} and a list of matching sources to produce the same diagnostics we did on our reference catalog.} While the scores cannot be directly compared due to the different testing areas, the number of detected sources is relatively similar to our reference catalog. The purity is even slightly improved by 0.70\%. The histogram of the apparent flux of the detected sources is presented in Fig.~\ref{fig:alt_flux_distribution_full_false_true}, exhibiting a similar distribution to our reference result. \textcolor{black}{The average source score is also slightly higher, thanks to the presence of sources from all the primary beam regimes in the training sample. Looking at individual fields, this new model produces results that are visually similar to our reference model. However, Fig.~\ref{fig:alt_pred_match_false_field_map} shows that the distribution of false positive detections over the full image field is more homogeneous. It demonstrates that this alternative training sample allows for better generalization without requiring extra noise regions.}

\subsection{Larger architecture and bootstrap training}
\label{sec:appendix:training_area:larger_and_bootstrap}

\textcolor{black}{An interesting side effect of this alternative training area is that the network training is more stable. We tried providing} a bit of extra expressivity with less regularization by replacing the second last layer of our default architecture with two $1{\times}1$ convolutional layers, the first one with 3072 filters and a dropout of 30\% and the second one with 2048 filters, constructing the YOLO-CIANNA-alt-train-large model. Training this new architecture with our alternative training area only modestly improves the score, as presented in Table~\ref{table:alt_train_arch_comparison}. This improvement can not be considered significant since it is of the same order as our typical multiple-training-identical-setup variability.

The combination of a more representative training sample and a stronger network makes it possible to use a bootstrap approach for refining our selection function. \textcolor{black}{Instead of deciding which objects from the True catalog should be considered detectable, we land this decision on the detector itself. Using a model trained with a hand-made selection function for a few hundred epochs, we can produce a catalog of predicted sources with an objectness score over the training region. This catalog can then be cross-matched with the underlying list of target objects from the simulation to construct a new training sample. Targets that the detector fails to detect can be removed from the training sample, and targets that are detected outside our hand-made selection function can be added back.} We stress that this process is never used to directly inject a detector prediction as a new target. The properties from the True catalog sources that were successfully detected are used to define the new training sample. This approach requires stable training, as selecting a prediction before network convergence or when the network overtrains can harm the self-constrained selection function. This approach can be repeated several times using the successively trained detectors to refine the selection function for the next training.

We tested this approach with our reference architecture using the default SDC1 training area, but it did not improve the result significantly. With the default training area, any change in our hand-made selection function results in significant variations in the best achievable score. This variability mainly comes from the intermediate primary beam regime in which the detector has to interpolate from the discontinuous training area. This effect is corrected with the alternative training area, and the bootstrap training approach actually becomes useful. \textcolor{black}{It is especially relevant in this context where the visibility of a source is strongly dependent on the beam regime, which was less the case for the central training area with less sensitivity variation. In such a context, having the detector provide information on which sources should be detectable is especially useful.}

Using our alternative training area and our larger architecture alternative, the bootstrap approach significantly improves the results after only a single step. We present the resulting selection function over the apparent flux histogram in Fig.~\ref{fig:alt_select_function_dist_flux}. Compared to our handmade custom function, it removes targets for almost all apparent flux values. The most plausible explanation is that it removes blended sources \textcolor{black}{that the detector cannot separate}. Another effect is to smooth the low-cut flux in a wider flux interval. The score of the resulting model is indicated in Table~\ref{table:alt_train_arch_comparison}, and the predicted sources flux distribution in Fig.~\ref{fig:alt_flux_distribution_full_false_true}. While all the diagnostics show results that are very similar to what we obtain with our reference architecture on our alternative training sample, we observe a significative score improvement of more than $3\%$, correlated with a better flux characterization score with $\bar{s}^{\textrm{flux}}=0.680$. More improvement would likely be achievable by searching the optimum architecture and hyperparameters setup with this new training sample.

Overall, this approach removes the necessity for a well-defined selection function. This could be useful in a highly dimensional input space where defining a selection function can be difficult. With this approach, the starting selection function could be the catalog of a less performant detection method. Still, it requires access to all the possible targets regardless of their visibility, so only real detections are kept at each step. While a simulated dataset is a best-case scenario for this approach, it could still be employed on observed datasets by confirming the proposed new detections with additional observations.

\section{Impact of simultaneous detection and characterization on detection-only perfomances}
\label{sec:appendix:characterization_impact}

\textcolor{black}{In Sect.~\ref{sec:challenge_and_score_disc:alternative_metrics}, we introduce the idea that merging the detection and characterization inside a single loss could result in suboptimal performances. In practice, it is possible to separate both tasks by having a pure object detector and a dedicated characterization network. We could even train an independent characterization network for each parameter to predict. However, we observed that predicting correlated parameters in a single network produces better results for both parameters. For this reason, we expect a combined detection and characterization network to achieve better results as long as the parameter to predict correlates with the detectability of the object. The aim of this section is to verify this point for our method over the SDC1.}

\textcolor{black}{To compare models on a detection-only task, we can use the purely geometrical metric defined in Sect.~\ref{sec:challenge_and_score_disc:alternative_metrics} based on DIoU matching criteria. To evaluate the results at different detection quality requirements we use multiple DIoU threshold values of 0.1, 0.5, and 0.9. To reduce the matching time, predictions below a too-small objectness threshold will be removed. This threshold depends on the detection units and is set to 0.1 for the six small-size regimes and 0.05 for the three intermediate and large-size regimes. This will result in lower AP values than in Sect.~\ref{sec:challenge_and_score_disc:alternative_metrics}, but the respective performances of different models will remain comparable. The results are compiled in Table~\ref{table:no_param_refine}.}

\textcolor{black}{One issue with a purely geometrical match is that it increases the chances of a random match compared to the SDC1 match criterium that accounts for the flux. Before training any model, we evaluate the chances of random matches. For this, we build a density map from the raw $\sim$950000 detections from our reference model. This map is used as a prior for drawing an equivalent number of sources with a random position, size, and objectness score that all follow the distributions of our reference model. The resulting ``random 950k'' catalog is passed through the scoring pipeline, and the results are reported in the same table. The number of random matches for $\textrm{DIoU} > 0.5$ is near $3\%$. However, this is not a good representation of our detection catalog as it comprises many detections overlapping over the same object. As stated in Sect.~\ref{sec:results:patched_results}}, almost 80\% of the detections are removed by the NMS process that expects interdependency between the detections. Therefore, we build a second random catalog with only 110000 random sources, corresponding to the typical number of objects from our catalog that are not removed and do not match at the end of the post-process pipeline for a DIoU threshold of 0.1. This time, the obtained AP values are likely good approximations of the uncertainty of the AP values over the detected sources catalogs at their respective thresholds.

\textcolor{black}{All the metric scores are reported for our default model with the prediction of the extra parameters as a reference (With param. reference). Our first test deactivates the optimization of the extra-parameters part of the loss while keeping the same architecture and training process (No param.), which is just a toggle switch in our CIANNA framework. As expected, the geometrical score decreases for all DIoU matching values, and the number of detected sources is also reduced. One alternative explanation would be that the network expressivity is now too high for the simplified task, which would also explain why the optimal epoch is found earlier than for our reference model. To test this hypothesis, we train two new models that slightly alter our reference architecture: one where we increase the dropout rate on the before-last layer (No param. more regular), and a second where we reduce the number of filters in the same layer by about 30\% (No param. light). Both models perform lower than the simple ``No param'' case at most DIoU thresholds. This indicates that predicting the extra parameters does have a direct positive impact. Still, we must verify if this effect is generic or if we mitigated the possible negative impact with our quality conditional fitting that results in the extra-parameter cascading loss (Sect.~\ref{sec:method_loss}). We test this hypothesis by retraining our reference model with the extra-parameter prediction activated but setting $L^{{\rm fIoU}}_{p} = -1$ (With param. no cascade). The results are lower than with the parameter deactivated, so predicting parameters for poorly detected sources has a negative impact. In summary, we conclude that extra-parameters can have a positive impact when predicting correlated quantities but only when the negative contribution from unreliable detection is mitigated in the corresponding loss subpart.}

\textcolor{black}{Finally, we perform a few extra tests to verify if the lower performance observed when not predicting the extra parameters was only caused by a bad input normalization or augmentation design. For this, we use a per-input patch normalization instead of our global normalization. We train two models with (With param. patch norm) and without (No param. patch norm) the parameter prediction enabled. As expected, the performance of the model with the parameter prediction is lowered, while the performance of the model without the parameter prediction is improved. However, our reference model, which combines parameter prediction and global normalization, remains better. As a last test, we use two input channels corresponding to the two normalization types and train new models with and without the parameter prediction (With param. dual norm. and No param. dual norm.). The detection performances are improved for both models and even surpass our reference model on the detection-only metric, but again, keeping the parameter prediction is better. Still, from the SDC1 metric perspective, our reference score remains higher than with the dual normalization input. This is likely the result of a lack of expressivity in the upper layers of the network that now have to handle two inputs with a limited number of parameters, negatively impacting the source characterization. These results strongly demonstrate that simultaneously predicting correlated parameters with a dynamic loss is a viable approach for improving detection-only performances.}

\textcolor{black}{Still, all these results were obtained for a single architecture with only small variations, a single set of hyperparameters, and one dataset. Empirical verification on a per-application basis is advised. Also, we acknowledge that global normalization is only possible in the context of a constant noise image from a single acquisition on a unique instrument. If the objective were to construct a generic source-detector, we would either have to i) input more context information about the instrumental setup or ii) adopt the per-input-patch normalization that forces the detector to base its detection on morphological contrast between the objects and the noise structure or artifacts. The resulting model would likely lose performances in a specific setup but would be more generic and resilient to new structures in the images.}

\begin{table*}
\centering
\caption{\label{table:no_param_refine} Detection-only score metrics for various approaches with and without the extra-parameter part of the loss.}
\begin{tabular}{ l c c c c c c c | c c c}
 \hline
 \hline
 Approach & Epoch &$\textrm{AP}_{0.1}$ & $\textrm{AP}_{0.5}$ & $\textrm{AP}_{0.9}$ & $N^{m}_{0.1}$ & $N^{m}_{0.5}$ & $N^{m}_{0.9}$ & Score & $\textrm{AP}_{SDC1}$ & $N^{m}_{SDC1}$\\
 \hline
With param. reference   & 3000 & 15.71 & 13.57 & 0.70 & 835789 & 737056 & 120467 & 480439 & 14.47 & 679778 \\
With param. no cascade  & 4200 & 14.94 & 13.12 & 0.69 & 791421 & 708597 & 117005 & 476943 & 14.24 & 676288\\
With param. patch norm. & 3200 & 15.25 & 13.05 & 0.71 & 811351 & 711355 & 117859 & 461417 & 14.08 & 654838\\
With param. dual norm.  & 3600 & 15.96 & 13.74 & 0.73 & 857652 & 751529 & 126662 & 473758 & 14.35 & 657853\\
 \hline
No param.               & 2600 & 15.12 & 13.05 & 0.58 & 801494 & 709853 & 108664 & \textrm{N/A} & \textrm{N/A} & \textrm{N/A} \\
No param. more regul.   & 3000 & 14.91 & 13.06 & 0.73 & 789935 & 707086 & 119189 & \textrm{N/A} & \textrm{N/A} & \textrm{N/A} \\
No param. light         & 3000 & 14.93 & 12.89 & 0.67 & 791478 & 699148 & 114422 & \textrm{N/A} & \textrm{N/A} & \textrm{N/A} \\
No param. patch norm.   & 2600 & 15.62 & 13.18 & 0.64 & 836355 & 723201 & 113519 & \textrm{N/A} & \textrm{N/A} & \textrm{N/A} \\
No param. dual norm.    & 2600 & 15.84 & 13.52 & 0.62 & 847780 & 740967 & 116617 & \textrm{N/A} & \textrm{N/A} & \textrm{N/A} \\
 \hline
Random 950k     & \textrm{N/A} &  0.40 &  0.01 & 0.00 & 137877 &  26416 &    633 & \textrm{N/A} & \textrm{N/A} & \textrm{N/A} \\
Random 110k     & \textrm{N/A} &  0.05 &  0.00 & 0.00 &  17152 &   3306 &     75 & \textrm{N/A} & \textrm{N/A} & \textrm{N/A} \\
\hline
\end{tabular}
\end{table*}

\end{document}